\documentclass[fleqn,usenatbib]{mnras}

\usepackage{newtxtext,newtxmath}

\usepackage[T1]{fontenc}

\DeclareRobustCommand{\VAN}[3]{#2}
\let\VANthebibliography\thebibliography
\def\thebibliography{\DeclareRobustCommand{\VAN}[3]{##3}\VANthebibliography}

\usepackage{graphicx}	
\usepackage{amsmath}	
\usepackage{bm}

	\newcommand{\msun}{\,{\rm M}_{\odot}}

    \newcommand{\Myr}{\,{\rm Myr}}
    \newcommand{\Gyr}{\,{\rm Gyr}}
    
    \newcommand{\Msun}{\,\mathrm{M}_{\odot}}

    \newcommand{\J}{\mathcal{J}}
    \newcommand{\Ln}{\hat{\bm{\L}}}
    \renewcommand{\L}{\bm{L}}

    \newcommand{\Out}{\rm out}

    \newcommand{\IMBH}{\bullet}
    \newcommand{\MSMBH}{M} 
    \newcommand{\orb}{\rm orb}    
	\def\gapprox{\;\rlap{\lower 3.0pt                       
			\hbox{$\sim$}}\raise 2.5pt\hbox{$>$}\;}
	\def\lapprox{\;\rlap{\lower 3.1pt                       
			\hbox{$\sim$}}\raise 2.7pt\hbox{$<$}\;}
	
\title[Stellar discs and IMBHs in galactic nuclei]
{Stellar discs and intermediate-mass black holes in galactic nuclei --\\I. 
Fragmenting the disc in an isotropic stellar potential
}

\author[Panamarev, Zou \& Kocsis]{
Taras Panamarev,$^{1,2}$\thanks{Corresponding author email: panamarevt@gmail.com} 
  Xiang Zou,$^{1,3}$ 
  and Bence Kocsis$^1$ \\
   \\
$^1$ Rudolf Peierls Centre for Theoretical Physics, Parks Road, OX1 3PU, Oxford, UK\\
$^2$ Fesenkov Astrophysical Institute, Observatory 23, 050020 Almaty, Kazakhstan\\
$^3$ Department of Physics, University of Toronto, 60 St. George Street, Toronto, ON, M5S 1A7, Canada\\
}

\date{Accepted XXX. Received YYY; in original form ZZZ}

\pubyear{2024}

\begin{document}
\label{firstpage}
\pagerange{\pageref{firstpage}--\pageref{lastpage}}
\maketitle

\begin{abstract}
The origin of the complex orbital structure of young massive stars at the Galactic centre remains an open question. If these stars formed in a single episode from a gaseous accretion disc, 
they may initially have constituted a single, coherently rotating stellar disc.
We investigate whether perturbations from an unseen intermediate-mass black hole (IMBH) could fragment and/or disrupt such a disc into the multiple orbital components 
observed today. First, we derive a theoretical criterion for when and where the IMBH's torque overcomes the disc's self-torque and tears it apart. We then test this picture with direct $N$-body simulations of a stellar disc interacting with an inclined IMBH around a central supermassive black hole.
We find that the outcome depends strongly on the IMBH's orbit and mass. A prograde IMBH rapidly aligns with the stellar disc, while a massive retrograde IMBH ($m_{\IMBH} \simeq 0.67\,M_{\rm d}$) 
anti-aligns relative to the radially overlapping stars and efficiently fragments the original disc into three components in angular-momentum space: an inner disc, a misaligned overlapping region, and an unperturbed outer disc. The IMBH also excites eccentricities in the overlapping region, driving stars away from initially circular orbits. These features emerge for an IMBH mass of $2000\,\Msun$ and a disc mass of $3000\,\Msun$ within 10--20 Myr, a timescale comparable to the age of the young Galactic centre stellar population, and provide a plausible explanation for the observed multiple orbital planes, warped geometry, and broad eccentricity distribution.
\end{abstract}




\section{Introduction}\label{sec:INTRO}

The existence of stellar-mass black holes ($m_{\IMBH} < 100 \msun$) and supermassive black holes suggests the presence of intermediate-mass black holes (IMBHs, $100$--$10^5 \msun$, see \citealt{Greene2020} for a review). An IMBH may form through hyper-Eddington growth of Population III remnant black holes in early
protogalaxies \citep{Madau_Rees2001}, through runaway collisions in a dense star cluster \citep{PortegiesZwart_McMillan2002,Portegies_Zwart+2006,O'Leary+2006,Tagawa+2020}, dynamical interactions of binaries containing a stellar-mass black hole \citep{Giersz+2015,Rizzuto2021}, gas fragmentation, accretion, and mergers in active galactic nuclei \citep{2004ApJ...608..108G,McKernan2012,McKernan+2014}. In galactic nuclei, the densest known stellar systems in the Universe, stellar-mass black holes can evolve into IMBHs through frequent collisions with surrounding main-sequence stars, potentially forming IMBHs as massive as $10^4 \Msun$ \citep{Rose2022}. Alternatively, they might form elsewhere and subsequently be transported and deposited into galactic nuclei. One mechanism for this pathway is through dynamical friction-induced migration of IMBHs initially formed in globular clusters, which can be tidally disrupted and absorbed into the nuclei of galaxies \citep{Tremaine1976, Capuzzo-Dolcetta1993, Capuzzo-Dolcetta2008, Ishchenko2023, Ishchenko2024}. 

Many of these formation scenarios suggest the possibility that IMBHs may be found in the centre of the Milky Way. While there is no direct evidence of an IMBH in the Galactic centre, observations by \citet{Tsuboi2017} have shed light on the IRS13E complex, an infrared object near Sagittarius A*, which might harbour an IMBH. The IRS13E complex is estimated to be at a projected distance of 0.13 pc from the centre \citep{Tsuboi2020}\footnote{who in fact argue the 3D distance is $\gtrsim 0.4$ pc.}.

The study utilised the ALMA telescope to detect ionised gas with a large velocity width and compact size within IRS13E, suggesting a high-eccentricity Keplerian orbit around a mass of approximately $10^4 \msun$. Additionally, \citet{Kaneko2023} reported on another potential IMBH candidate, located further away from Sgr~A*. This candidate is associated with a peculiar compact cloud, featuring a steep velocity gradient and a distinct head-tail structure. The kinematics of this cloud suggest a Keplerian motion around a central point-like object of a mass of $1 \times 10^5 \msun$.

On the other hand, recent studies have provided constraints on the presence of IMBHs in the Galactic centre.
The absence of a detectable `wobble' in the proper motion of Sgr~A* has effectively excluded the presence of IMBH companions heavier than approximately \(3 \times 10^4 \msun\) within \(0.003\) to \(0.1\) pc of Sgr~A* \citep{Reid2004, Reid2020, Oyama2024}. Additionally, \citet{Gravity2022} constrain the \emph{extended} mass enclosed within S2's orbit of apocentre $9.4$ mpc to $\lesssim 3000\,\msun$ ($1\sigma$). 
The presence of an IMBH within the S-star cluster could lead to significant perturbations in the orbital motions of the S-stars. These perturbations include von Zeipel--Lidov--Kozai oscillations \citep{2016ARA&A..54..441N}, the scattering of stars into loss-cone orbits \citep{Perets2007}, and the ejection of stars from the region \citep{YuTremaine2003}.
Thus, from the orbital motion of stars in the S-star cluster \citep{GualandrisMerritt2009, Gravity2023}, 
in particular S2 \citep{Gualandris2010, Naoz2020, Will2023}, one can further constrain the parameter space for IMBHs. 
\citet{Gravity2023} used S2's orbit to rule out IMBHs with masses $\gtrsim 2000\,\msun$ on orbits with 
semi-major axes smaller than S2's.\footnote{S2 has semimajor axis $a_{\rm S2}\simeq 5~\mathrm{mpc}$; and eccentricity $e_{\rm S2}\simeq 0.87$} 

Similarly, \citet{Will2023} analysed 23 years of astrometric and radial-velocity data on S2's orbit to exclude IMBHs with masses between \(10^3\) and \(10^5 \msun\) at orbital separations of $\sim200$--$4000$ au ($\sim1$--$20$ mpc) from Sgr~A*. \citet{Evans2023} used observations of hypervelocity stars from Gaia data to place an upper limit on the mass of a possible companion to Sgr~A*. Their analysis rules out a companion more massive than 1000 \(\msun\) within one milliparsec of Sgr A*, contributing to the overall constraints on the IMBH parameter space in the Galactic centre. Additionally, a close massive companion would have merged with Sgr~A* by emitting gravitational radiation \citep{GualandrisMerritt2009}. In the future, the Laser Interferometer Space Antenna (LISA) could further refine these constraints, particularly for IMBHs with masses between \(10^3\msun\) and \(10^5\msun\) at distances of 0.1--2 mpc from the SMBH \citep{Strokov2023}. 

Previous results summarised above effectively exclude a broad spectrum of IMBH masses and orbits within the S-star cluster, extending up to 0.04 pc. Beyond this region, IMBHs with masses exceeding \(10^4 \msun\) within a 0.1 pc radius are precluded, and IMBHs of lower masses remain unconstrained in these extended regions. Crucially, adjacent to the S-star cluster lies a disc of young massive stars, spanning distances from 0.04 to 0.5 parsecs \citep{LevinBeloborodov2003, Yelda2014, vonFellenberg2022, Jia2023}, offering a fertile ground for further investigation into the dynamical interactions between a putative IMBH and the stellar disc. Given the estimated age range of these stars -- approximately 6--10 Myr \citep{LevinBeloborodov2003, PaumardEtAl2006, Habibi2017, Bartko2010} -- any dynamical interaction with an IMBH would likely manifest within this timespan. The dominant mechanism in this context is vector resonant relaxation (VRR), which governs the evolution of angular momentum vector orientations within the system \citep{Rauch1996, Kocsis2011, Kocsis2015, Panamarev2022}. Of particular relevance is the concept of resonant dynamical friction (RDF), a process in which the orbit of an IMBH gradually aligns with the stellar disc. This alignment rate is predominantly influenced by the IMBH's mass, its initial orbital inclination, and the local density profile of the stellar disc \citep{Szolgyen2021,Ginat2023}. In the case of the Milky Way's disc of young stars, an IMBH of approximately 250 \(\msun\) with an orbital inclination of 45$^\circ$ is predicted to align within 5--6 Myr \citep{Szolgyen2021}. Furthermore, \citet{Ginat2023} demonstrated that such a perturber induces a notable alignment in the longitudes of the ascending nodes for stars within the disc, ultimately leading to the orbital angular momentum alignment of the IMBH with the stellar disc itself.

The inference of a stellar disc's presence in the Galactic centre has been primarily derived from the distribution of the angular momentum vector directions of young massive stars, as projected onto the sky \citep{LevinBeloborodov2003, Yelda2014}. Recent observations, however, suggest a more complex distribution, revealing several substructures within the projected map of angular momentum vectors \citep{Ali2020, vonFellenberg2022, Jia2023}. 
Exploring the possible influence of an IMBH on this complex stellar distribution allows us both to assess whether such an object could be driving the observed dynamical behaviour and to refine the constraints on its orbital parameters. By linking the IMBH's dynamical imprint to the measured properties of the disc, we can either uncover evidence for its presence or further restrict the range of viable orbital configurations in the Galactic centre.

In this study, we conduct a comprehensive series of direct $N$-body simulations, focusing on the interactions between an IMBH and the stellar disc in the inner region of the Milky Way's Nuclear Star Cluster. Our investigation encompasses a broad parameter space, including various orbital orientations, eccentricities, and IMBH masses. This research builds upon and extends the works of \citet{Szolgyen2021}, \citet{Panamarev2022}, \citet{Ginat2023}, and \citet{Panamarev2025} by exploring a much larger parameter space and significantly enhancing the realism of the models.
We initiate our study with a basic semi-analytical model to develop an intuitive understanding of the expected dynamics. Following this, we employ idealised $N$-body simulations to evaluate the semi-analytical model and thoroughly investigate the parameter space. 

The structure of the paper is as follows: 
Section~\ref{sec:RDF} revisits the concepts of resonant and non-resonant dynamical friction (DF), and outlines the semi-analytical model for the IMBH's interaction with the stellar disc. Section~\ref{sec:simulations} details the numerical methods employed and the initial conditions of the simulations. In Section~\ref{sec:imbh-evol}, we focus on the influence of the stellar disc on the evolution of the IMBH's orbital parameters. Section~\ref{sec:disc-imbh} investigates the effects of the IMBH on the orbital parameters of the stellar disc and discusses our results in the context of observations and previous work, and outlines caveats and future directions. Finally, Section~\ref{sec:SUM} summarises our findings and conclusions.

\section{Resonant relaxation and resonant dynamical friction}
\label{sec:RDF}

\subsection{Hamiltonian of the system}
\label{subsec:hamiltonian}

The VRR Hamiltonian, which represents the gravitational interaction energy of the ring $i$ with all other rings is given by \citep{Kocsis2015}:
\begin{equation}\label{eq:H_VRR}
 H_{i} = -\sum_{j=1, i\neq j}^{N}\sum_{\ell=0}^{\infty} \J_{ij\ell}\,P_{\ell}\big( \Ln_{i}\cdot \Ln_{j} \big),
\end{equation}
where $P_{\ell}$ represents the Legendre polynomial of degree $\ell$, and the unit vectors $\Ln_{i} = \L_{i}/L_{i}$ normal to the orbital planes, $N$ is the total number of rings in the system. The term $\J_{ij\ell}$ represents the constant coupling coefficients, defined as
\begin{equation}\label{eq:Jijl}
\J_{ij\ell} = \frac{ G m_i m_j}{a_{\Out}}
 P_{\ell}(0)^2\, s_{ i  j \ell}\, \alpha_{ij}^{\ell}\,.
\end{equation}
where $\alpha_{ij} = a_{\mathrm{in}} / a_{\mathrm{out}}$, $a$ is the semimajor axis with indices `${\mathrm{out}}$' and `${\mathrm{in}}$' denoting the larger and smaller semi-major axes among $i$ and $j$, respectively. The coefficient $s_{i j \ell}$ is set to 1 for circular orbits.\footnote{\label{fn:chi}For eccentric orbits where the periapsis of the outer orbit is outside the apoapsis of the inner orbit, $s_{ij\ell}\alpha_{ij}^{\ell}=a_{\rm in}^{-1}b_{\rm in}^{\ell+1}b_{\rm out}^{-\ell}P_{\ell+1}(\chi_{\rm in})P_{\ell-1}(\chi_{\rm out})$ where $b_i=a_i\sqrt{1-e_i^2}$ and $\chi_i=a_i/b_i$ and see \citet{Kocsis2015} for the general case.} The Legendre polynomials at zero (in terms of Gamma function) are given by
\begin{equation} \label{eq:pnzero}
P_{2n}(0) = (-1)^n\frac{\Gamma(2n+1)}{2^{2n}[\Gamma(n+1)]^2}, \quad P_{2n+1}(0) = 0.
\end{equation}

Note that the Hamiltonian given by Eq.~\eqref{eq:H_VRR} represents double orbit averaging: over the orbital period \( t_{\orb} \) and the apsidal precession period \( t_{\mathrm{prec}} \gg t_{\orb} \). As a result, semimajor axes and eccentricities of the orbits are conserved, leading to the conservation of the angular momentum magnitudes, but the directions of the angular momentum vectors may change.  This approach is justified for the time-scales of interest: \( t_{\mathrm{prec}} \ll t \ll t_{\mathrm{rel}} \), where \( t_{\mathrm{rel}}\gtrsim 10^8 \)yr is the two-body relaxation time.

Thus, in a galactic nucleus which harbours a central SMBH, a stellar disc, a nuclear star cluster, and an IMBH, the dynamical evolution of the angular momentum vector of star \(i\) can be described by the Hamiltonian:
\begin{equation}
H_i = H_{i,\IMBH} + H_{i,\mathrm{disc}} + H_{i,\mathrm{sphere}},
\label{eq:H_net}
\end{equation}
where:
\begin{itemize}
    \item \(H_{i,\IMBH}\) represents the interaction between star \(i\) and the IMBH.
    \item \(H_{i,\mathrm{disc}}\) denotes the integrated interactions between star \(i\) and all $N_{\text{disc}}$ stars in the disc.
    \item \(H_{i,\mathrm{sphere}}\) is the Hamiltonian for the interactions between star \(i\) and $N_{\text{sphere}}$ stars in the spherical component of the cluster.
\end{itemize}

Following \citet{Kocsis2015}, we can write the equations of motion for the angular momentum vectors:
\begin{align}
 \dot{\L }_i &=   \bm{\Omega}_i \times \L _i, \\
\bm{\Omega}_i &= \bm{\Omega}_{i,\IMBH} + \bm{\Omega}_{i,\mathrm{disc}} + \bm{\Omega}_{i,\mathrm{sphere}} \nonumber\\
&= -\sum_{j=1, }^{N}\sum_{\ell=0}^{\infty} \frac{\J_{ij\ell}}{L_i L_j} P'_{\ell}\big(\Ln_i\cdot \Ln_j\big)\, \L _j .
\label{eq:EOM}
\end{align}

The vector $\bm{\Omega}_i$ is the  angular velocity of the
precession of the angular-momentum vector of star $i$ which is decomposed into the contributions from the three components, i.e. the IMBH, the disc stars, and the stars in the spherical component, respectively. 
$P'_\ell(x)$ is the derivative of the Legendre polynomial. 

In this paper we approximate the spherical component as isotropically distributed (approximated by a smooth analytical potential), neglect a net angular momentum for the spherical component, and neglect fluctuations from isotropy. Fluctuations drive a root-mean-square (RMS) angular momentum precession with rate of order $\Omega_{i,\rm sphere} \sim [M_{i, \rm sphere}/(N_{i, \rm sphere}^{1/2} \MSMBH)] t_{i,\orb}^{-1}$ where $M_{i, \rm sphere}=4\pi r_i^3 \rho_{\rm sphere}(r_i)$ where $\rho_{\rm sphere}$ is the density of the spherical component and $N_{i, \rm sphere} = M_{i, \rm sphere}/m_{\rm RMS}$ where $m_{\rm RMS}$ is the RMS mass of stars in this component \citep{Kocsis2015,Panamarev2022}, and $\MSMBH$ is the SMBH mass. We also neglect a net rotation for the spherical component, which would cause a precession rate of order $\Omega_{i,\rm sphere} \sim a_{\rm rotation} (M_{i, \rm sphere}/\MSMBH) t_{i,\orb}^{-1}$, where $0<a_{\rm rotation}<1$ is the dimensionless angular momentum of the sphere relative to that of the case where the angular momentum vectors of all  stars in the sphere were parallel. Thus, we neglect the sphere terms in Eq.~\eqref{eq:H_net} and ~\eqref{eq:EOM}. In reality the torque generated by the sphere can be important \citep{Kocsis2011, Panamarev2022}. However, isolating the disc -- IMBH interaction provides a clearer understanding of the torque-driven effects caused by the IMBH and the self-interaction of the disc. We address the impact of the spherical component in our companion paper (Paper II).

Below, we analyse leading order effects of the torques and discuss possible trends in the evolution of the system.

\subsection{Torque-driven evolution in the disc -- IMBH system}
\label{subsec:torques}

The evolution of the system will depend on the relative masses of the stellar disc and the IMBH. Everywhere below we assume $\MSMBH \gg m_{\IMBH} \gg m_i$.  Let us start with the case where $m_{\IMBH} \ll M_\mathrm{d}$ which was studied numerically by \citet{Szolgyen2021} and analytically by \citet{Ginat2023}.

\subsubsection{Resonant dynamical friction. $m_{\IMBH} \ll M_\mathrm{d}$.}

\citet{Szolgyen2021} showed with direct $N$-body and orbit-averaged (\textsc{$N$-Ring}) integrations that an IMBH on an initially inclined, prograde orbit with respect to the stellar disc rapidly reorients toward the disc plane via orbit-averaged torques, much faster than standard \citep{Chandra1943} dynamical friction; the semi-major axes of the IMBH and stars remain approximately conserved during this process, identifying the mechanism as {\it resonant} dynamical friction (RDF). Building on this, \citet{Ginat2023} derived an analytic description in the hierarchical limit, treating the IMBH as a perturber radially embedded in a thin stellar disc. Although the mass ratio is small, the disc's thinness makes the perturber--disc part of the double-averaged Hamiltonian dominate the short-term nodal dynamics, rendering the problem singular and introducing two distinct time-scales. Solving the orbit-averaged nodal equation with a Taylor expansion in the small parameter
\[
\delta \equiv \frac{\nu_{\IMBH}}{b_{i\IMBH}}\,, 
\]
where \(\nu_{\IMBH}\equiv \dot{\Psi}_{\IMBH}\) is the IMBH's nodal precession rate (here \(\Psi\) is the longitude of the ascending node) and \(b_{i\IMBH}\) is the star--IMBH secular coupling coefficient (see Eq.~(8) of \citealt{Ginat2023} for the explicit definition from Lagrange's planetary equations), they showed that all stars satisfying \(|b_{i\IMBH}|\gtrsim |\nu_{\IMBH}|\) (roughly \(a_i\lesssim 2a_{\IMBH}\)) \emph{node-lock} to the IMBH with phase offset $\simeq 90^\circ$. In this locked state the relative node obeys
\begin{equation}
\cos\!\big(\Psi_i-\Psi_{\IMBH}\big) \;=\; -\,\frac{\nu_{\IMBH}}{b_{i\IMBH}}\,,
\label{eq:nodelock}
\end{equation}
so that expanding to first order in \(\delta\) gives
\begin{equation}
\Psi_i \simeq \Psi_{\IMBH} + \frac{\pi}{2} + \frac{\nu_{\IMBH}}{b_{i\IMBH}} + \mathcal{O}\!\left(\delta^2\right).
\end{equation}

The coherent node-locking of the overlapping disc produces a net torque on the IMBH and damps its inclination on the RDF time-scale. For power-law discs, \citet{Ginat2023} obtain a closed-form early-time solution (their equations~(12)--(16)) that leads to an inclination-damping rate proportional to \(m_{\IMBH}\) and a total alignment time that \emph{decreases} with perturber mass (approximately \(t_{\rm RDF}\propto m_{\IMBH}^{-1/2}\) at fixed local disc mass). Their analytic predictions are validated by a suite of direct \(N\)-body experiments, which further demonstrate that node-locking persists until the IMBH finishes aligning with the disc.

\subsubsection{The test particle limit. $m_{\IMBH} \gg M_\mathrm{d}$.}

\begin{figure}
    \centering
        \includegraphics[width=\columnwidth]{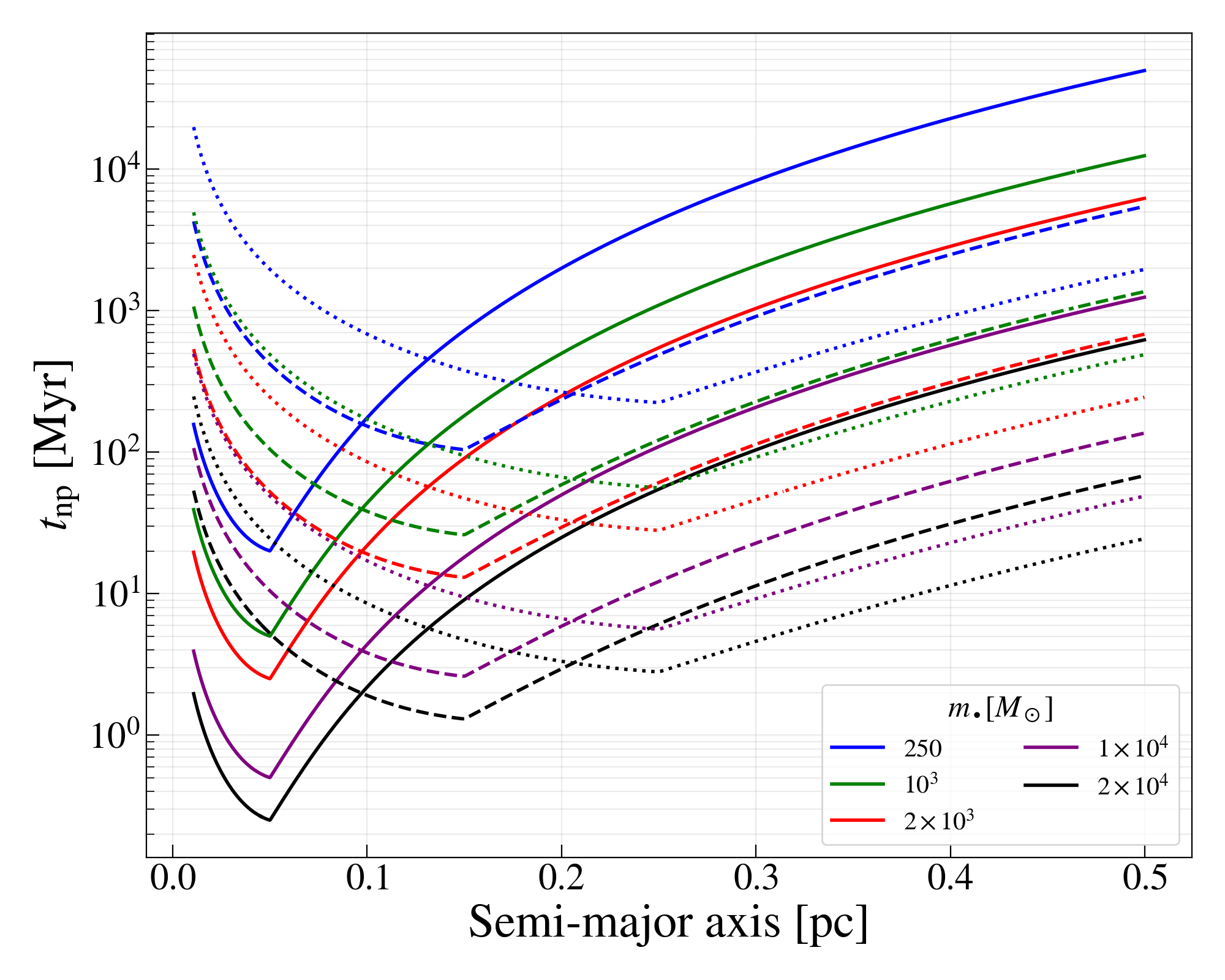}
    \caption{ Nodal precession period for a disc comprised of test particles induced by the IMBH of different masses as shown in the legend and different semimajor axes: 0.05 pc (solid lines), 0.15 pc (dashed lines) and 0.25 pc (dotted lines). All lines correspond to an inclination angle of 45$^\circ$ with respect to the IMBH (cf. Fig.~\ref{fig:t_prec_i}). Both disc stars and the IMBH are assumed to be on circular orbits. }
    \label{fig:t_prec_ma}
\end{figure}

\begin{figure}
    \centering
        \includegraphics[width=\columnwidth]{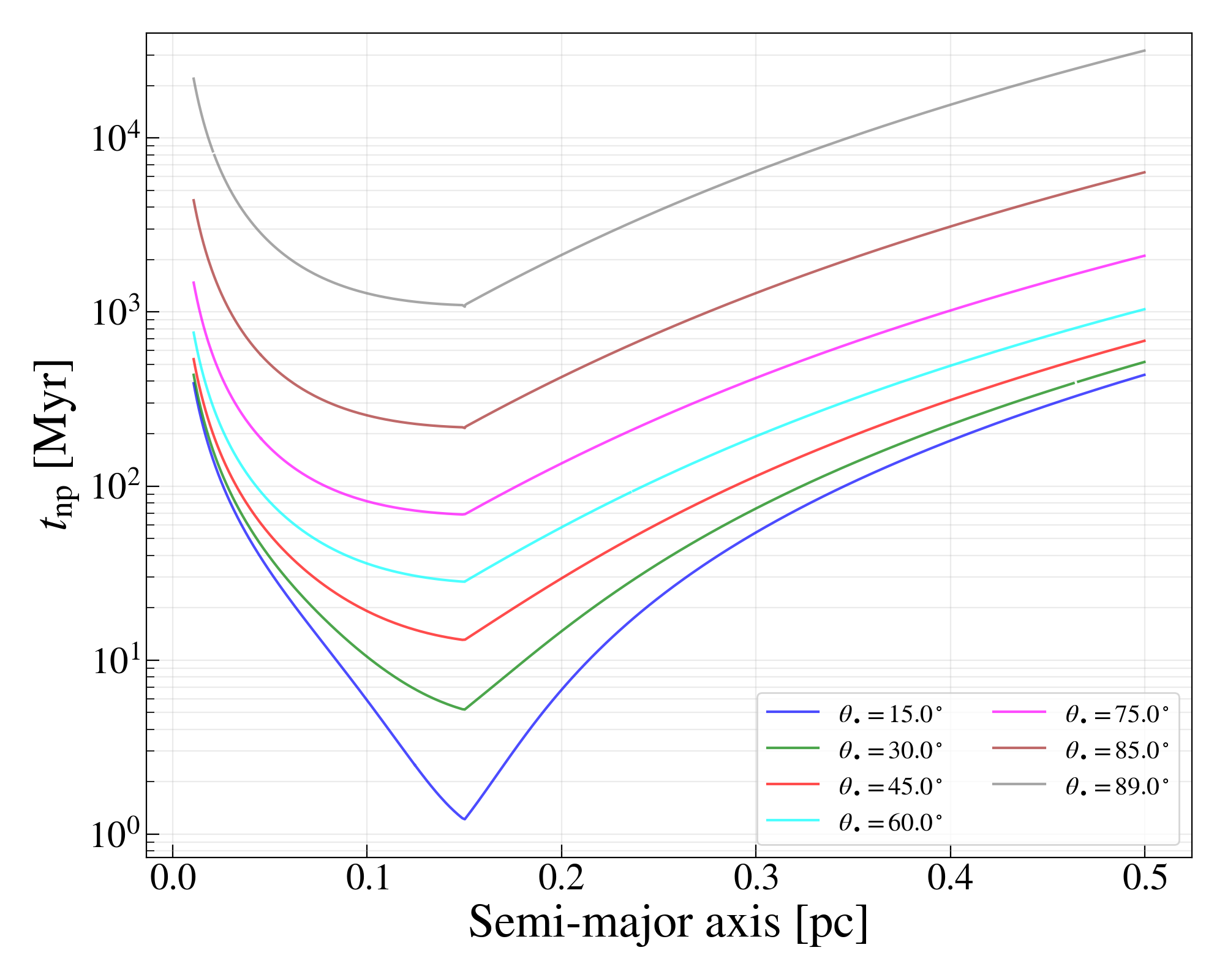}
    \caption{Similar to Figure~\ref{fig:t_prec_ma} showing the nodal precession period for disc particles induced by the IMBH but for a fixed mass $m_{\IMBH} = 2\times 10^3 \msun$, semimajor axis $a = 0.15$ pc and various inclination angles as shown in the legend.  Both disc stars and the IMBH are assumed to be on circular orbits. 
    }
    \label{fig:t_prec_i}
\end{figure}

In the opposite limit, when the mass of the IMBH significantly exceeds that of the disc, we can neglect the contribution from the disc stars. Therefore, the angular momentum evolution of the stars will be driven by the torque from the IMBH resulting in a precession of each individual disc star around the angular momentum vector of the IMBH:
\begin{align}
 \dot{\L }_i &=   \bm{\Omega}_{i,\IMBH} \times \L _i,  \\
\bm{\Omega}_{i,\IMBH} &=-\sum_{\ell=0}^{\infty} \frac{\J_{ij\ell}}{L_i L_{\IMBH}}
P'_{\ell}\big(\Ln_i\cdot \Ln_{\IMBH}\big)\, \L_{\IMBH} .
\label{eq:EOM-test}
\end{align}

If the orbital radius of the disc star is much larger or smaller than that of the IMBH, the precession frequency is dominated by the $\ell=2$ term. Keeping only $\ell=2$ in Eq.~\eqref{eq:EOM-test} and using $P_2(0)^2=1/4$ and $P_2'(\cos\theta)=3\cos\theta$, we obtain 
\begin{align} 
\Omega_{i,\IMBH} &\simeq 3\,\frac{\J_{i\IMBH,2}}{L_i}\,\cos\theta_{i,\IMBH}
= \frac{3}{4}\,\omega_{i}\,
\frac{m_{\IMBH}}{\MSMBH}\,
\frac{a_{i}a_{\rm in}^{2}}{a_{\rm out}^{3}}
\frac{\chi_{\rm out}^{3}P_{3}(\chi_{\rm in}) }{\chi_{\rm in}^{3}}\cos\theta_{i,\IMBH} \,,
\end{align}
where $\omega_i=(G\MSMBH/a_i^3)^{1/2}$ is the orbital frequency around the SMBH and $\chi=1/\sqrt{1-e^2}$,  and the `in' and `out' index denote the orbits with smaller and larger semimajor axis, respectively \citep{Kocsis2015}. 

When the test star is on a nearly circular orbit ($e_i \to 0$), the expression above can be explicitly written as:
\begin{equation}
\Omega_{i,\IMBH} = \frac{3}{4}\,\omega_{i}\,
\frac{m_{\IMBH}}{\MSMBH}\,
\cos\theta_{i,\IMBH}\;
\begin{cases}
\displaystyle
\left(\frac{r_i}{a_{\IMBH}}\right)^{3}
\frac{1}{(1-e_{\IMBH}^2)^{3/2}},
& r_i \ll a_{\IMBH}(1-e_{\IMBH}),\\[1.5em]
\displaystyle
\left(\frac{a_{\IMBH}}{r_i}\right)^{2}
\left(1 + \frac{3}{2}e_{\IMBH}^2\right),
& r_i \gg a_{\IMBH}(1+e_{\IMBH}),
\end{cases}
\label{eq:quad_nonover_circ}
\end{equation}
where $r_i$ is the orbital radius of the test star.

In the opposite limit when the test star (on a circular orbit) and the IMBH have overlapping orbits
\begin{equation}
\label{eq:asymp_over}
    \Omega_{i,\IMBH}  \simeq \frac{2}{\pi^2}\omega_i\frac{m_{\IMBH}}{\MSMBH} 
    \frac{r_i^2}{a_{\IMBH}\sqrt{(r_i - r_{\mathrm{p},\IMBH})(r_{\mathrm{a},\IMBH} - r_i)}}
     \cot\theta_{i,\IMBH},
\end{equation}  
where $r_{\mathrm{p},\IMBH}$ and $r_{\mathrm{a},\IMBH}$ are the periapsis and apoapsis of the IMBH, respectively. The expression is valid when $r_{\mathrm{p},\IMBH} < r_i < r_{\mathrm{a},\IMBH}$, i.e. the test star is radially embedded within the orbit of the IMBH \citep{Panamarev2025}. While this expression diverges at the IMBH peri- and apocentre, it provides a first-order approximation in the inner embedded region, becoming more accurate for more eccentric and more nearly coplanar IMBH orbits.

Thus, the time evolution of the angular momentum vector of the star $i$ is given by the solution of Eq.~\eqref{eq:EOM-test} \citep{Kocsis2015} which may be written symbolically as
\begin{equation}
\L_i(t) = \exp\left(\bm{\Omega}_{i,\IMBH} \Delta t\right) \L_i(t_0),
\label{eq:l(t)_test}
\end{equation}
Here, $\bm{\Omega}_{i,\IMBH}$ is interpreted as a skew-symmetric matrix that represents rotation. Therefore, to determine the coordinates of the angular momentum vectors for each particle $i$ at time $t$, we rotate the initial angular momentum vector $\L_i(t_0)$ by the angle $\phi_{i,\IMBH}=\Omega_{i,\IMBH}\Delta t$ about the angular momentum vector of the IMBH where $\Delta t=t-t_0$, while the IMBH's position remains unchanged, i.e. (using Rodrigues's rotation formula) 
\begin{align}
    \L_i(t) =& \L_i(t_0) \cos \phi_{i,\IMBH} + (\hat{\bm{\Omega}}_{i,\IMBH} \times \L_i(t_0)) \sin\phi_{i,\IMBH} \nonumber\\ &+ \hat{\bm{\Omega}}_{i,\IMBH} ~[\hat{\bm{\Omega}}_{i,\IMBH} \cdot \L_i(t_0)] (1 - \cos\phi_{i,\IMBH})
    \nonumber\\
    \hat{\L}_i(t)=&    
    \hat{\bm{x}}\,\sin\theta_{i,\IMBH} \cos \phi_{i,\IMBH}   + \hat{\bm{y}}\, \sin\theta_{i,\IMBH} \sin\phi_{i,\IMBH}  +\hat{\bm{z}} \cos\theta_{i,\IMBH}
\,.
    \label{eq:rotation_Rodriguez}
\end{align}
where $\hat{\bm{x}}=\hat{\bm{y}}\times \hat{\bm{z}}$, $\hat{\bm{y}}=(\hat{\L}_{\IMBH} \times \L_i(t_0))/\|\hat{\L}_{\IMBH} \times \L_i(t_0)\|$, $\hat{\bm{z}}=\hat{\L}_{\IMBH}$, and $ \cos\theta_{i,\IMBH}=\L_{\IMBH}\cdot\L_i$ is constant in time.

Substituting the above expressions for the angular velocity vector \( \bm{\Omega}_{i,\IMBH} \), the nodal precession period \( T_i \) for star $i$ driven by an IMBH is computed as \( T_i = 2\pi \, |\bm{\Omega}_{i,\IMBH}|^{-1} \). For the Galactic centre, which is known to host a \( 4 \times 10^6 \) M\(_\odot\) SMBH \citep{Ghez2005, Gillessen2017}, we calculate these precession periods using Eq.~\eqref{eq:EOM-test} with $\ell_{\max}=1000$ 

for various IMBH masses and semimajor axes for circular orbits. Figure~\ref{fig:t_prec_ma} demonstrates how these periods vary with IMBH mass (as specified in the figure legend) and semimajor axes of 0.05, 0.15, and 0.25 pc, all set at an inclination of \( 45^{\circ} \) to the disc. The figure reveals that IMBHs with masses of the order of \(10^4 \Msun\) induce a high precession rate, significantly altering the angular momentum vector distribution within the disc in less than a few Myr depending on the IMBH orbital radius. Conversely, an IMBH weighing \(250 \Msun\) has only a marginal impact on the distribution for more than 20 Myr, even when positioned on an inner orbit with \(a = 0.05\) pc.

Figure \ref{fig:t_prec_i} illustrates the precession times induced by an IMBH with a mass of \(2 \times 10^3 \Msun\) at various orbital inclinations relative to the disc for circular orbits. It is evident that the precession period is strongly dependent on inclination: the period ranges between approximately 1 \Myr\ and 1 \Gyr\ for inclinations \(15^\circ\) to \(89^\circ\), for stars with similar semimajor axes to the IMBH. Consequently, the precession period tends towards infinity as the inclination approaches \(90^\circ\), and approaches zero for nearly coplanar orbits. However the amplitude of the warp is approximately zero in the coplanar limit, see Eq.~\eqref{eq:rotation_Rodriguez} with $\bm{\Omega}_{i,\IMBH} \times \L_i(t_0)=0$. 

\begin{figure}
    \centering
    \includegraphics[width=\columnwidth]{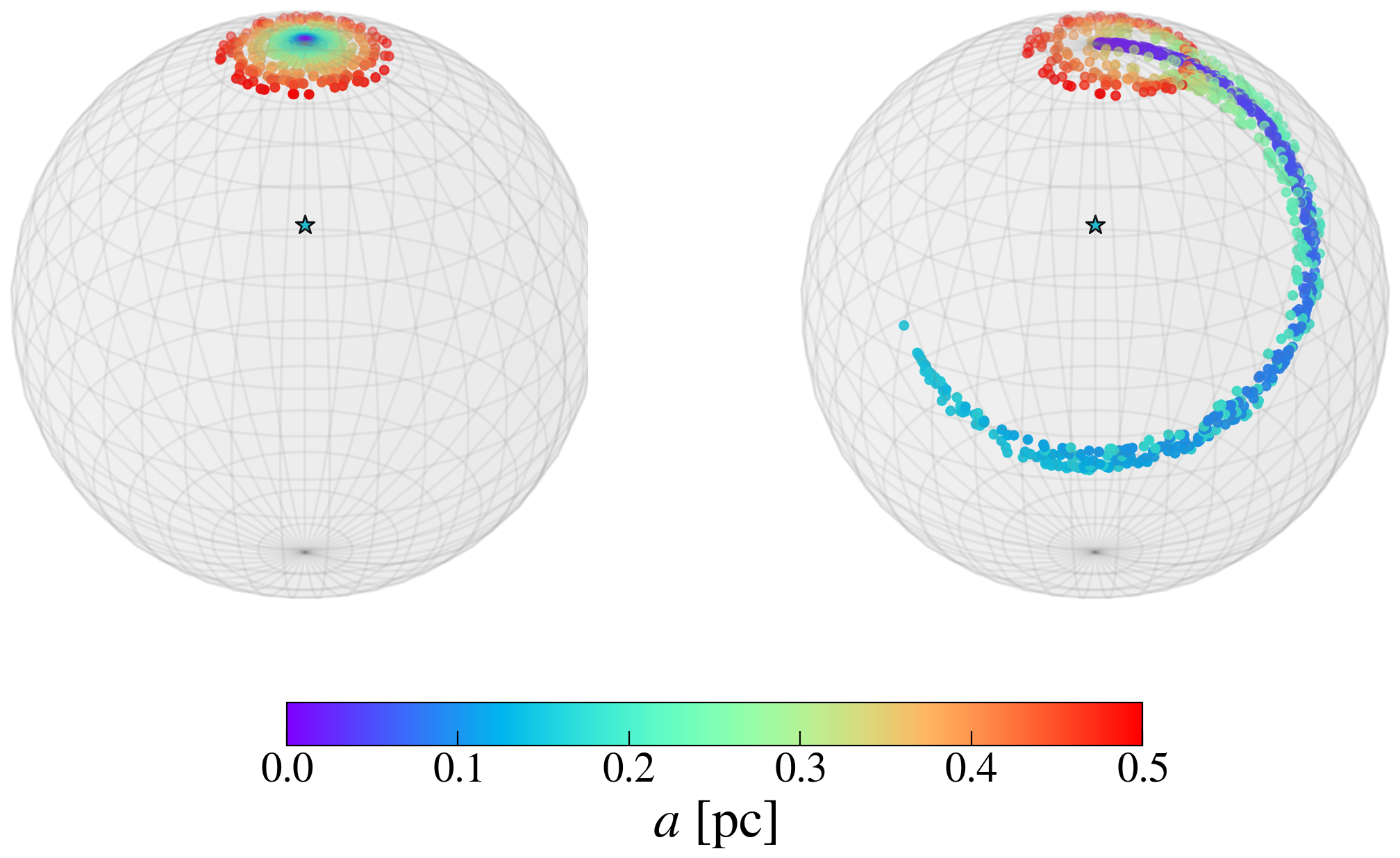}
    \caption{Angular momentum direction vectors on the unit sphere at t = 0 (left) and at t $\simeq 8.0$ Myr (right). Colour coding shows semimajor axes of stars in the disc. Star symbol indicates the IMBH of mass $m_{\IMBH} = 2\times10^3 \msun$, semimajor axis 0.15 pc, eccentricity 0.1, and initial inclination angle $45^\circ$ with respect to the disc. Note that the disc mass is set to zero here (cf. Fig.~\ref{fig:L_unit150}), and the thickness of the disc in angular-momentum space increases with semimajor axis.
}
    \label{fig:L_unit}
\end{figure}

To explore the behaviour of the system in a more realistic setting, let us use Eq.~\eqref{eq:l(t)_test} to compute angular momentum vectors at time $t$ for a stellar disc that could have originated from star--disc interactions in an active galactic nucleus (\citealt{Panamarev2018}, cf.~Sec.~\ref{subsec:stellar-disc}). We take positions and velocities of such a disc at time $t=0$ but set the total mass of the disc to zero.
Figure~\ref{fig:L_unit} shows the angular momentum vectors of the disc particles on a unit sphere with an IMBH of mass $m_{\IMBH} = 2\times10^3 \msun$ (denoted as a star symbol) and inclination angle of $45^\circ$ with respect to the disc. We can see that the angular momentum vectors form a ring centred around the IMBH. As we see from the right panel, it takes $\simeq 8.0$ Myr for the stars close to the IMBH to move 3/4 of the precession period at a distance of $0.15$ pc. The ring forms due to the differential rotation given that the precession time varies with orbital radius of the star relative to the IMBH, as indicated by Fig.~\ref{fig:t_prec_i}. The warp angle at some given time $t$ (such as the age of the clockwise disc, 6 Myr) is $\Omega_{i, \IMBH} t= 2 \pi t/t_{\rm np}$ where $t_{\rm np}$ is the nodal precession period shown in Figures~\ref{fig:t_prec_ma} and \ref{fig:t_prec_i}.

\subsubsection{Intermediate mass range. $m_{\IMBH} \simeq M_\mathrm{d}$.}

When the mass of an IMBH is comparable to that of a stellar disc, it becomes necessary to account for the torque exerted by the disc. The principles of angular momentum conservation and the geometry of the interaction suggest a resultant precession of both the IMBH and the disc around the total angular momentum vector of the system. In general, the precession speed may not be uniform resulting in a combination of precession and nutation. However, since our focus is not on the detailed nutation dynamics, we simplify by considering an average angular velocity contribution from the disc stars to the IMBH's motion over a time step $\Delta t = t - t_0$:
\begin{align}
&\L_{\IMBH}(t) = \exp\left( \bm{\Omega}_{\mathrm{\IMBH,disc}}\Delta t\right) \L_{\IMBH}(t_0),\nonumber \\
&\L_i(t) = \exp\left(\bm{\Omega}_{i, \IMBH}\Delta t \,\right) \exp\left(\bm{\Omega}_{i, \mathrm{disc}}\Delta t \right) \L_i(t_0),
\label{eq:l(t)}
\end{align}
where 
$\exp\left(\bm{\Omega}_{\mathrm{\IMBH,disc}}\Delta t\right)$ and 
$\exp\left(\bm{\Omega}_{i,\mathrm{disc}}\Delta t\right)$ are rotation operators by finite angles defined similar to Eq.~\eqref{eq:rotation_Rodriguez}.
Here we denote disc particles by the index $i$, and $\bm{\Omega}_{i,\mathrm{disc}}$ denotes the average angular velocity of the particle with index $i$ over $\Delta t$ due to all stars in the disc. The vector $\bm{\Omega}_{\mathrm{\IMBH,disc}}$ describes the average angular velocity of the IMBH around the total angular momentum of the disc -- IMBH system, so that the corresponding finite rotation angles are $|\bm{\Omega}_{\mathrm{\IMBH,disc}}|\Delta t$ for the IMBH and $|\bm{\Omega}_{i,\mathrm{disc}}|\Delta t$ for the disc particle with index $i$.

\begin{figure}
    \centering   \includegraphics[width=\columnwidth]{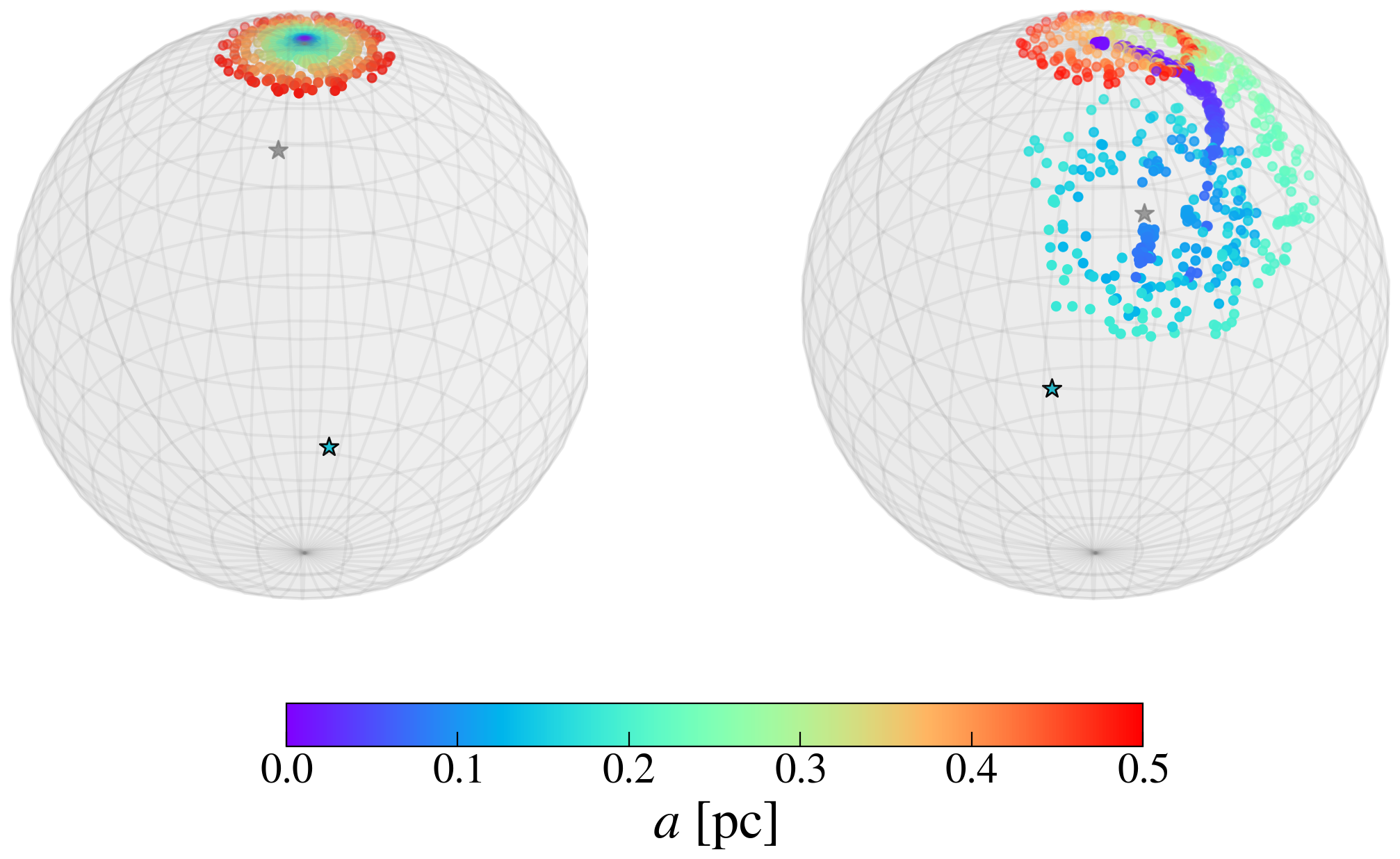}
    \caption{Same as Figure~\ref{fig:L_unit} showing the angular momentum vectors' directions on a unit sphere but for a massive disc ($M_\mathrm{d} =  3\times10^3\Msun \approx 1.5\,m_{\IMBH}$) and a retrograde IMBH (initial inclination angle $150^\circ$) at t = 0 (left) and at t = 2.5 Myr (right). Colour coding shows semimajor axes of stars in the disc. The grey star symbol shows the direction of $-\L_{\IMBH}$ which lies on the front side of the sphere both in the initial time and at 2.5 Myr and is the centre of circulation for the stars' angular momenta on radially overlapping orbits (light green).}
    \label{fig:L_unit150}
\end{figure}

The rotation of disc stars involves two consecutive rotations: one around the IMBH's angular momentum vector and another around the disc's angular momentum. Given the non-linear nature of these finite rotations, which in general do not commute, we perform them in the order determined by the magnitudes of the angular velocities for the disc stars. Another approach to minimise errors from this non-linearity is to alternate the order of the rotations over subsequent half-time steps \citep{Kocsis2015}. We neglect this minimisation procedure here and instead use the basic integrator described above, with $\Delta t$ chosen to be sufficiently small. In this simplified treatment, the disc is treated as a single component: at each time step, each particle is rotated only twice, once with respect to the total angular momentum of the disc and once with respect to the angular momentum of the IMBH. This procedure was used to obtain the right panel of Fig.~\ref{fig:L_unit150}, as described below.

Now that the mathematical framework is established, let us now examine the evolution of the disc and the IMBH numerically, by evolving the system according to the orbit-averaged VRR interaction, Eq.~\eqref{eq:l(t)}. We set up the disc discussed previously, composed of 1000 particles (i.e. angular momentum vectors) with a total mass of \( M_\mathrm{d} \approx 3000 \, \Msun \). An IMBH of comparable mass, \( m_{\IMBH} = 2000 \, \Msun \approx 0.67\,M_\mathrm{d} \), is introduced to explore two configurations: the IMBH on a prograde orbit and the IMBH on a retrograde orbit relative to the disc.

In the prograde scenario, the mutual torques tend to align the angular momentum vectors of the IMBH and the stellar disc \citep{Szolgyen2021,Ginat2023}. First, the stars on orbits similar to that of the IMBH will align with the IMBH followed by mutual alignment. Consequently, stars sharing similar semimajor axes with the IMBH will contribute to forming a thicker disc structure, with a maximum opening angle that reflects their initial relative orientation to the IMBH.

Conversely, in the retrograde scenario, such alignment is absent. Figure~\ref{fig:L_unit150} shows an example of the dynamics induced by an IMBH with initial inclination of $150^\circ$ with respect to the disc. Here, disc stars proximal to the IMBH behave like test particles forming a ring of angular momentum vectors. However, the ring is centred at the negative angular momentum vector of the IMBH (denoted as a grey star symbol in the figure) and it moves with the IMBH. This interaction could fragment the original stellar disc into three distinct components based on their semimajor axes: an outer disc largely unaffected within 5 to 10 Myr, a middle disc that becomes anti-aligned with the IMBH, and a stream of stars connecting two discs. We determine the condition for the disruption analytically below.

Although, this model is very approximate, we will see that it qualitatively describes the evolution of angular momentum vector direction of the disc particles during the first few Myr for the case when the IMBH mass is comparable to that of the disc even when we include a live spherical stellar component (cf. Sec.~\ref{sec:disc-imbh}).

\subsection{Disruption of the disc by an IMBH}
\label{subsec:disc disruption}

\subsubsection{The disruption condition}
\label{subsubsec:disruption-condition}

\begin{figure*}
    \centering
    \includegraphics[width=0.8\textwidth]{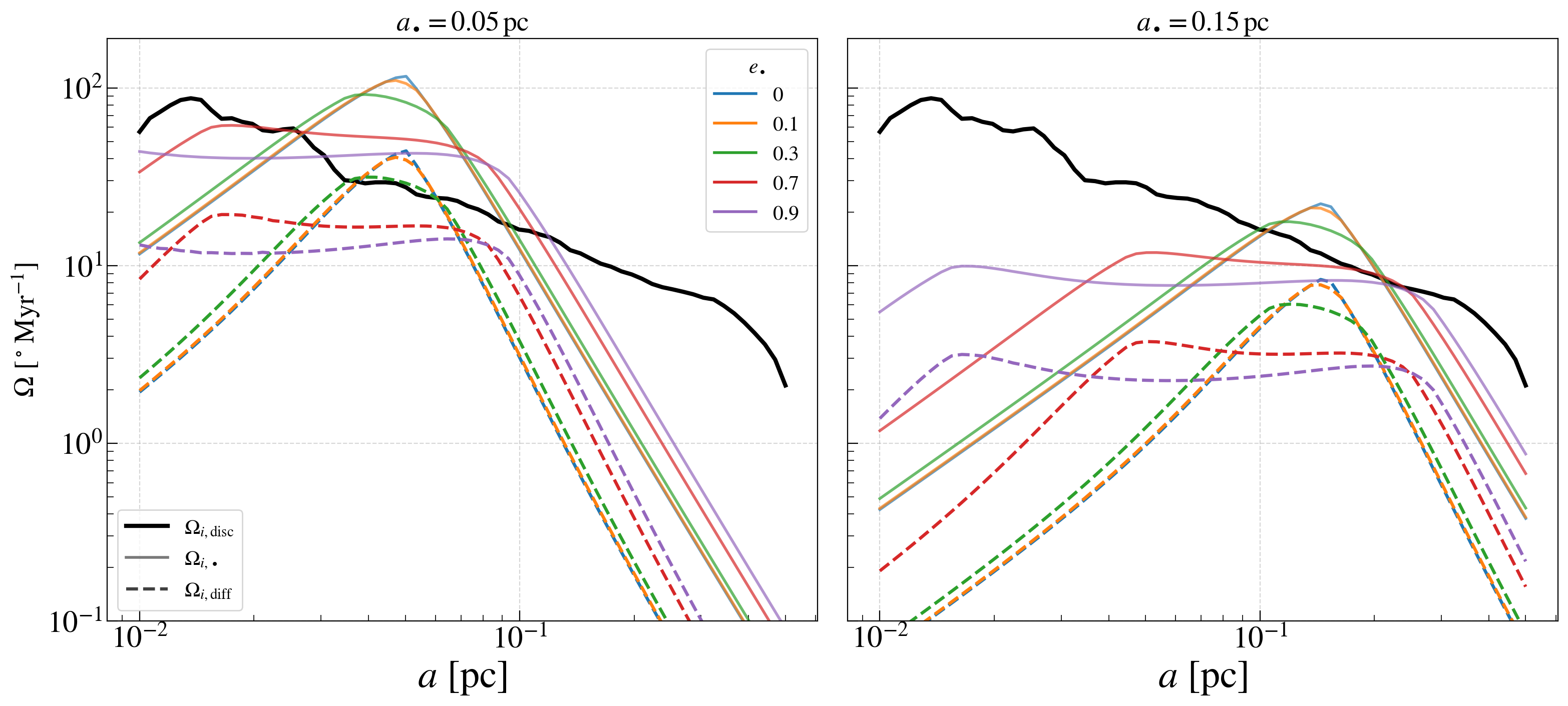}
    \caption{Disruption of the stellar disc by an IMBH with mass $m_\IMBH = 2000\msun$, semimajor axes $a_{\IMBH}=0.05$ pc (left panel) and $a_{\IMBH}=0.15$ pc (right panel), with various eccentricities (see legend) and an inclination of $\theta_\IMBH = 135^{\circ}$. The thick black line shows the precession rate of each disc particle due to the net torque from the disc, $|\mathbf{\Omega}_{i,\mathrm{disc}}\times\hat\L_i|$, the left-hand side of Eq.~\eqref{eqn:fundamental criterion}. The right-hand side, $\Omega_{i,\mathrm{diff}}$, the IMBH induced differential precession rate, is shown by coloured dashed lines. Finally $|\mathbf{\Omega}_{i,\IMBH}\times\hat\L_i|$, the net IMBH-induced rate, is shown by coloured solid lines. A fragment of the disc breaks off where the thick black line is below the coloured solid line but above the dashed line. The disc disperses
    if the black thick line is below both the solid and dashed coloured lines. 
    }
    \label{fig:disc-disruption}
\end{figure*}

The disc stars are initially bound together in angular momentum space by the self-gravity of the disc. However, an IMBH may perturb this coupling in two related ways. First, the IMBH can dominate the precession of the angular-momentum vector of a disc star, causing the affected part of the disc to precess mainly under the IMBH torque. Second, if the IMBH-induced precession varies sufficiently strongly across the finite opening angle of the disc, this differential precession can overcome the self-coupling of the disc and disperse it. The latter effect gives the formal disruption condition \citep{Panamarev2025}, which is obtained by comparing the precession rate produced by the self-gravity of the disc with the $\Omega_{i,\mathrm{diff}}$ differential precession produced by the IMBH:
\begin{align}\label{eqn:fundamental criterion}
    &\left|\mathbf{\Omega}_{i,\mathrm{disc}}\times\hat\L_i\right| \leq 
    \Omega_{i,\mathrm{diff}} \\\nonumber
    &\Omega_{i,\mathrm{diff}}\equiv
    \left| \left(\mathbf{\Omega}_{i,\IMBH}(\theta_\IMBH+\Delta\theta)\times\hat\L_i - \mathbf{\Omega}_{i,\IMBH}(\theta_\IMBH-\Delta\theta)\times\hat\L_i \right) \cdot \Delta\hat{\L}\right|,
\end{align}
where in the $\mathbf{\Omega}_{i,\IMBH}$ argument $\Delta\theta$ is the angle between the angular momentum vector of particle $i$ and the total angular momentum vector of the disc, i.e. the half-opening angle of the disc. We take two test directions on the disc opening-angle circle. They lie in the azimuthal plane of the IMBH and are separated by $180^\circ$ around the disc angular-momentum axis. Then $\Delta\hat{\L}$ is the unit vector pointing from one of these directions to the other. Here, we treat $i$ as a test particle; the torque difference on the right-hand side is calculated for the same particle at opposite locations with respect to the disc's angular momentum during a precession cycle. This corresponds to the hierarchical mass ratio case discussed in \citet{Panamarev2025}\footnote{See also \citet{Ginat2025}, where the authors discuss the possible disruption of a disc by an axion dark matter background.}. Generally, the condition in Eq.~\eqref{eqn:fundamental criterion} must be evaluated numerically using the individual positions and velocities of all disc particles and the IMBH in Eq.~\eqref{eq:EOM-test}.

Fig.~\ref{fig:disc-disruption} compares the precession frequency due to the torque from the entire disc, $\left|\mathbf{\Omega}_{i,\mathrm{disc}}\times\hat\L_i\right|$ as a function of $a_i$, with the right-hand side of the inequality in Eq.~\eqref{eqn:fundamental criterion}. The coloured dashed lines show the right-hand side of Eq.~\eqref{eqn:fundamental criterion} corresponding to an IMBH with semimajor axes $a_{\IMBH}=0.05$ pc (left panel) and $a_{\IMBH}=0.15$ pc (right panel), and eccentricities of $e_{\IMBH}=0, 0.1, 0.3, 0.7,$ and $0.9$ as labelled. The solid coloured lines in each panel show the precession rate of a disc star due to the IMBH. For the disruption to occur, the differential precession induced by the IMBH must exceed the precession produced by the self-gravity of the disc. However, whether the affected stars form a coherent fragment or are fully dispersed depends also on the IMBH-induced precession rate itself. If the precession due to the IMBH dominates over the precession from within the disc, while the differential IMBH-induced precession remains smaller than the disc-induced precession, the affected part of the disc may break off as a coherently precessing fragment. In this case, the fragment will precess as a whole with respect to the IMBH. By contrast, when the differential torque from the IMBH exceeds the net torque from the disc, the disruption condition is met and the disc is expected to be fully dispersed: each individual particle of the disc will independently precess with respect to the IMBH. To compute the net torque from the disc, we generated stars arranged in a thin disc with an $r^{-\Gamma}$ surface density, where the power-law exponent is $\Gamma=1.3$ \citep{Bartko2010}. We computed the torque on a test particle with $\Delta\theta=10^\circ$. To compute the torque difference on the right-hand side of Eq.~\eqref{eqn:fundamental criterion}, we rotated the test particle by 180$^\circ$ with respect to the net angular momentum vector of the stellar disc.

\begin{figure*}
    \centering
    \includegraphics[width=0.8\linewidth]{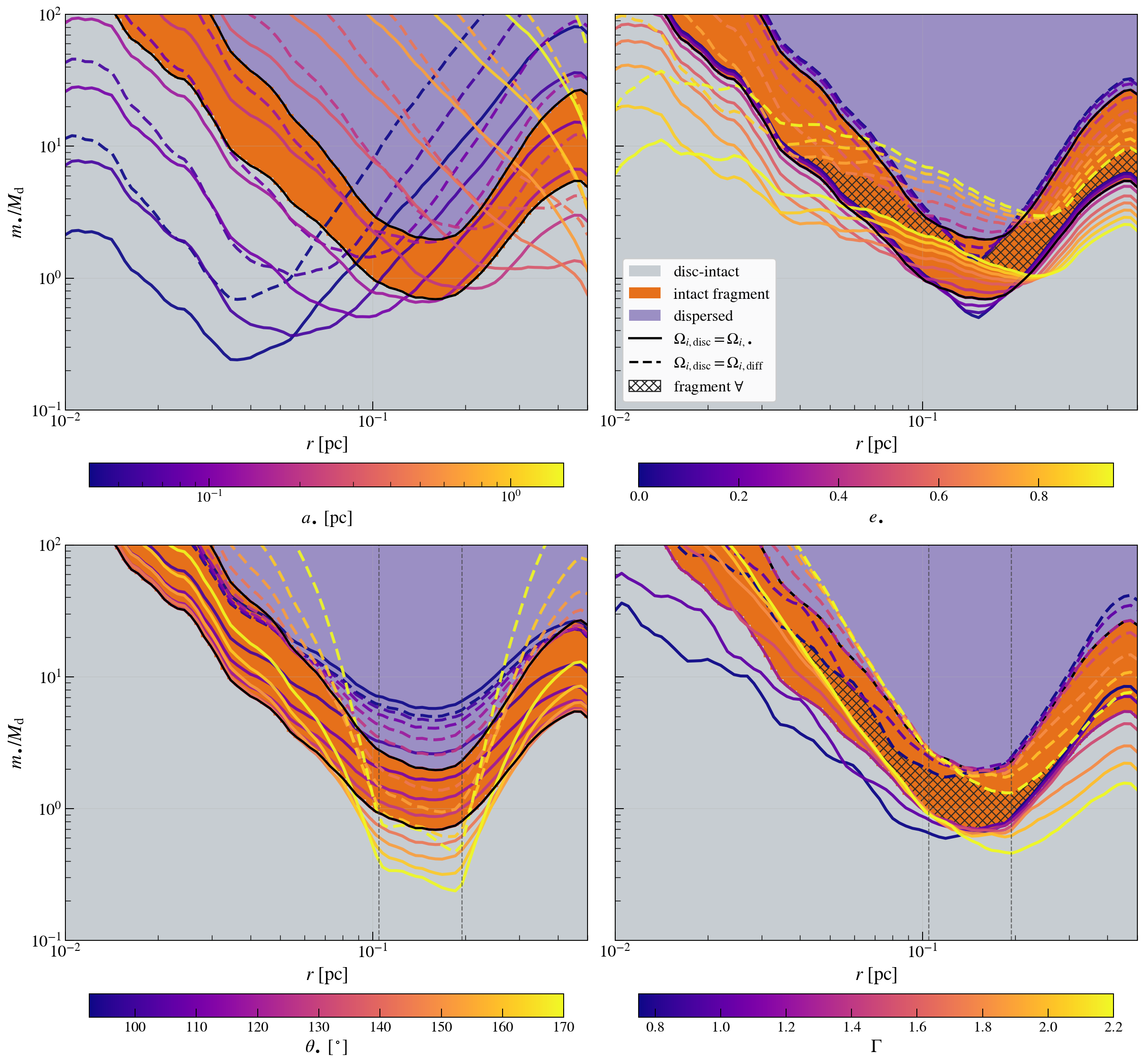}
\caption{The disruption parameter space for a disc star at radius $r$ perturbed by an IMBH, shown as a function of the IMBH-to-disc mass ratio $m_\IMBH/M_{\rm d}$. Shaded regions show different outcomes: (i) grey -- disc remains intact, (ii) orange -- disc fragments into clumps, and (iii) purple -- disc disperses.  The shaded regions are the same in all panels having an IMBH with $a_\IMBH=0.15$ pc, $e_\IMBH=0.3$, $\theta_\IMBH=135^\circ$, and a disc with $\Gamma=1.3$. 
The regimes are determined by comparing the disc self-torque ($|\mathbf{\Omega}_{i,\mathrm{disc}}\times\hat\L_i|$), the total precession  rate of the IMBH ($|\mathbf{\Omega}_{i,\IMBH}\times\hat\L_i|$), and the differential precession rate due to the IMBH ($\Omega_{i,\mathrm{diff}}$). 
In (i) the grey region the disc self-torque dominates; in (ii) the orange region the overall precession is set by the IMBH but the disc self-torque is still larger than the differential IMBH torque, and in (iii) the  purple region the differential IMBH torque exceeds the disc self-torque. The black solid and dashed curves show the boundaries where  $|\mathbf{\Omega}_{i,\mathrm{disc}}\times\hat\L_i|=|\mathbf{\Omega}_{i,\IMBH}\times\hat\L_i|$ and $|\mathbf{\Omega}_{i,\mathrm{disc}}\times\hat\L_i|=\Omega_{i,\mathrm{diff}}$, respectively. The four panels show these boundaries with coloured lines when changing $a_\IMBH$, $e_\IMBH$, $\theta_\IMBH$, or $\Gamma$, respectively, while keeping the other parameters unchanged at their fiducial values. The cross-hatched regions mark where an intact fragment inevitably forms for all parameter values shown in that panel, i.e. the intersection of the intact fragment regions over the corresponding parameter sweep. Rates are computed using the exact even-Legendre torque series in Eq.~\eqref{eq:EOM-test} with $\ell_{\max}=120$ for a synthetic disc with an $r^{-\Gamma}$ surface density.}
    \label{fig:disruption-regimes}
\end{figure*}

In certain regimes, the condition can be simplified analytically. For a thin disc ($\Delta\theta\ll 1$), the disruption condition due to the variation in angle with respect to the IMBH simplifies to \citep{Panamarev2025}:
\begin{align}
    \Omega_{i,\mathrm{disc}} \leq \kappa\frac{\partial \Omega_{\IMBH}}{\partial \theta} \sin\theta_\IMBH,
    \label{eq:torques_mag}
\end{align}
where $\kappa$ is a factor of order unity that accounts for the change in orientation of the pair's relative angular momentum vector with respect to the IMBH.

For a thin disc of stars on circular orbits, we can analytically estimate the initial disc torque by replacing the contribution of each individual star with the average torque (see Appendix~\ref{App:torques analytical} for details). For the torque due to the IMBH, we use the asymptotic expression for a circular orbit embedded within that of the IMBH (see Eq.~\ref{eq:asymp_over}) with eccentricity $e_\IMBH$ and semimajor axis $a_\IMBH$ \citep{Panamarev2025}. Thus, the inequality in Eq.~\eqref{eq:torques_mag} becomes:
\begin{align}
\label{eq:finite-circ-disc-overlap}
    & (2-\Gamma) \Bigg\{ z^{2-\Gamma} \left[ \frac{2}{\pi}\cot\Delta\theta + \mathcal{C}(\Gamma) \right]
    - \frac{z}{\pi} \left[ \frac{z^2}{1-z^2} - \frac{1}{2}\ln(1-z^2) \right] \notag \\
    & \quad - z \bigg[ 3\left(\frac{1}{4(1+\Gamma)} - \frac{1}{2\pi}\right)z^2 + 10\left(\frac{9}{64(3+\Gamma)} - \frac{1}{8\pi}\right)z^4 \bigg] \Bigg\} \notag \\
    & \quad \leq \frac{2}{\pi^2}\frac{m_\IMBH}{M_{\rm d}}\frac{r^2_i}{a_\IMBH} 
    \frac{1}{\sqrt{(r_i - r_{\mathrm{p},\IMBH})(r_{\mathrm{a},\IMBH} - r_i)}}\frac{1}{|\sin\theta_\IMBH|},
\end{align}
where $z=r_i/r_{\rm d}$, $r_{\rm d}$ is the maximum radius of the disc and $\mathcal{C}(\Gamma)$ is a function of $\Gamma$ expressed in terms of simple fractions, as defined in Eq.~\eqref{eq:C_gamma}.

If the disc extends to infinity, it is more convenient to work with the local disc mass per logarithmic radius interval at $r_i$,
\begin{equation}
  M_{\rm d,loc}(r_i)
  \equiv \frac{{\rm d}M_{\rm d}}{{\rm d}\ln r}\Big|_{r_i}
  = 2\pi r_i^2\,\Sigma(r_i).
\end{equation}
In this limit, the disruption condition~\eqref{eq:torques_mag} becomes
\begin{align}
    \frac{2}{\pi}\cot\Delta\theta + \mathcal{C}(\Gamma)
   \leq
    \frac{2}{\pi^2}\frac{m_\IMBH}{M_{\rm{d, loc}}}\frac{r^2_i}{a_\IMBH}
    \frac{1}{\sqrt{(r_i - r_{\mathrm{p},\IMBH})(r_{\mathrm{a},\IMBH} - r_i)}}\frac{1}{|\sin\theta_\IMBH|}.
    \label{eq:infinite disc}
\end{align}
The conditions above are valid for the case of overlapping orbits of a test star with the IMBH. In the opposite limit, for a distant IMBH ($r_i/r_{\mathrm{p},\IMBH}\lesssim 0.3$), one can use the quadrupole approximation for the star--IMBH torque (Eq.~\ref{eq:quad_nonover_circ}), and the right-hand side of the disruption condition (Eq.~\ref{eqn:fundamental criterion}) can be written as: 
\begin{align}
    \frac{2}{\pi}\cot\Delta\theta + \mathcal{C}(\Gamma)
   \leq  \frac{3}{4}\,
\frac{m_{\IMBH}}{M_{\rm d,loc}}\,
\left(\frac{r}{a_{\IMBH}}\right)^{3}
\frac{\sin^2\theta_\IMBH}{(1-e_{\IMBH}^2)^{3/2}}.
\label{eq:disrupt-quad}
\end{align}
For a finite disc, the same quadrupole criterion follows by expressing the local mass through the total disc mass, $M_{\rm d,loc}=(2-\Gamma)\,M_{\rm d}\,(r/r_{\rm d})^{2-\Gamma}$.

\subsubsection{The disc disruption parameter space}
\label{subsubsec:disruption-parameters}

The examples in Fig.~\ref{fig:disc-disruption} with
$(m_\IMBH/M_{\rm d}, a_{\IMBH}, e_{\IMBH}, \theta_\IMBH) = (0.67, 0.05\,\mathrm{pc}, 0.3, 135^{\circ})$ and $(0.67, 0.15\,\mathrm{pc}, 0.3, 135^{\circ})$, respectively, can be generalised across the IMBH -- disc parameter space. At fixed IMBH orbital parameters and disc structure, the IMBH-induced rates, $|\mathbf{\Omega}_{i,\IMBH}\times\hat\L_i|$ and $\Omega_{i,\mathrm{diff}}$, scale with the IMBH mass, whereas the disc self-torque rate, $|\mathbf{\Omega}_{i,\mathrm{disc}}\times\hat\L_i|$, scales with the disc mass. Therefore, for a given $(a_\IMBH,e_\IMBH,\theta_\IMBH,\Gamma)$, the boundaries between the disruption regimes can be expressed in terms of the mass ratio $m_\IMBH/M_{\rm d}$. The outcome also depends on the IMBH orbit through $a_\IMBH$, $e_\IMBH$ and $\theta_\IMBH$, and on the disc surface-density slope $\Gamma$. Fig.~\ref{fig:disruption-regimes} shows the resulting regimes in the $(r,m_\IMBH/M_{\rm d})$ plane. The shaded regions correspond to the fiducial model ($a_\IMBH=0.15$ pc, $e_\IMBH=0.3$, $\theta_\IMBH=135^\circ$ and $\Gamma=1.3$), while the coloured curves show how the regime boundaries shift when $a_\IMBH$, $e_\IMBH$, $\theta_\IMBH$, or $\Gamma$ is varied in the four panels, respectively. The companion figures in Appendix~\ref{App:disruption regimes} show these fragmentation/disruption regions for different eccentricity and inclination.

For the fiducial model, in the radial range closest to the IMBH orbit the disc remains intact for $m_\IMBH/M_{\rm d}\lesssim0.7$--$1$, forms an intact fragment for $m_\IMBH/M_{\rm d}\sim1$--$3$, and becomes dispersed for $m_\IMBH/M_{\rm d}\gtrsim2$--$5$, with the exact value depending on radius. Away from the IMBH orbit the mass ratio required for fragmentation increases: at $r\lesssim0.05$ pc and $r\gtrsim0.3$ pc the intact-fragment band moves to $m_\IMBH/M_{\rm d}\sim10$--$100$. The strongest shift is produced by changing $a_\IMBH$, which moves the fragmenting region to the radial part of the disc closest to the IMBH orbit. For $a_\IMBH\simeq0.4$ pc, comparable to the possible three-dimensional position of IRS~13 \citep{Tsuboi2020}, the  intact-fragment region is shifted to the outer disc; fragmentation is expected mainly at $r\gtrsim0.1$--$0.2$ pc and requires mass ratios from a few to several tens, while the inner disc remains intact unless the mass ratio is much larger. By contrast, for $a_\IMBH\simeq1.5$ pc, comparable to the possible location of the gaseous circumnuclear disc (CND), a low- or moderate-eccentricity massive perturber\footnote{Note that it does not have to be literally an IMBH, but rather any perturber for which the double-orbit average approximation holds, which is the case for the CND \citep{Kocsis2011}.} orbit lies mostly outside the stellar disc; the  intact-fragment region is then pushed to the outer edge and to very large perturber-mass to stellar-disc-mass ratios, typically $m_\IMBH/M_{\rm d}\gtrsim10^2$ in the $e_\IMBH=0.3$ case, so most of the disc remains  intact over the physically relevant range of mass ratios considered here. For a given mass-ratio, a CND-scale perturber can affect the disc more efficiently only if its eccentricity is large enough for the pericentre to enter the disc, as in the high-eccentricity cases shown in Appendix~\ref{App:disruption regimes}. Increasing $e_\IMBH$ generally spreads the affected region over a wider radial range, while changing $\theta_\IMBH$ mainly shifts the fragment band vertically in $m_\IMBH/M_{\rm d}$ and changes its width. Varying $\Gamma$ produces a more moderate shift of the boundaries: for shallower profiles, $\Gamma\lesssim1$, it is easier to disrupt the inner regions, whereas for steeper profiles it is easier to disrupt the outer regions. Thus, disc fragmentation at large radii by a massive perturber at $a_\IMBH\simeq1.5$ pc may suggest a steeper surface-density profile for the disc.

The goal of this paper is to test these VRR-model predictions using direct $N$-body simulations to understand when disc-fragmentation or dispersion takes place, and to determine the conditions necessary to reproduce the observations. While the analytical VRR-model fragmentation/dispersion conditions presented here provide a basic understanding of the complex phenomena at play, the direct $N$-body simulations are useful to test the validity of the simplified VRR model and allow us to test the relevance of other relaxation processes which are neglected by this simple VRR model. Such processes include scalar resonant relaxation and two-body relaxation. Thus, in the following sections, we shift our focus to direct $N$-body simulations.

\section{Simulations and initial conditions}
\label{sec:simulations}

In this section, we describe the computational framework employed to investigate the dynamics of galactic nuclei. The simulations include four components: a central supermassive black hole (SMBH), a nuclear star cluster (NSC), a stellar disc, and an IMBH placed on an inclined orbit with respect to the disc. The following subsections provide detailed descriptions of each component and the numerical methods used in our simulations.

\subsection{The code}
\label{subsec:code}

We use a modified version of the direct $N$-body code phi-GRAPE \citep{HarfstEtAl2007}, which employs a 4th-order Hermite integration method to solve the equation of motion. The code was originally designed for the GRAPE cards and now utilises an emulation library to run on modern GPUs. This modified version of the code includes the gravitational interaction between stars and a massive central object, implemented as an external point-mass potential, as well as the accretion of stars onto the central object. The equation of motion is:

\begin{equation}\label{eq:motion}
\ddot{\vec{r}}_{i}=-\sum_{i\not= j}\frac{Gm_{j}\vec r_{ij}}{(r_{ij}^2 + \epsilon_\mathrm{ss}^2)^{3/2}} -
\frac{G\MSMBH\vec r_{i}}{r_{i}^3} \,, 
\end{equation}
where $\vec r_{ij}=\vec r_{i}-\vec r_{j}$ with $\vec r_{i}$,
$\vec r_{j}$ the positions of stars $i$ and $j$, respectively, $\epsilon_\mathrm{ss} = 5.0\times10^{-5}$ pc is the stellar softening parameter to prevent singularities when the objects are very close to each other. 

The central massive black hole can grow by consuming stars. To determine which stars can be accreted, we use a distance threshold based on the tidal disruption radius of a $2R_\odot$ star by a $4\times10^6\msun$ black hole. When a star passes within this distance, it is absorbed by the black hole and its mass is added to the black hole's total mass. 

In our simulations, we use a system of units where $G=\MSMBH=R_\mathrm{out}=1$, where $R_\mathrm{out}$ is the outer radius of the stellar system which is defined as the orbital semi-major axis of the outermost star in the system. We convert these units to physical units by assuming $R_\mathrm{out}=0.5$~pc and $\MSMBH=4\times10^6\msun$.

\subsection{The nuclear star cluster -- a fixed isotropic potential}
\label{subsec:NSC}

The galactic nuclei considered here are comprised of a stellar disc, a spherically symmetric NSC, an IMBH, and an SMBH.

Direct $N$-body summation of this full system is computationally expensive at the relevant distance scales. To span a broad parameter range at reasonable cost, we model the spherical stellar component with a fixed Plummer potential \citep{Plummer1911} of total mass $M_\mathrm{Pl}=2\times10^5\,\msun$ and scale radius $r_0=0.5~\mathrm{pc}$, chosen to match the enclosed mass at $0.5~\mathrm{pc}$ in the `live' sphere models of \citet{Panamarev2022}. This non-rotating spherically symmetric background drives a rapid apsidal precession for the disc stars and the IMBH, but -- being isotropic without stochastic fluctuations -- it does not generate VRR torques and so it does not reorient the orbital planes. We emphasise that we neglect the stochastic fluctuations of the spherical component in this work.

\begin{table}
\caption{List of simulations with IMBHs}
\label{tab:runs}
\centering
\begin{tabular}{|c|c|c|c|c|c|}
\hline
\( \theta_\IMBH \) [\( ^\circ \)] & \( a_\IMBH \) [pc] & \( e_\IMBH \) & \( m_\IMBH \) [\( \msun \)] & \( m_\IMBH/M_\mathrm{d} \) & thermal \\
\hline
0   & 0.05 & 0.33 & 500   & 0.17 &  \\
45  & 0.05 & 0.00 & 500   & 0.17 &  \\
45  & 0.05 & 0.33 & 250   & 0.08 &  \\
45  & 0.05 & 0.33 & 500   & 0.17 &  \\
45  & 0.05 & 0.33 & 2000  & 0.67 &  \\
45  & 0.05 & 0.70 & 500   & 0.17 &  \\
45  & 0.05 & 0.90 & 500   & 0.17 &  \\
45  & 0.15 & 0.33 & 2000  & 0.67 & \checkmark \\
90  & 0.15 & 0.33 & 2000  & 0.67 &  \\
135 & 0.05 & 0.10 & 2000  & 0.67 &  \\
135 & 0.05 & 0.33 & 2000  & 0.67 &  \\
135 & 0.05 & 0.33 & 500   & 0.17 &  \\
135 & 0.05 & 0.70 & 2000  & 0.67 &  \\
135 & 0.15 & 0.10 & 250   & 0.08 &  \\
135 & 0.15 & 0.10 & 2000  & 0.67 &  \\
135 & 0.15 & 0.33 & 250   & 0.09 &  \\
135 & 0.15 & 0.33 & 500   & 0.17 &  \\
135 & 0.15 & 0.33 & 2000  & 0.67 &  \\
135 & 0.15 & 0.70 & 250   & 0.08 &  \\
135 & 0.15 & 0.70 & 500   & 0.17 &  \\
135 & 0.15 & 0.70 & 2000  & 0.67 & \checkmark \\
135 & 0.15 & 0.90 & 2000  & 0.67 &  \\
150 & 0.05 & 0.10 & 500   & 0.17 &  \\
150 & 0.15 & 0.10 & 2000  & 0.67 &  \\
150 & 0.15 & 0.33 & 2000  & 0.67 &  \\
150 & 0.15 & 0.33 & 500   & 0.17 &  \\
150 & 0.15 & 0.70 & 2000  & 0.67 &  \\
150 & 0.15 & 0.90 & 500   & 0.17 &  \\
150 & 0.15 & 0.90 & 2000  & 0.67 &  \\
180 & 0.05 & 0.33 & 2000  & 0.67 &  \\
\hline
\end{tabular}
\par\medskip
\begin{flushleft}\textbf{Notes.} The IMBH parameters in the simulated models in this paper. Columns denote: \(\theta_\IMBH\) the initial inclination angle of the IMBH relative to the plane of the stellar disc, measured in degrees; \(a_\IMBH\) the initial semi-major axis expressed in parsecs; \(e_\IMBH\) the eccentricity; \(m_\IMBH\) the mass of the IMBH in solar masses; \(m_\IMBH/M_\mathrm{d}\) the ratio of the IMBH's mass to the total mass of the stellar disc. The column thermal designates models that were executed with a `thermal' stellar disc. The mass of the disc was set to $M_\mathrm{d}\simeq3000\Msun$ in all models.\end{flushleft}
\end{table}

\subsection{The stellar disc}
\label{subsec:stellar-disc}

We examine the formation of the stellar disc from interactions between stars and a gaseous accretion disc in a previously active galactic nucleus. In this scenario, the drag force experienced by stars during crossings of the gaseous disc dissipates their kinetic energy, leading to their capture within the gaseous disc. This process results in the subsequent formation of a stellar disc \citep{Panamarev2018}. Our simulations begin at the point when the gaseous disc has disappeared, leaving behind the disc of stars. Under these conditions, the stars initially follow nearly circular orbits aligned with the plane of the former gaseous disc \citep{Panamarev2018}. The disc exhibits an outer warp and a correlation of inclination and eccentricity with semimajor axis: stars closer to the SMBH are on more circular, lower-inclination orbits.\footnote{\label{footnote:stardisk} In the nomenclature of \citet{Panamarev2022}, this is the \textsc{stardisc} model; cf. their Fig. 3.} 

Additionally, we explore selected models featuring an alternative set of initial conditions for the stellar disc. In these models, the eccentricities of the stars follow a thermal distribution, and their inclination angles are uniformly sampled between \(\cos 0\) and \(\cos10^\circ\).\footnote{This variant is termed \textsc{thermal} in \citet{Panamarev2022}.} 
Both the \textsc{stardisc} and \textsc{thermal} models feature a power-law 3D density slope \(\rho \propto r^{-2.3}\) and adhere to a \citet{Kroupa2001} top-heavy initial mass function (IMF). For a comprehensive description of the initial conditions, including IMF power-law indices and break points, as well as a graphical representation of the initial disc models, we refer the reader to our previous work \citep{Panamarev2022}.

Alternative pathways for the formation of the stellar disc include the fragmentation of the gaseous accretion disc, which also results in stars predominantly on circular orbits \citep{LevinBeloborodov2003, Levin2007}. Another scenario involves a giant molecular cloud being accreted by the SMBH, leading to a stellar disc with stars initially on more eccentric orbits \citep{Bonnell2008, Generozov2022}. While the present work does not explicitly investigate these alternative formation pathways, it is worth noting that the \textsc{stardisc} model may serve as a reasonable approximation for the disc fragmentation scenario due to its emphasis on circular orbits, whereas the \textsc{thermal} model could capture some characteristics of the molecular cloud scenario, given its population of more eccentric orbits.

We also note that the observed eccentricity distribution of young stars in the Galactic centre represents a superthermal distribution for the stars within 0.04 pc (the so-called S-cluster) and a distribution peaked around $e\sim 0.3$ for the young stars between 0.04 pc and 0.4 pc (including the clockwise disc and the counterclockwise structure). The \textsc{stardisc} initial condition leads to an eccentricity distribution reminiscent of the latter population after 6 Myr, the approximate age of that component \citep{Genzel2010, Habibi2017}.

\subsection{The intermediate-mass black hole}
\label{subsec:models}

In our simulations, the IMBH is modelled as a distinct particle within the direct $N$-body simulation framework. It is subject to the same gravitational interactions as the other particles, albeit with a significantly higher mass. We continuously trace the IMBH's evolution throughout the simulation to analyse its dynamical behaviour. The IMBH's initial conditions are varied to explore a range of orbital parameters. Specifically, we adjust the inclination angles relative to the plane of the stellar disc to examine both prograde and retrograde orbits. Additionally, we vary the eccentricities and semi-major axes of the IMBH's orbit. The assumed mass ratio between the IMBH and the stellar disc ranges from \(\approx 0.67\) (corresponding to \(m_{\IMBH} = 2000 \msun\)) to \(\approx 0.08\) (\(m_{\IMBH} = 250 \msun\)). A comprehensive list of all the models employed in this study is presented in Table~\ref{tab:runs}.

While several formation scenarios for IMBHs have been proposed as listed in the introduction, the focus of this study is not on the IMBH's formation mechanisms. Instead, we assume an IMBH was already present in the Galactic centre when the stellar disc formed. This assumption allows us to explore a broad parameter space, as detailed in Table~\ref{tab:runs}, thereby capturing a wide array of potential IMBH formation scenarios.

\section{Effect of the disc on an IMBH: resonant dynamical friction}
\label{sec:imbh-evol}

Following \citet{Szolgyen2021}, we categorise the stars in the disc into three regions based on their orbital parameters relative to those of the IMBH. Let \( r_{\rm a,*} \) and \( r_{\rm p,*} \) denote the apocentre and pericentre of a star, respectively, and \( r_{\rm a,\IMBH} \) and \( r_{\rm p,\IMBH} \) denote the apocentre and pericentre of the IMBH. We distinguish three distinct regions of the stellar disc based on the radial range of the orbits as follows:
\begin{equation}
\label{eq:orb-cond}
\begin{aligned}
&\text{Inner:} \quad && r_{\rm a,*} < r_{\rm p,\IMBH}, \\
&\text{Overlapping:} \quad && r_{\rm a,*} \geq r_{\rm p,\IMBH} \text{~and~} r_{\rm p,*} \leq r_{\rm a,\IMBH}, \\
&\text{Outer:} \quad && r_{\rm p,*} > r_{\rm a,\IMBH}.
\end{aligned}
\end{equation}
In this notation:
\begin{itemize}
    \item The `inner' region contains stars with orbits entirely within the IMBH's orbit.
    \item The `radially overlapping' (middle) region contains stars with orbits that intersect the IMBH's orbit.
    \item The `outer' region contains stars with orbits entirely outside the IMBH's orbit.
\end{itemize}

\begin{figure*}
    \centering
    \includegraphics[width=0.8\textwidth]{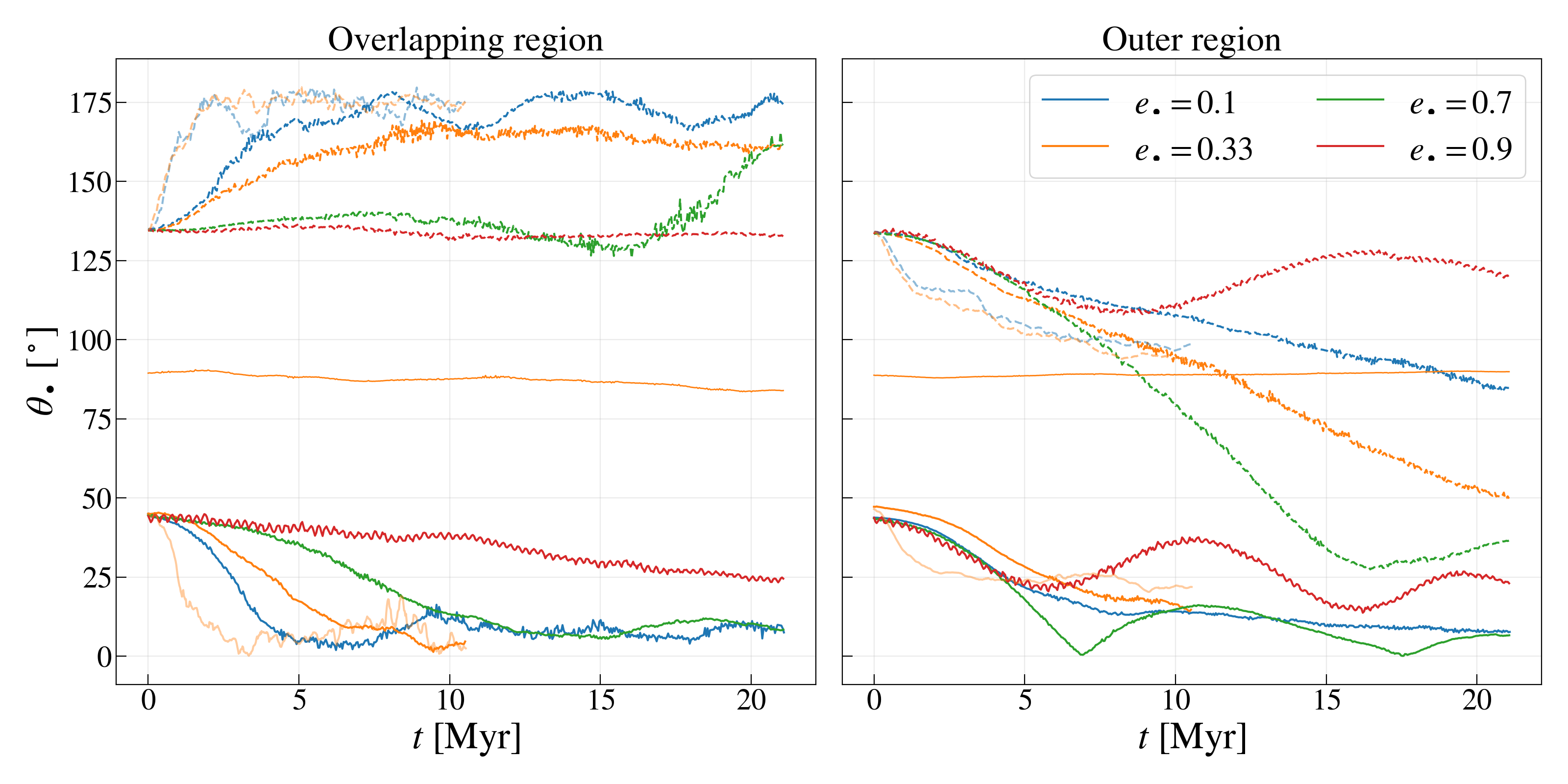}
    \caption{Inclination angles of an IMBH on prograde ($\theta_\IMBH=45^\circ$, solid lines), retrograde ($\theta_\IMBH=135^\circ$, dashed lines), and orthogonal orbit ($\theta_\IMBH=90^\circ$, solid orange line) with respect to the stellar disc over time. Models are shown for an initial semimajor axis of \(a_{\IMBH} = 0.15\) pc and mass of \(m_{\IMBH} = 2000\,\msun\). Additionally, three faint lines represent models with an initial semimajor axis of \(a_\IMBH=0.05\) pc. The left panel displays the inclination angles relative to the angular momentum vector of the stars in the overlapping region of the disc. The right panel shows the inclination angles of the IMBH with respect to the outer part of the disc. Lines of different colours correspond to different values of the IMBH's initial eccentricity, as indicated in the legend.  The figure demonstrates that in the prograde models the angular momentum vectors of the IMBH and the stars in the overlapping region align; in the retrograde scenario, the stars in the overlapping region \textit{anti}-align with the IMBH (except for the highly eccentric orbits where the evolution is much slower). In models with $e_\IMBH\leq0.33$, the IMBH also aligns with respect to the outer region of the disc, while models with higher eccentricity show oscillatory behaviour indicating precession of the IMBH with respect to the outer region of the disc in the prograde case. Models with initially retrograde IMBH approach $90^\circ$ with respect to the outer region around 10 Myr. This is the \textsc{stardisc} model$^{\ref{footnote:stardisk}}$ with nearly circular initial eccentricity for the disc.}
    \label{fig:inclination_angles}
\end{figure*}

In this section, we extend the investigation of \citet{Szolgyen2021}, by considering an IMBH with a mass of the order of the disc mass and our analysis encompasses a range of eccentricities for both prograde and retrograde orbits relative to the disc.

\subsection{Alignment and anti-alignment}
The process of vector resonant relaxation is expected to facilitate the alignment of an IMBH with the stellar disc, particularly when the mass of the IMBH is significantly less than that of the disc (\( m_{\IMBH} \ll M_\mathrm{d} \)) and it is on a prograde orbit \citep{Szolgyen2021, Ginat2023}. Our simulations show that this alignment effect persists even when the mass of the IMBH is of the same order as the disc mass, as depicted in Figure~\ref{fig:inclination_angles}. This is evident from the lines in the lower part of the left panel, which indicate pronounced alignment within the overlapping region of the disc. However, the right panel shows that the IMBH maintains a slightly larger inclination angle with respect to the outer region of the disc and exhibits oscillations when on eccentric orbits. These oscillations suggest that the IMBH's angular momentum vector precesses around that of the outer disc. This indicates that the inner disc or the overlapping region is misaligned relative to the outer parts of the disc. This is consistent with the conservation of angular momentum: the IMBH, with its significant angular momentum, influences the dynamics within the overlapping region, promoting the alignment of the stellar disc and the IMBH with respect to their total angular momentum vector. Consequently, the relative angle between the inner and outer discs is less than the initial orbital inclination angle of the IMBH on a prograde orbit with respect to the disc.

In contrast, an IMBH on retrograde orbits, as indicated by the upper lines in Figure~\ref{fig:inclination_angles}, promotes an anti-alignment between its angular momentum vector and that of the stars in the overlapping region of the disc. In this configuration, the angular momenta of the disc's stars become redistributed into a common plane, albeit with directions opposite to that of the IMBH. This is expected from our semi-analytical model presented in Sec.~\ref{subsec:torques}: the radially overlapping stars tend to behave as test particles and precess about the negative angular momentum vector of the IMBH, as shown in Fig.~\ref{fig:L_unit150}. Notably, for models with lower eccentricities (\( e_{\IMBH}=0.33 \) and \( e_{\IMBH}=0.1 \)), the progression towards anti-alignment proceeds at a rate comparable to the alignment in the prograde scenario. However, this process is much slower for a higher IMBH eccentricity (or lower IMBH mass, or angular momentum): anti-alignment starts only after 15 Myr of the evolution for \( e_{\IMBH}=0.7 \), culminating in a nearly coplanar arrangement at around 20 Myr.\footnote{In this model, the stellar disc fragments into several substructures; in particular, the inner and middle discs start to drift away from each other in angular momentum space (cf. Figure~\ref{fig:skymaps_plum_to20}).} The IMBH with \( e_{\IMBH}=0.9 \) does not show any anti-alignment throughout the 20 Myr simulation, maintaining a nearly constant inclination relative to the overlapping region, precessing around the angular momentum vector of the outer disc. Furthermore, the relative angle between the inner and outer regions is significantly greater than that observed for prograde orbits, indicating a clear separation between the inner and outer discs, which we quantify below in Section~\ref{sec:disc-imbh}. Finally, the inclination angle of the IMBH on an initially orthogonal orbit with respect to the disc maintains this configuration with both the inner and outer regions, suggesting that the mutual torques, which drive angular momentum exchange in the other models, are in equilibrium. This was also seen in the semi-analytical model; the precession period tends towards infinity as the inclination approaches $90^\circ$.

\begin{figure*}
    \centering
    \includegraphics[width=\textwidth]{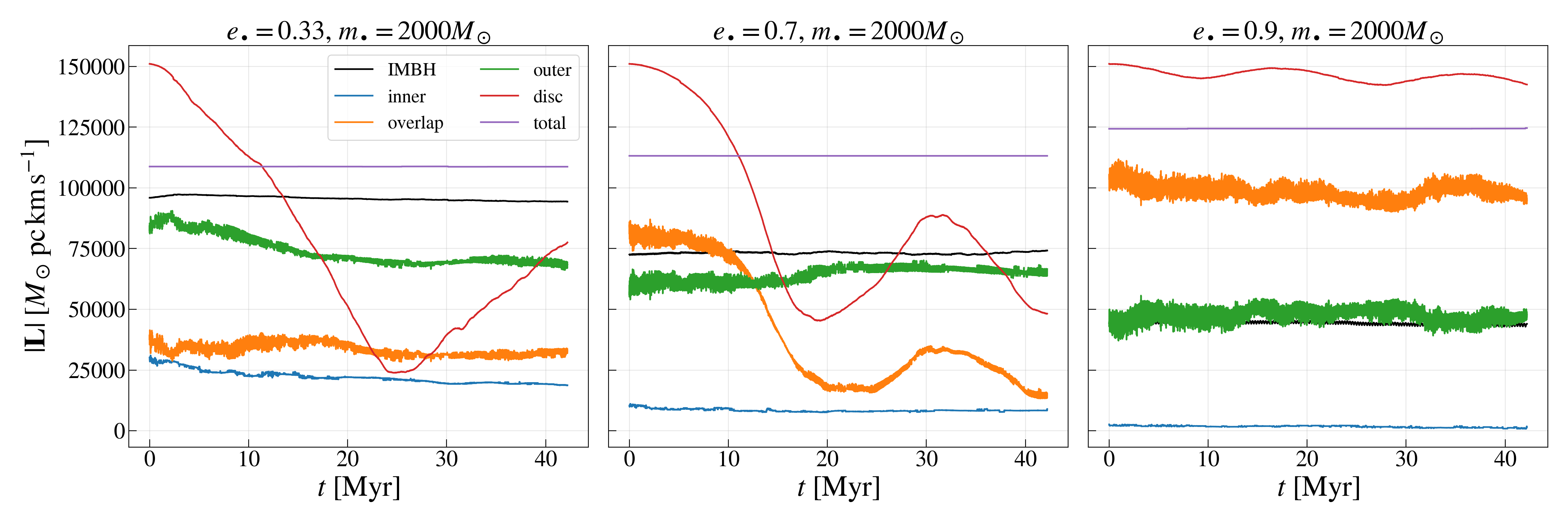}
    \caption{Evolution of the angular momenta of the IMBH and the stellar disc over time for a retrograde IMBH model with $m_\IMBH=2000\msun$, $a_\IMBH = 0.15$ pc, $\theta_\IMBH=135^\circ$, and different values of the IMBH's initial eccentricity as labelled. The black line represents the angular momentum of the IMBH; the blue, orange, and green lines represent the angular momenta of the inner, middle, and outer regions of the stellar disc, respectively (see Eq.~\ref{eq:orb-cond}); the red line represents the magnitude of the total angular momentum of the disc (typically misaligned with respect to the IMBH); and the purple line represents the total angular momentum of the system (IMBH and stellar disc showing angular momentum conservation). In the left panel ($e_\IMBH=0.33$), the magnitude of the disc's angular momentum drastically changes, while that of the inner, middle, and outer components stays roughly constant, indicating that the components maintain their thin structure but 
    become misaligned with each other. In the middle panel ($e_\IMBH=0.7$), the middle (overlapping) region fragments into separate substructures (this is where the orange line starts decreasing). After the angular momentum magnitude crosses a critical value, the IMBH starts dominating the evolution of the stars in the overlapping region and drives anti-alignment with itself. In contrast, the right panel ($e_\IMBH=0.9$) shows that for low IMBH angular momentum the entire disc can generally preserve its structure, showing only weak oscillations in the total angular momentum value indicating a low amount of internal precession.}
    \label{fig:angular_mom_mag}
\end{figure*}

\begin{figure}
    \centering
    \includegraphics[width=\columnwidth]{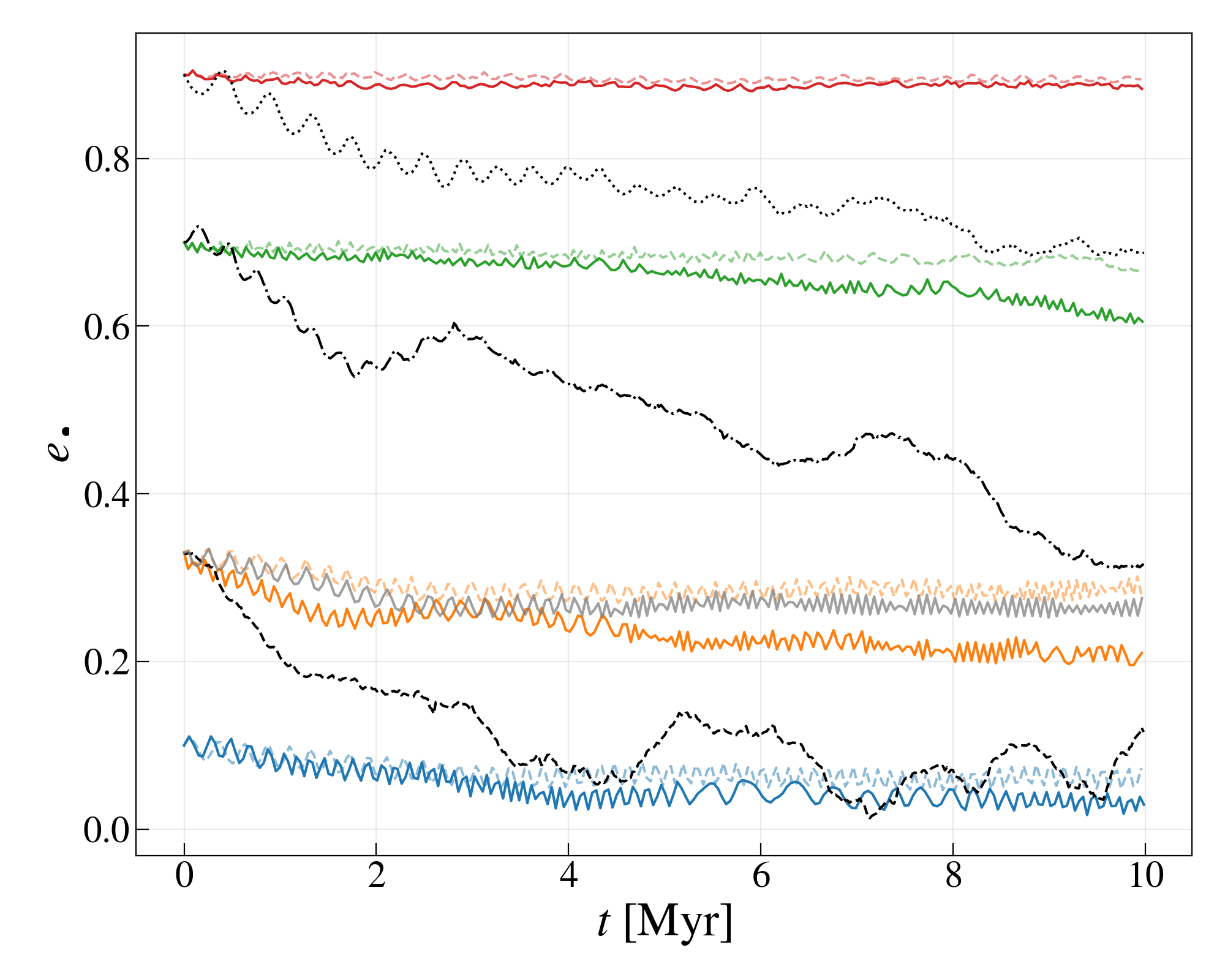}
\caption{Evolution of eccentricities for the IMBH with $m_{\IMBH}=2000\msun$ (coloured lines) and with $m_{\IMBH}=500\msun$ (dashed, dotted, and dash-dotted black lines). Solid coloured lines correspond to prograde orbits (initial inclination angle of $45^{\circ}$ with respect to the disc), while faint dashed lines represent retrograde orbits ($\theta_\IMBH = 135^{\circ}$). Different colours show different initial eccentricity. The grey line indicates the orthogonal orbit $\theta_\IMBH=90^{\circ}$.}
    \label{fig:ecc_evol}
\end{figure}

\subsection{Alignment vs. angular momentum hierarchy}

Our simulations suggest that the IMBH must possess a substantial amount of angular momentum in order to drive anti-alignment between its angular-momentum vector and that of the overlapping region. Figure~\ref{fig:angular_mom_mag} shows the time evolution of the angular momenta of the IMBH and the stellar disc for different IMBH masses and eccentricities. In each panel, the IMBH angular momentum (initially $\theta_\IMBH=135^\circ$) is shown in black, while the angular momenta of the inner, overlapping, and outer disc regions are shown in blue, orange, and green, respectively. The total angular momentum of the disc is shown in red, and the total angular momentum of the system (IMBH plus stellar disc) is shown in purple. Each panel corresponds to a different model. The evolution is followed for an extended period of 40 Myr.

The figure indicates that anti-alignment occurs when the IMBH angular-momentum magnitude exceeds that of the overlapping disc region.

The left panel illustrates a model in which the IMBH angular momentum dominates (the black line lies above the orange line). In this case, alignment is achieved within $\sim 6$ Myr (cf. the left panel of Fig.~\ref{fig:inclination_angles}). By contrast, the right panel shows a model in which the angular momentum of the overlapping region dominates over that of the IMBH throughout the simulation. In this case, the inclination angle of the IMBH relative to the overlapping region remains approximately constant in time (cf. Fig.~\ref{fig:inclination_angles} left panel red dashed line).

Finally, the middle panel in Figure~\ref{fig:angular_mom_mag} shows an intermediate case in which the IMBH and overlapping-region angular momenta are initially comparable. Over time, however, the angular momentum of the overlapping region (and of the disc as a whole) decreases due to internal differential precession (cf. Fig.~\ref{fig:skymaps_plum_to20} below). As a result, after $\sim 10$ Myr the IMBH angular momentum becomes larger than that of the overlapping region, and its dominance continues to grow. Once the ratio crosses a critical threshold (around $\sim 15$ Myr; see the left panel of Fig.~\ref{fig:inclination_angles}), the IMBH is able to drive anti-alignment. Overall, the decrease in the disc's total angular momentum is primarily due to differential precession, which progressively misaligns stellar angular momenta, and only to a much lesser extent to disc thickening and the increase in eccentricity.

\subsection{IMBH eccentricity evolution}

The simulations by \citet{Szolgyen2021} demonstrated that the orbits of \emph{prograde} IMBHs undergo circularisation during the alignment phase. Figure~\ref{fig:ecc_evol} shows the time evolution of the eccentricities in our models for $m_{\IMBH}=2000\,\msun=0.67 M_{\rm d}$ (coloured lines) and $m_{\IMBH}=500\,\msun=0.17 M_{\rm d}$ (black lines). 

\begin{figure*}
    \centering
    \includegraphics[width=\textwidth]{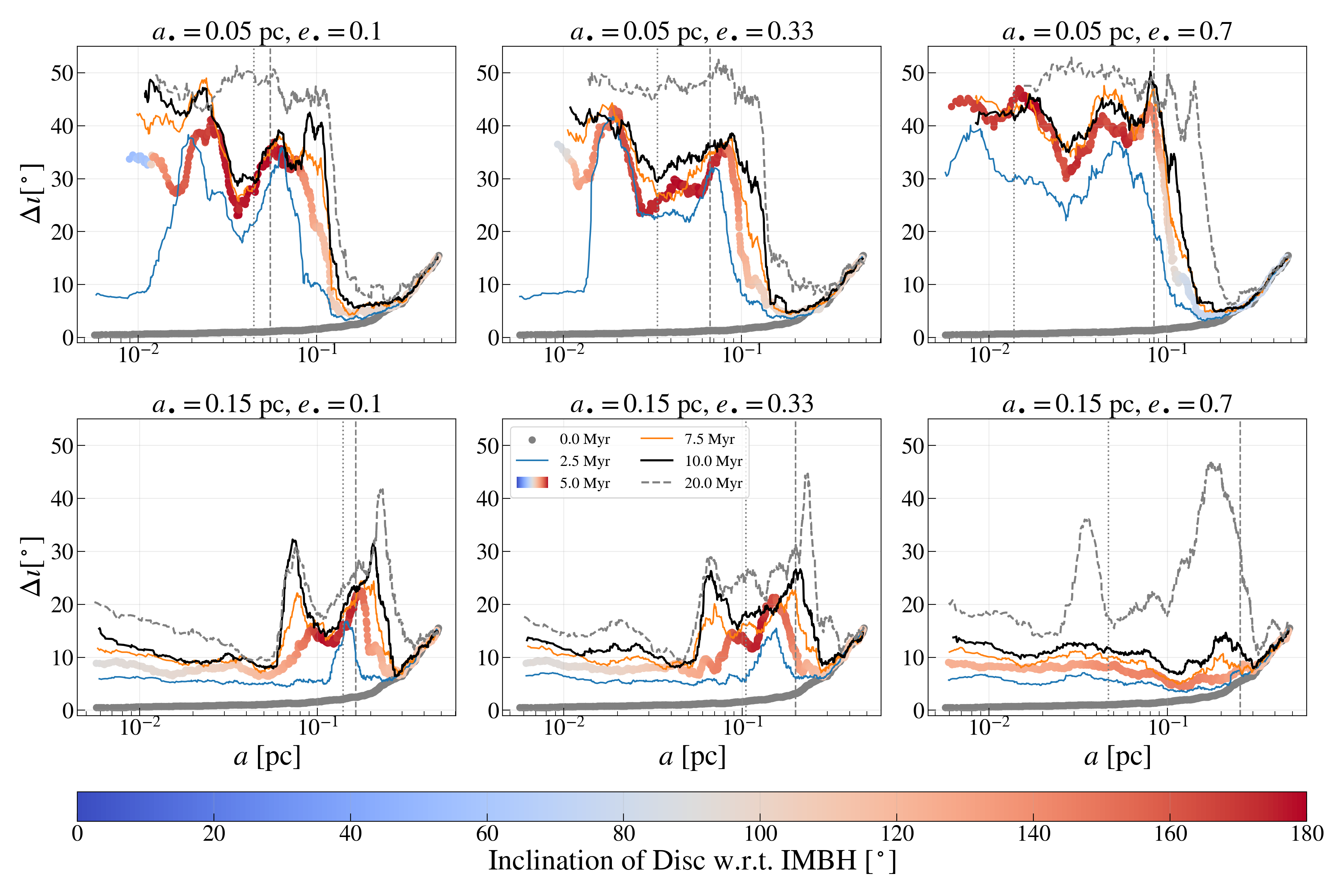}
    \caption{The local thickness and warp of the stellar disc ($\Delta \iota$, Eq.~\ref{eq:iota}) as a function of semimajor axis for \textit{retrograde} models with initial inclination $\theta_\IMBH=135^{\circ}$, for different IMBH's initial semimajor axis (0.05 pc in top row panels, and 0.15 pc in bottom row) and eccentricity (0.1, 0.33, 0.7 from left to right panels). The evolution is traced at time intervals of 0, 2.5, 5, 7.5, 10, and 20 Myr, with each line representing the local disc thickness at these epochs (see legend). The 5 Myr lines (thick curves) are overlaid with a gradient of shades, with blue tones indicating alignment (prograde motion) and red tones indicating anti-alignment (retrograde motion) of the disc's angular momentum vector with respect to the IMBH; the thick grey curves show the initial conditions, and the dashed grey curves show the state at 20 Myr. Each data point shows the disc thickness for 50 stars and their median semimajor axis (corresponding to a range of approximately a factor of 2 in semimajor axis). Vertical lines within each panel mark the peri- and apocentre positions of the IMBH's orbit. 
    }
    \label{fig:warp_plum}
\end{figure*}

For \emph{prograde} orbits, our results are in agreement with \citet{Szolgyen2021} for the low-mass IMBH case, $m_{\IMBH}=500\,\msun$ ($m_{\IMBH}/M_\mathrm{d}=0.17$): the eccentricity decreases efficiently on Myr time-scales. In particular, the orbit with initial eccentricity $e_{\IMBH}=0.33$ is circularised by $\sim 5$ Myr, while the orbit starting at $e_{\IMBH}=0.7$ decreases to $e_{\IMBH}\simeq 0.3$. In contrast, for the higher-mass prograde IMBH, $m_{\IMBH}=2000\,\msun$ ($m_{\IMBH}/M_\mathrm{d}=0.67$), circularisation is much less efficient: for low and intermediate initial $e_{\IMBH}$ the eccentricity shows a weak but systematic decrease, while at the highest eccentricity it remains nearly constant over 10 Myr.

For \emph{retrograde} orbits, \citet{Szolgyen2021} found that the eccentricity remains nearly constant. We find the same behaviour in our simulations, and this agreement persists even for the higher IMBH mass: the retrograde eccentricities show minimal evolution over the 10 Myr time-scale (Figure~\ref{fig:ecc_evol}). Finally, the \emph{orthogonal} configuration (grey line) also exhibits little eccentricity evolution, closely resembling the retrograde case. Overall, for $m_{\IMBH}=2000\,\msun$ the prograde and retrograde eccentricity evolution differs mainly at low and intermediate $e_{\IMBH}$: the solid coloured curves (prograde) show a systematic decrease in $e_{\IMBH}$, whereas the corresponding faint dashed curves (retrograde) remain nearly constant; at the highest eccentricity (red curves) both orbit configurations show little evolution over 10 Myr.

\section{Effects of an IMBH on the properties of the stellar disc}
\label{sec:disc-imbh}

\subsection{Thickness and warp}
Adopting the approach of \citet{Szolgyen2021}, we characterise the disc's structure through the quadrupole moment tensor formed from the angular momenta as
\begin{equation}\label{eq:Q}
Q_{\alpha\beta}=\dfrac{\sum_{i=1}^N L_{i\alpha}L_{i\beta}}{\sum_{i=1}^N{|\boldsymbol{L}_{i}|^2}},
\end{equation}
where \( \boldsymbol{L}_{i} \) denotes the angular momentum of the \( i^{\rm th} \) star and $\alpha$ and $\beta$ are Cartesian component indices. The principal eigenvalue \( q \) of \( Q_{\alpha\beta} \) quantifies the disc's flatness, which may also be calculated as the weighted mean cosine squared of the stellar inclination angles \( \iota_i \) with respect to the principal eigenvector\footnote{The principal eigenvector does not necessarily align with the total angular momentum vector, but for highly rotating systems they nearly align.}:
\begin{equation}\label{eq:q}
    q = \langle \cos^2 \iota \rangle = \dfrac{\sum_{i=1}^N |\boldsymbol{L}_{i}|^2 \cos^2 \iota_i}{\sum_{i=1}^N |\boldsymbol{L}_{i}|^2}.
\end{equation}
The disc's warp and thickness are inferred from the root-mean-square warp angle \( \Delta \iota \):
\begin{equation}\label{eq:iota}
    \Delta \iota = \cos^{-1}(\sqrt{q}).
\end{equation}
This angle is zero for an ideally planar disc and approaches \( 54.7^\circ \) in an isotropic distribution. Note that this angle is independent of the disc's overall tilt, quantifying its thickness or warp. If $q$ and $\Delta \iota$ are calculated for a restricted narrow range of orbital radii (or semimajor axis) rather than over all stars, then these quantities represent the disc thickness at the given location.

Here we focus on initially retrograde models with $m_\IMBH=2000\msun$. The observed anti-alignment between the IMBH and the overlapping region of the disc, as discussed in the previous subsection, implies that the disc breaks into fragments. To quantify this fragmentation, we analyse the warp angle across the stellar disc's semimajor axis at specific snapshots for selected retrograde models with $\theta_\IMBH = 135^\circ$. We sort the stars by their semimajor axes and calculate a rolling value of the warp angle (Eq.~\ref{eq:iota}) over windows of 50 stars. Minima in the resulting plot indicate the thinnest regions of the disc, while maxima identify potential points where the disc might fragment. The total count of these minima provides the number of distinct disc segments.

Figure~\ref{fig:warp_plum} presents the warp angle evolution for models with the initial semimajor axis of the IMBH set at \( a_\IMBH = 0.05 \) pc (top panels) and \( a_\IMBH = 0.15 \) pc (bottom panels), each with different eccentricities (\( e_\IMBH=0.1, 0.33, \) and \( 0.7 \), from left to right). Each panel captures the warp angle at discrete time intervals: \( t=0, 2.5, 5.0, 7.5, 10, \) and \( 20 \) Myr. Notably, the 5 Myr line is enhanced with a colour map that indicates the inclination angle of the disc region relative to the IMBH. The majority of the models delineate three distinct disc segments (with the exception of the model at \( a_\IMBH = 0.15 \) pc and \( e_\IMBH=0.7 \)), and those with the closer IMBH (\( a_\IMBH=0.05 \) pc) demonstrate markedly thicker inner discs. This is anticipated due to the IMBH's shorter orbital period, exerting a more pronounced effect on the inner disc regions and accelerating the dynamical evolution. Conversely, models with the more distant IMBH (\( a_\IMBH = 0.15 \) pc) show a much thinner inner disc, as the IMBH's influence is insufficient to cause significant disruption within the observed timeframe. The thick red lines in the figure indicate a counter-alignment of the IMBH with respect to the local disc orientation. Across all models, the disc's outermost region remains unaltered by the IMBH's presence within 10 Myr. The model with \( a_\IMBH = 0.15 \) pc and \( e_\IMBH=0.7 \) fragments more slowly: at 10 Myr it has not yet developed the three-segment structure seen in the other models,
but by 20 Myr it shows comparable fragmentation, in line with the trends noted in Figure~\ref{fig:inclination_angles}. The dashed grey curves in each panel show the disc state at 20~Myr, confirming that by this time even this model ($a_\IMBH=0.15$ pc, $e_\IMBH=0.7$) develops comparable disc fragmentation within the disruption zone.

\begin{figure*}
    \centering
    \includegraphics[width=\textwidth]{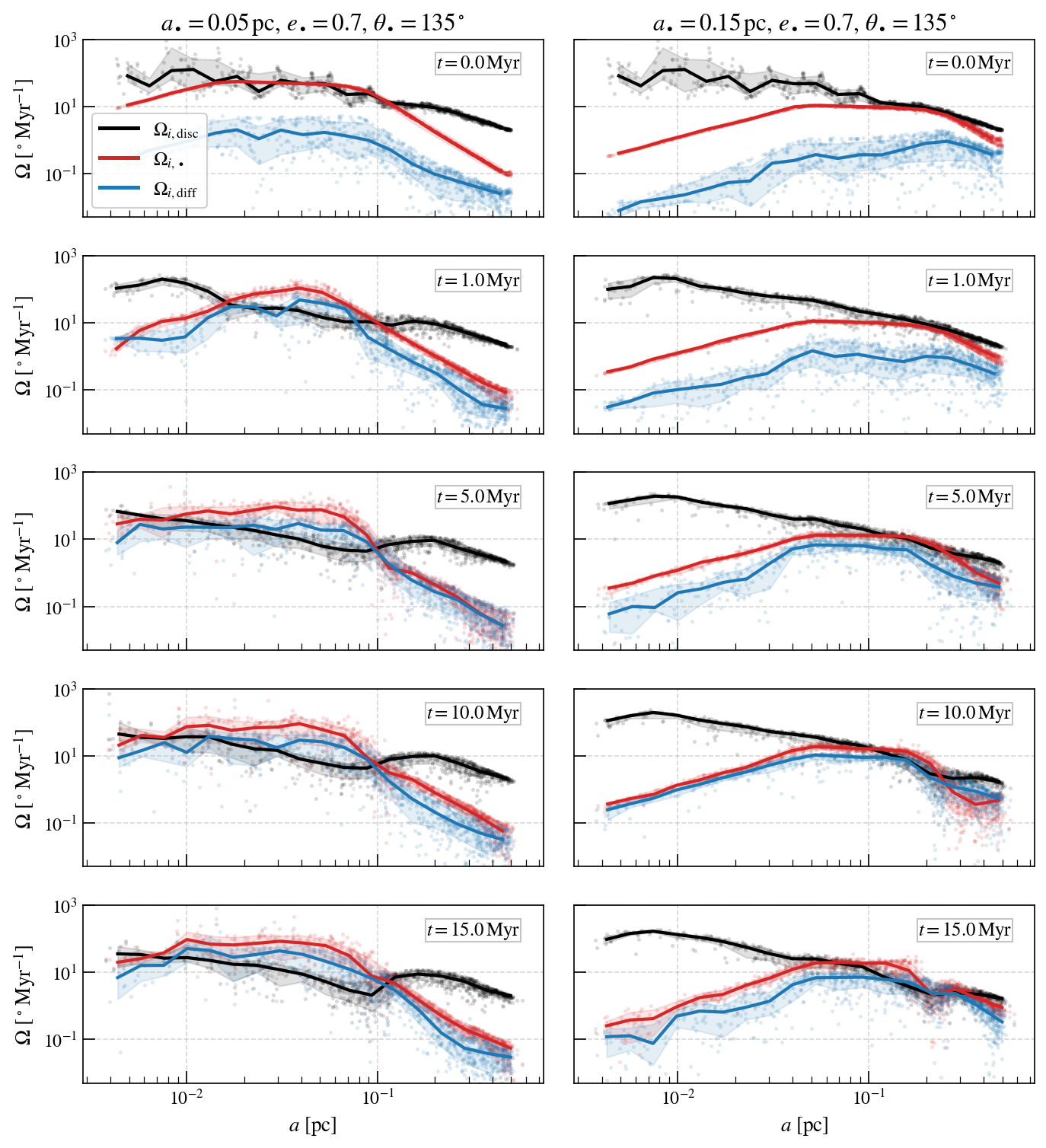}
    \caption{Precession rates governing the relevant components of disc disruption as in Fig.~\ref{fig:disc-disruption} but evaluated directly for the $N$-body simulation snapshots of the retrograde models with inclination $\theta_\IMBH=135^\circ$ and eccentricity $e_\IMBH=0.7$, as a function of stellar semimajor axis $a$. Columns correspond to IMBH semimajor axes $a_\IMBH=0.05$\,pc (left) and $a_\IMBH=0.15$\,pc (right); rows show snapshots at $t=0,1,5,10,$ and $15$\,Myr (top to bottom). In each panel the three rates are the disc self-torque $|\mathbf{\Omega}_{i,\mathrm{disc}}\times\hat\L_i|$ (black), the direct IMBH rate $|\mathbf{\Omega}_{i,\IMBH}\times\hat\L_i|$ (red), and the IMBH differential precession rate $\Omega_{i,\mathrm{diff}}$ (blue, Eq.~\eqref{eqn:fundamental criterion}). Faint points show individual disc stars; solid lines are their running medians and shaded bands the 16th--84th percentile range. The disc is locally disrupted at radii where the blue curve rises above the black one.}
    \label{fig:snap_torques}
\end{figure*}

To test the disc disruption criterion of Eq.~\eqref{eqn:fundamental criterion} using the simulated stellar discs, Fig.~\ref{fig:snap_torques} shows three precession rates, computed on the actual $N$-body snapshots of the retrograde ($\theta_\IMBH=135^\circ$, $e_\IMBH=0.7$) models. The disc self-torque rate $|\mathbf{\Omega}_{i,\mathrm{disc}}\times\hat\L_i|$ (black) measures how strongly the disc holds itself together in angular-momentum space, whereas the IMBH differential rate $\Omega_{i,\mathrm{diff}}$ (blue, the right-hand side of Eq.~\eqref{eqn:fundamental criterion}) measures how strongly the IMBH shears neighbouring disc orbits apart; the direct IMBH rate $|\mathbf{\Omega}_{i,\IMBH}\times\hat\L_i|$ (red) is shown for reference. The disc is locally disrupted wherever $\Omega_{i,\mathrm{diff}}\gtrsim|\mathbf{\Omega}_{i,\mathrm{disc}}\times\hat\L_i|$. 

Initially ($t=0$) the disc self-torque exceeds the IMBH differential torque across essentially all radii, so the disc remains intact. 
As the system evolves, $\Omega_{i,\mathrm{diff}}$ overtakes $|\mathbf{\Omega}_{i,\mathrm{disc}}\times\hat\L_i|$ over an inner-to-intermediate range of semimajor axes, which is the same range over which the warp grows and the disc fragments in Fig.~\ref{fig:warp_plum}. The closer IMBH ($a_\IMBH=0.05$\,pc, left column) disrupts a broad inner region within a few Myr, whereas the more distant IMBH ($a_\IMBH=0.15$\,pc, right column) disrupts a zone centred near its own orbital radius; in both cases the outermost disc, where $|\mathbf{\Omega}_{i,\mathrm{disc}}\times\hat\L_i|$ remains dominant, stays coherent over the $15$--$20$\,Myr shown. Note that for the IMBH with $a_\IMBH=0.05$\,pc our model predicts the disruption already at 1 Myr, while for the IMBH with $a_\IMBH=0.15$\,pc the disruption is predicted only after 10 Myr explaining the disc coherence at 10 Myr (cf. lower right panel of Fig.~\ref{fig:warp_plum}).

\subsection{Sky maps of the angular momentum directions}

\begin{figure*}
    \centering
    \includegraphics[width=\linewidth]{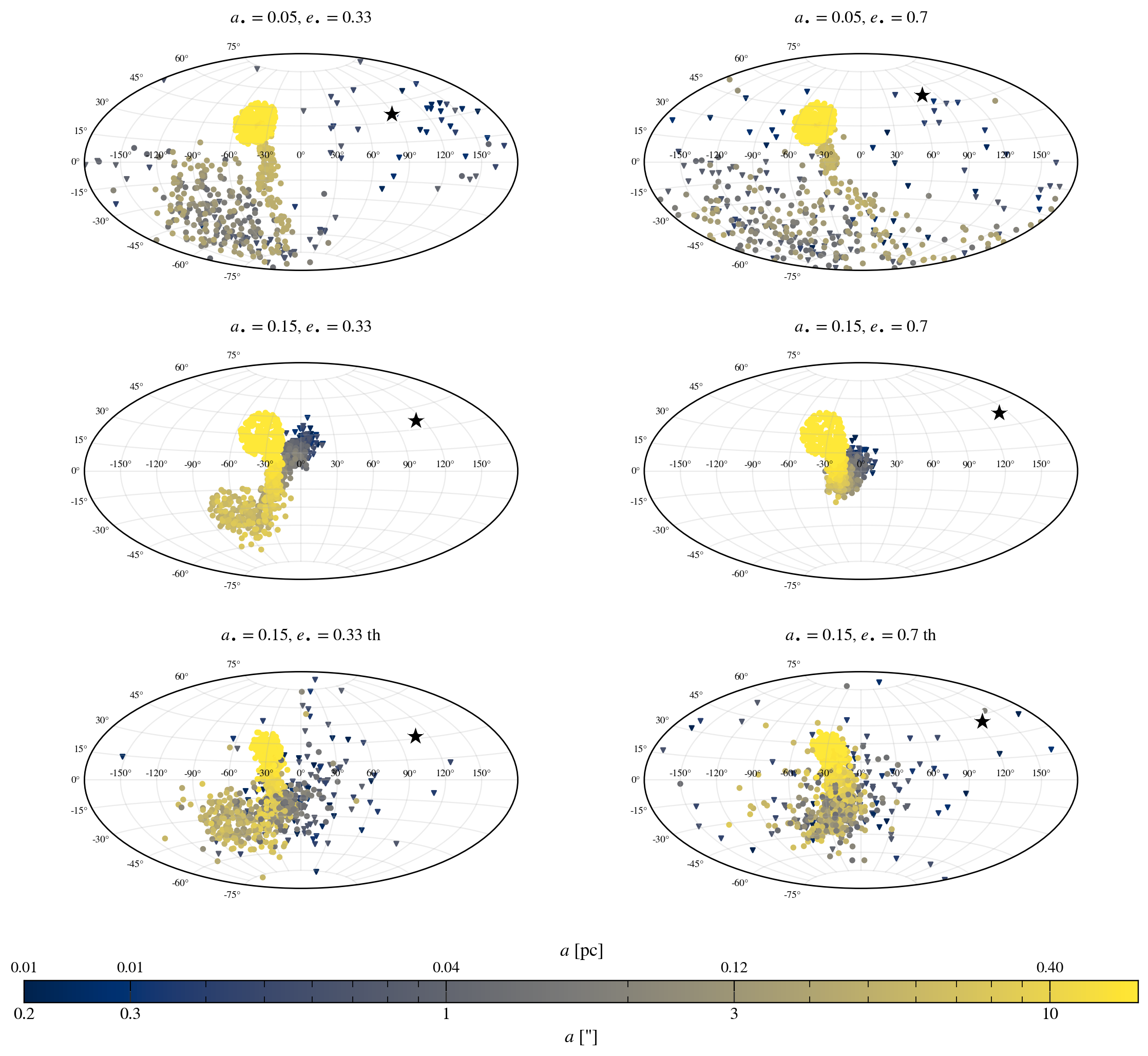}
    \caption{Angular momentum vector directions on the Aitoff projection at t = 5 Myr for the selected models of $m_{\IMBH} = 2000\,\msun=0.67\,M_{\rm d}$ and initial inclination of the IMBH orbit relative to the disc, $\theta_\IMBH=135^\circ$. Each panel title indicates the IMBH's semimajor axis (in parsecs) and eccentricity (`th' stands for \textit{thermal} otherwise \textit{stardisc} distribution$^{\text{\ref{footnote:stardisk}}}$). Colour coding represents the semimajor axes of the disc stars. Triangles indicate the inner stars ($a< 0.04$ pc). IMBH is shown with a black star symbol.}
    \label{fig:skymaps_plum}
\end{figure*}

\begin{figure*}
    \centering
    \includegraphics[width=\linewidth]{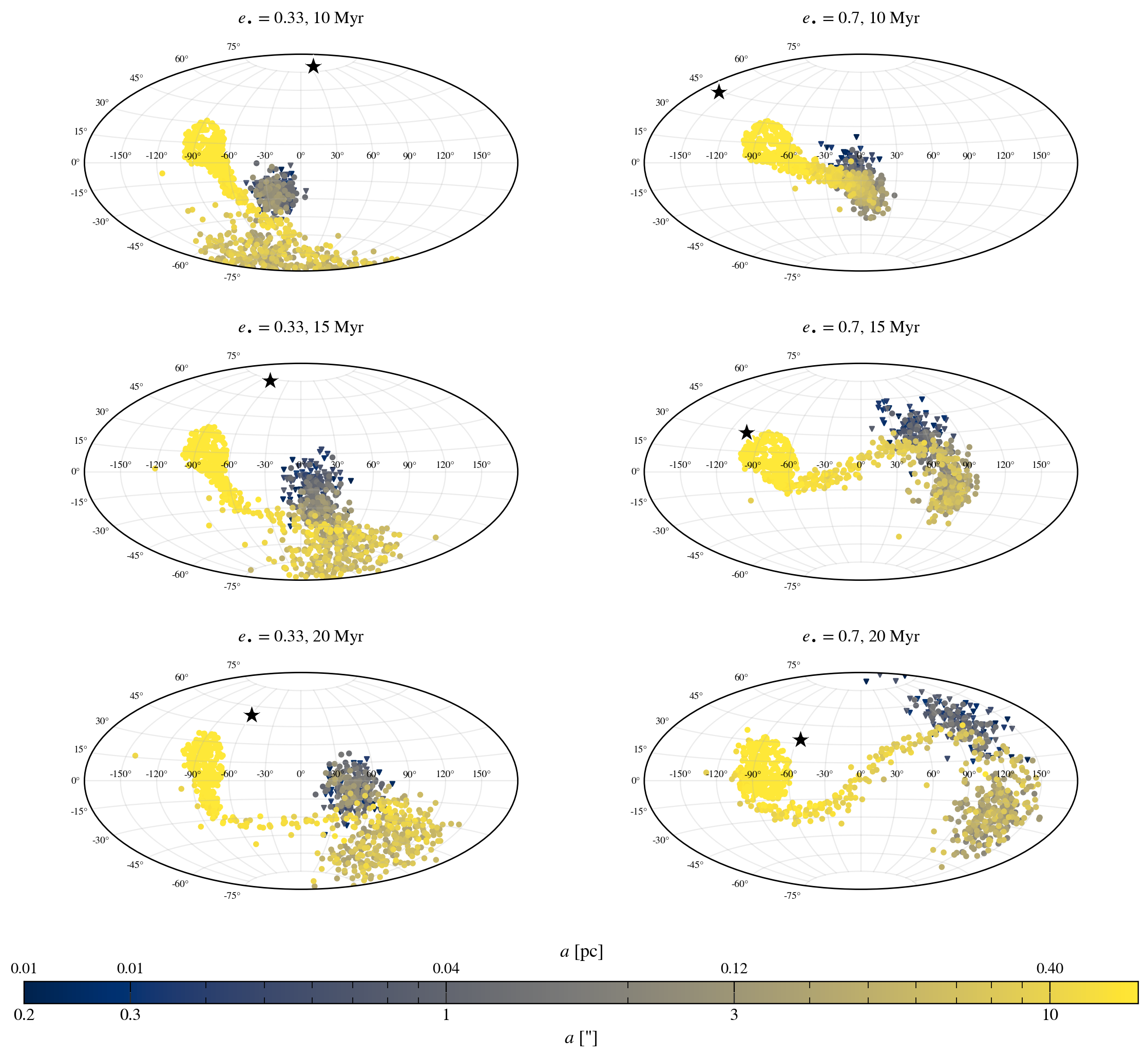}
    \caption{Same as the middle panels in Fig.~\ref{fig:skymaps_plum}, but showing sky projections for the models with $a_\IMBH=0.15$ pc, $e=0.33$ (left column), $e=0.7$ (right column) at the time intervals of 10 Myr (upper row), 15 Myr (middle row) and 20 Myr (bottom row).}
    \label{fig:skymaps_plum_to20}
\end{figure*}

\begin{figure*}
    \centering
    \includegraphics[width=\linewidth]{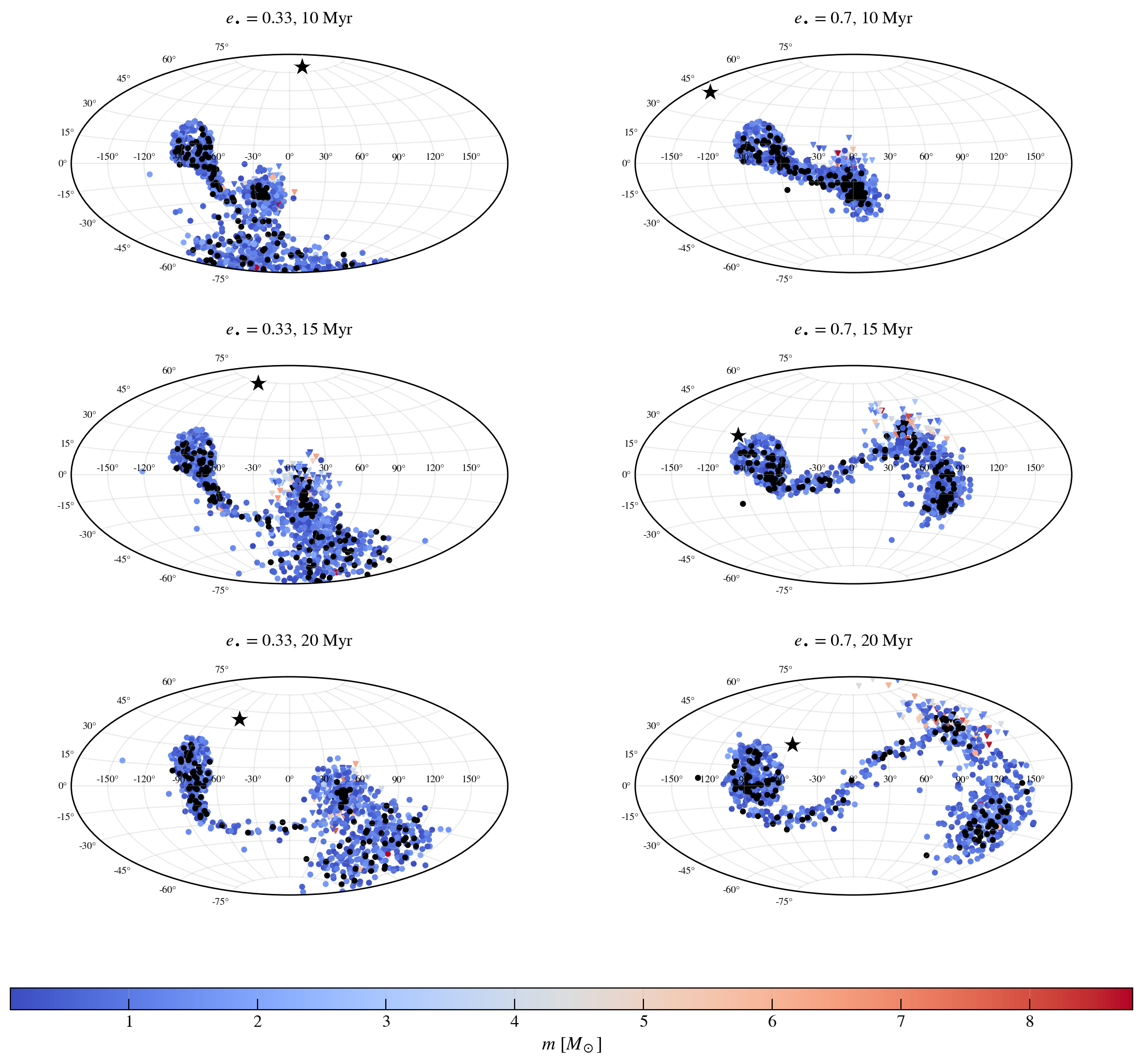}
    \caption{Same as Fig.~\ref{fig:skymaps_plum_to20}, but with colours coded by masses of the objects. Black colour represents stellar-mass black holes ($m\geq10\Msun$).}
    \label{fig:skymaps_plum_to20mass}
\end{figure*}

\begin{figure*}
    \centering
    \includegraphics[width=\linewidth]{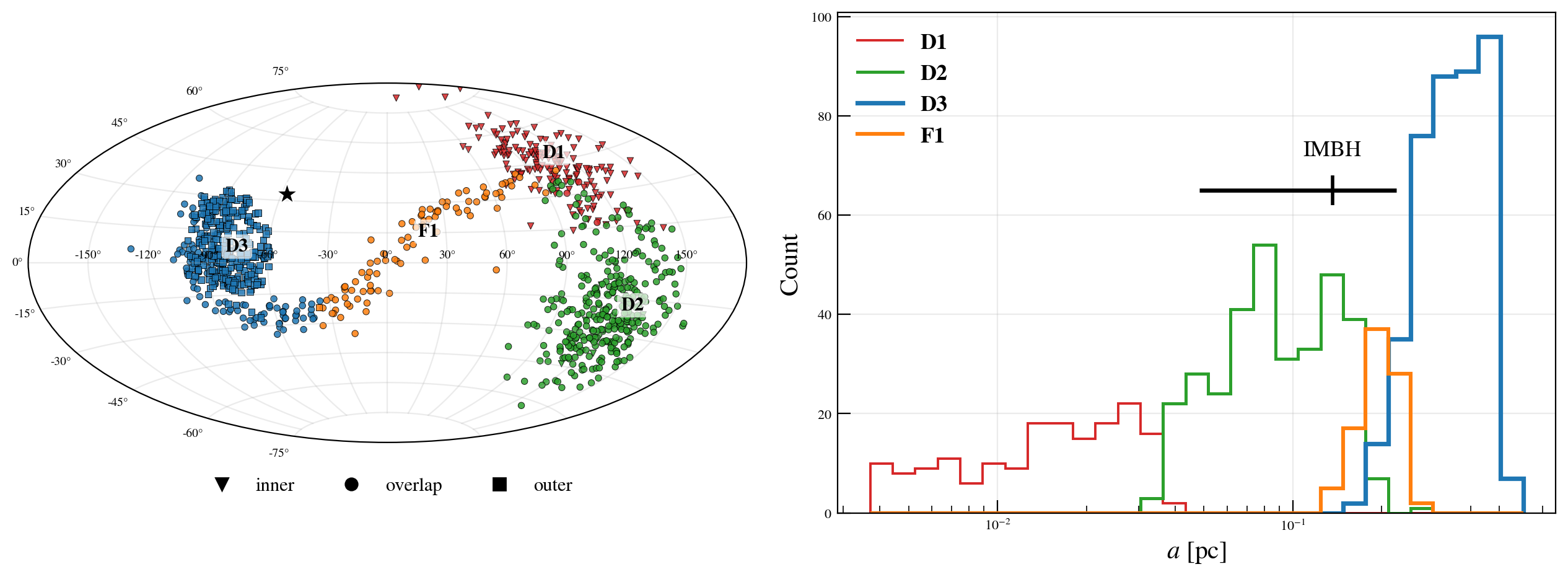}
    \caption{Structural decomposition of the stellar disc at $t=20$~Myr for the model with $e_\IMBH=0.7$, $a_\IMBH=0.15$~pc, $m_\IMBH=2000\,\Msun$, and $\theta_\IMBH=135^\circ$. \textit{Left panel:} Angular momentum vector directions on the Aitoff projection, colour-coded by membership in four structures identified via K-means clustering: D1 (red), D2 (green), D3 (blue), and F1 (orange), corresponding to three discs and one filament. Symbols denote the orbital regime with respect to the IMBH: inner (triangles, $a < r_{\rm peri}$), overlapping (circles), and outer (squares, $a > r_{\rm apo}$), following Eq.~\eqref{eq:orb-cond}. The IMBH is shown as a black star. Most overlapping stars reside in D2 or F1, inner stars are concentrated in D1, and outer stars predominantly belong to D3. \textit{Right panel:} Semimajor axis distributions for each structure, colour-coded consistently with the left panel, showing that the identified structures occupy distinct radial ranges.}
    \label{fig:skymaps_with_semi}
\end{figure*}

\begin{figure*}
    \centering
    \includegraphics[width=\linewidth]{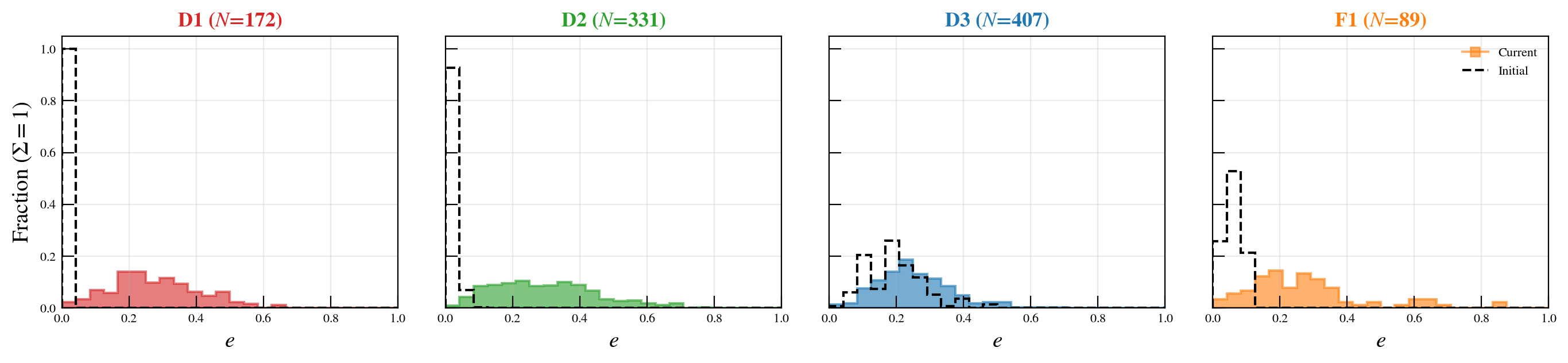}
    \caption{Distribution of eccentricities within the structures identified in the simulations shown in  Fig.~\ref{fig:skymaps_with_semi}. Coloured histograms show the eccentricity distributions at 20 Myr, while dashed black lines show the initial eccentricity distribution for the stars residing in the corresponding structures.}
    \label{fig:structure_ecc}
\end{figure*}

A clear way to visualise the evolution of the disc and the IMBH is to plot the sky map of the stars' angular momentum vector directions at different time snapshots using an Aitoff projection. These sky maps may also be compared to similar maps showing the observed distribution of young stars at the centre of the Milky Way \citep{vonFellenberg2022,Jia2023}. In particular,  \citet{vonFellenberg2022} showed that the distribution of young stars 
 in the central 0.5 pc is described by a complex, multi-component angular momentum vector distribution representing at least two counter-rotating discs and additional filamentary structures. Below, we show that in our simulations an initially thin stellar disc fragments into similar substructures in angular momentum direction space, possibly resembling the observed distribution. These maps also demonstrate that the outermost part of the disc remains intact in the simulations suggesting that it keeps the `memory' of the initial conditions. In order to compare our results with observations, we rotate the coordinate system such that the outermost structure in our simulation matches the corresponding observed outermost structure of \citet{vonFellenberg2022} denoted by \( F2 \) there.

\subsubsection{Sky maps at 5 Myr}

Figure~\ref{fig:skymaps_plum} illustrates the angular momentum vector projections for models with an $m_{\IMBH} = 2000 \msun=0.67 M_{\rm d}$ IMBH on an initially retrograde orbit of $\theta_\IMBH=135^\circ$ with respect to the disc.
The figure shows the projections at 5 Myr. The IMBH is shown as a black star symbol in all panels. For comparison, the corresponding sky-map evolution for a model with a prograde IMBH is shown in Appendix~\ref{App:skymap i45}.

The first row corresponds to the models with $a_\IMBH=0.05$ pc, while the other two rows correspond to $a_\IMBH=0.15$ pc. In the plotted $a_\IMBH=0.05$ pc models, the IMBH strongly thickens the inner part of the initially thin disc , while the outer region remains intact, consistent with the disruption criterion. The inner and overlapping regions of the model with $e_\IMBH=0.33$ can be treated as a thick disc ($\Delta\iota\gtrsim25^\circ$; cf. upper middle panel in Fig.~\ref{fig:warp_plum}). The model with $e_\IMBH=0.7$ features an even thicker disc in the same region ($\Delta\iota\gtrsim30^\circ$; cf. upper right panel in Fig.~\ref{fig:warp_plum}). Such thick inner discs already appear to be inconsistent with the observed stellar distribution, and by 20 Myr these structures are fully dispersed, as shown in Fig.~\ref{fig:warp_plum}. This disagreement provides a compelling argument against the presence of an IMBH with $a_\IMBH=0.05$ pc and a mass of $2000\,\Msun$ in the Galactic centre. This argument is further strengthened by the fact that our simulations neglect torques due to the stochastic anisotropy of the background NSC and do not include two-body relaxation with the NSC: these additional processes are expected to add angular-momentum diffusion and further disperse the disc, as shown by \citet{Panamarev2022}.

The second row of Fig.~\ref{fig:skymaps_plum} shows models with $a_\IMBH=0.15$ pc and initial eccentricities $e_\IMBH=0.33$ (left) and $e_\IMBH=0.7$ (right). These plots illustrate how the IMBH starts fragmenting the disc by 5 Myr: into three components in the first case, and into two components in the more eccentric case. The resulting discs remain relatively thin ($\Delta\iota\lesssim10^\circ$ in the inner region). The fragments are connected by streams of stars, from the outer region to the middle region and from the middle region to the inner region. Thus, while the first row shows the onset of the full disruption predicted by the disruption criterion in the inner disc, the models with $a_\IMBH=0.15$ pc occupy the region of the disruption parameter space where the disc forms intact fragments (cf. Sec.~\ref{subsubsec:disruption-parameters}). In this regime, the IMBH breaks the disc into several components, but these fragments remain intact in angular-momentum space.

The third row shows models with the same IMBH parameters as in the second row, but with a different initial disc model, corresponding to the \textit{thermal} model.While the \textit{stardisc} models feature nearly circular orbits, an outer warp, and a correlation between orbital parameters, with lower eccentricities and inclinations for the innermost stars (see Sec.~\ref{subsec:stellar-disc} for a detailed description, and Figs.~2--3 of \citet{Panamarev2022} for an illustration), the \textit{thermal} model initially has a thermal eccentricity distribution, inclinations drawn uniformly in $\cos\theta$ over $[\cos(10^\circ),\cos(0^\circ)]$, and no initial warp. These different initial conditions lead to a clear difference in the angular-momentum distribution after 5 Myr. In the \textit{thermal} models, the inner and middle regions tend to disperse, and the discs are significantly thicker than in the corresponding \textit{stardisc} models. Therefore, for the IMBH-to-disc mass ratio considered here, $m_\IMBH/M_{\rm d}\simeq0.67$, we conclude that reproducing the observed distribution of angular-momentum vector directions favours a stellar-disc formation scenario with initially colder, nearly circular orbits.

\subsubsection{Sky maps at later times}

While the IMBH ultimately succeeds in fragmenting the disc in the simulation, the angular distances between the disc structures at 5 Myr are smaller in the simulation than in the observed distribution in several cases. However, the evolutionary time-scale in the simulation is sensitive to the assumed disc mass and the semimajor axis of the IMBH. Moreover, in a more realistic situation, the evolution may proceed faster because torques from the stochastic anisotropy of the NSC would add angular-momentum diffusion, while dynamical friction against the NSC may drive the IMBH inward, increasing the IMBH-induced precession rates and shortening the corresponding time-scales. Therefore, what occurs in our idealised simulations within 20 Myr may take place earlier in a more realistic model. It is therefore useful to compare the simulated sky maps at later times with the observations, as shown in Fig.~\ref{fig:skymaps_plum_to20}.

For the $a_\IMBH=0.15$ pc \textsc{stardisc} models ($e_\IMBH=0.33$ and $0.7$; second row of Fig.~\ref{fig:skymaps_plum}), the structures separate over longer time-scales. Figure~\ref{fig:skymaps_plum_to20} shows the evolution of the disc structures in angular-momentum space, similar to Fig.~\ref{fig:skymaps_plum}, but for the \textsc{stardisc} models with $e_\IMBH=0.33$ and $e_\IMBH=0.7$ at 10, 15 and 20 Myr. The model with $e_\IMBH=0.7$ shows the closest match with the observed distribution: we can identify three discs and a filament with (qualitatively) matching angular distances. Apart from the outermost structure, which is aligned with the observed $F2$ by construction, the location of the innermost region also matches the observed inner disc, whereas the location of the middle region differs. Qualitatively, the features are similar, but some details differ, particularly the location of the middle region and the absence of additional filaments. In particular, \citet{vonFellenberg2022} identified five features, although the significance of these additional structures in the observations is debated \citep{Jia2023}.

\subsubsection{Mass-dependent thickness of the disc fragments}

Given that disc fragmentation and the subsequent motion of angular momentum vectors are driven by VRR, we can check if the simulated distributions develop the expected features of the dynamical VRR equilibria. In particular, massive objects are expected to form thinner discs within the substructures with less scatter in angular momentum vector directions \citep{Szolgyen2018, Mathe2022, Gruzinov2020}. This process is known as anisotropic mass segregation.
Figure~\ref{fig:skymaps_plum_to20mass} shows the evolution of the angular momentum vectors in the simulations colour-coded by mass. This reveals that the most massive particles are indeed more concentrated towards the corresponding cluster centre especially in the innermost parts.

\subsubsection{Structural decomposition of the model closest to the observed sky map}

Let us now focus specifically on the $e_\IMBH=0.7$ model of $m_{\IMBH}=2000\,\Msun=0.67 M_{\rm d}$ at 20 Myr which shows the most similar angular momentum vector direction distribution compared to the observed distribution. We can quantitatively identify the constituent clusters using a basic clustering algorithm, $k$-means clustering, and weigh the distances on the unit sphere by semimajor axes. Using this algorithm, we identified 4 structures, which we denote as: D1, D2 and D3 (standing for discs 1 to 3), and F1 (filament-1). Fig.~\ref{fig:skymaps_with_semi} (left) illustrates this classification, where we map each cluster with red, green, blue, and orange colours respectively. Additionally, we show with symbols whether the particles belong to inner, overlapping or outer regions with respect to the IMBH's orbit (Eq.~\ref{eq:orb-cond}). Most overlapping stars reside in D2 or F1, while inner stars are in D1 and outer ones are in D3 mostly. The right panel of the same figure shows clear differences in the semimajor-axis distributions of the corresponding structures. D1 is dominated by inner stars, D3 by outer stars, while D2 and F1 are mainly populated by stars whose orbits overlap with the IMBH orbit. Thus, the clustering in angular-momentum space is also correlated with the radial structure of the disc: the IMBH separates the initially continuous disc into components that occupy different radial ranges. The semimajor-axis distributions of D1 and D2, and the difference between them, broadly match the observationally inferred semimajor axes of the CW1 and CW2 regions in \citet{vonFellenberg2022}, although some individual values differ; in particular, our model has more objects in the inner region with small semimajor axes.

\subsection{Distribution of eccentricities}

Another reasonable question to ask is whether the eccentricities in the respective structures match those inferred from the observations. In particular, \citet{vonFellenberg2022} measured eccentricities spanning a range of $\sim 0.2$--$0.9$, with median values of $\sim 0.3$--$0.4$ that vary among the identified disc substructures. Similarly, \citet{Jia2023} found broadly consistent eccentricity distributions, with some subgroups featuring high eccentricities. Our analysis (see Fig.~\ref{fig:structure_ecc}) shows that eccentricities do change from the initial distribution (compare the coloured histograms with the dashed black histogram). The structures in the overlapping and inner regions (D1, D2, and F1) develop broader eccentricity distributions with significant excitation, while the outer structure D3 retains a distribution closer to the initial conditions. The eccentricity distributions of D1 and D2 broadly match the observationally inferred eccentricities of the CW1 and CW2 regions in \citet{vonFellenberg2022}. In particular, both structures contain stars with moderate eccentricities rather than remaining on nearly circular orbits. The observations also suggest relatively high eccentricities in some outer components, whereas in our model the outer structure D3 remains closer to the initial distribution. Nevertheless, the IMBH naturally excites the eccentricities of the inner and overlapping structures to values comparable to those inferred for the observed clockwise components, while simultaneously producing the fragmented angular-momentum structure discussed above.

Having discussed the eccentricities of the identified fragments, we next use a different set of models to isolate how the IMBH excites eccentricities in different radial regions and how this depends on the relative orientation of the IMBH orbit. For this comparison, we consider models with a smaller IMBH semimajor axis, $a_\IMBH=0.05$ pc, and compare them to equivalent simulations without an IMBH. The impact of the IMBH on the stellar disc's geometry is notably significant when the IMBH is on an initially retrograde orbit. This raises the question of whether it also affects the eccentricity distribution of the disc stars. To address this, we compared the eccentricity distributions for the disc with prograde and retrograde IMBHs to those in a control simulation without an IMBH \citep{Panamarev2022}.

\begin{figure*}
    \centering
    \includegraphics[width=0.8\linewidth]{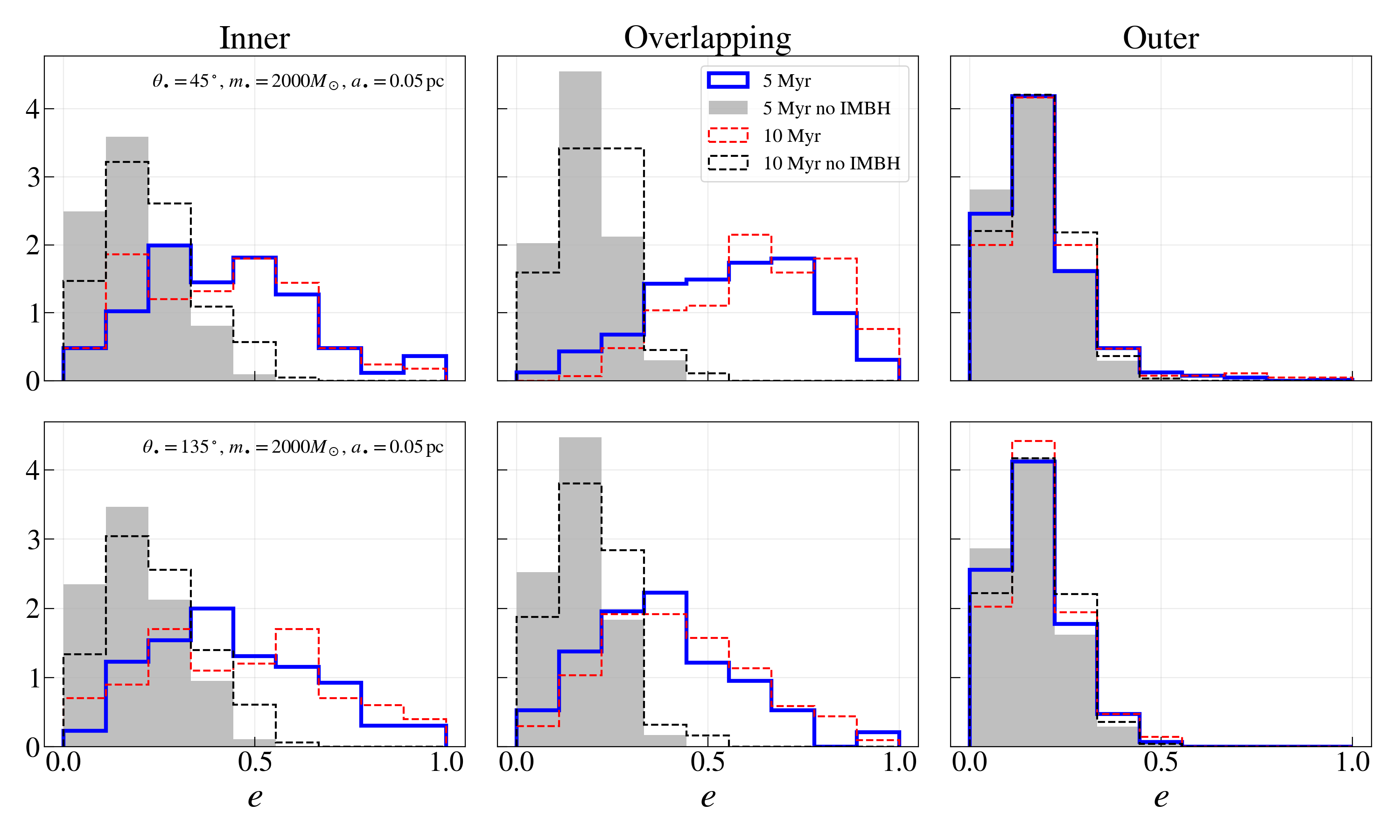}  
    \caption{Eccentricity distributions for the stellar disc influenced by an IMBH with a mass of \( m_{\IMBH} = 2000 \msun \), semimajor axis $a_\IMBH=0.05$ pc, and eccentricity \( e_\IMBH = 0.33 \). \textbf{Top row:} Prograde orbit ($\theta_\IMBH=45^\circ$). \textbf{Bottom row:} Retrograde orbit ($\theta_\IMBH=135^\circ$). Both rows display simulations with the same initial conditions. The shaded histograms and black dashed lines represent simulations without the IMBH. The blue and red lines represent the eccentricity distributions at evolutionary times of 5 Myr and 10 Myr, respectively. See Fig.~\ref{fig:structure_ecc} for the observationally matching fragment model, whose eccentricity distributions are closer to those inferred for the observed substructures.}
    \label{fig:ecc_plum}
\end{figure*}

Figure~\ref{fig:ecc_plum} presents histograms of the eccentricity distributions for the inner, overlapping and outer regions of the disc. The top panels illustrate the scenario with a prograde IMBH, while the bottom panels depict the retrograde case, with distributions shown at 5 and 10 Myr. The shaded areas in the histograms and the dashed black lines represent the eccentricity 
distributions for the same regions in the model without an IMBH.

The histograms show that the IMBH significantly excites the eccentricity distribution in the overlapping region, shifting it from nearly circular orbits in its absence to a broad distribution of eccentric orbits. The excitation is stronger for the prograde IMBH: in this case, the overlapping region develops a pronounced high-eccentricity tail, extending to $e\gtrsim0.7$--$0.9$, whereas the retrograde model produces a broader but less strongly shifted distribution. This increase in eccentricity is less pronounced in the inner region, but is still noticeable. In contrast, the eccentricity distribution in the outer region appears to be largely unaffected by the IMBH in both prograde and retrograde cases. This may explain the larger spread in orbital inclinations (thick discs) in the overlapping regions for models with the IMBH on orbits with \( a_\IMBH=0.05 \) pc. 

Together, Figs.~\ref{fig:structure_ecc} and \ref{fig:ecc_plum} show that an inclined IMBH can naturally excite initially cold stellar orbits to a broad range of moderate-to-high eccentricities. Reaching the highest eccentricities suggested by the observations, especially in the outer structures, may require  different model parameters or additional sources of eccentricity excitation. This is plausible, since our initial stellar disc is dynamically cold, with nearly circular orbits, and the simulations neglect additional processes such as two-body relaxation and torques from the stochastic anisotropy of the NSC. Different initial conditions for the stellar disc could also contribute to a broader eccentricity distribution. Nevertheless, this excitation is consistent with the observed spread, suggesting that the gravitational influence of an IMBH on an inclined orbit provides a plausible mechanism for producing a substantial part of the eccentricity distributions seen in the young stellar disc of the Galactic centre.

\section{Summary and discussion}
\label{sec:SUM}

We have studied the disruption of a stellar disc by an inclined IMBH with an analytic torque model and a suite of direct $N$-body simulations of stellar discs orbiting a central SMBH and embedded in a smooth spherical nuclear star cluster (modelled as an analytic Plummer potential). Our main results can be summarised as follows:

\begin{itemize}

\item We derived an analytic disruption condition (Eq.~\ref{eqn:fundamental criterion}) for the stellar disc to fragment or to disperse. The condition compares the disc-induced precession rate with the differential precession rate induced by the IMBH across the finite opening angle of the disc: regions where the latter exceeds the former are expected to disperse. The disrupted regions predicted by this condition agree with those measured in the $N$-body simulations.

\item The criterion defines three outcomes as a function of the IMBH-to-disc mass ratio and the IMBH--disc orbital configuration (Figs.~\ref{fig:disc-disruption}--\ref{fig:disruption-regimes}): the disc may warp and twist while remaining intact (\emph{intact-disc}), the disc fragments with individual fragments remaining intact and precess coherently about the IMBH, or the disc is fully \emph{dispersed}.

\item A retrograde IMBH with mass comparable to the disc ($m_\IMBH \simeq M_{\rm d}$) fragments an initially coherent stellar disc into well-separated components in angular-momentum space, separated by large angles on the unit sphere. For the fiducial model ($m_\IMBH=2000\msun=0.67M_{\rm d}$, $a_\IMBH=0.15$~pc, $e_\IMBH=0.7$), these structures develop by 20 Myr and comprise three disc-like components and a filament occupying different semimajor-axis ranges (Fig.~\ref{fig:skymaps_with_semi}), qualitatively resembling the multi-component kinematic structure of the young stellar population near Sgr~A* reported by \citet{vonFellenberg2022} and \citet{Jia2023}.

\item Moreover, results from direct $N$-body simulations show that the IMBH efficiently excites eccentricities in the radially overlapping region of the disc, transforming a nearly circular population into a broad distribution of moderate eccentricities. In the inner and overlapping substructures, these eccentricities are consistent with the values inferred from observations of the corresponding structures of the young stars at the Galactic centre \citep{vonFellenberg2022}.

\item Within the fragmented substructures, the more massive stars settle into thinner configurations with less scatter in their angular-momentum directions than lighter stars showing anisotropic mass segregation expected from VRR \citep{Szolgyen2018, Mathe2022, Gruzinov2020}.

\item For prograde configurations, the IMBH and the stellar disc tend to evolve toward alignment with the total angular-momentum vector of the system, consistent with \citet{Szolgyen2021, Ginat2023}.

\item For retrograde configurations, an initially inclined IMBH may anti-align with the angular momentum of the radially overlapping disc region. This anti-alignment is expected when the IMBH angular momentum exceeds that of the overlapping region; otherwise, the relative inclination in the overlap region remains approximately constant.

\item The IMBH's own orbital eccentricity circularises efficiently only in the low-mass prograde case ($m_\IMBH\leq500\,\msun$); for $m_\IMBH \simeq M_{\rm d}$, and for retrograde or orthogonal orbits, it shows little evolution over $\sim10$~Myr, extending the \citet{Szolgyen2021} circularisation result to the comparable-mass regime.

\end{itemize}

Taken together, an inclined IMBH of mass comparable to the disc may convert an initially coherent disc into well-separated components in angular-momentum space.

These similarities suggest that even a simple scenario with a single initially thin disc and one inclined perturber can reproduce several qualitative features of the young stellar population in the Galactic centre, including its multi-component angular-momentum structure, the apparent warp of the clockwise disc, and the excitation of eccentricities to values comparable to those inferred for the main observed components (Fig.~\ref{fig:structure_ecc}; cf. \citealt{vonFellenberg2022,Jia2023}).

A complementary scenario has recently been proposed by \citet{Zheng2026}. Their unified model also attributes the kinematics of the young Galactic-centre disc to an intermediate-mass companion, but it is more massive than in our models ($\sim10^4\,\msun$) and located outside the disc (semi-major axis $a\sim0.35$~pc), featuring a high orbital inclination between $40^\circ$ and $140^\circ$. In their simulations, a sweeping secular resonance driven by a depleting gaseous disc (modelled as an analytic potential) is the dominant mechanism raising the clockwise-disc stars to their observed moderate eccentricities ($e\sim0.3$). Because our gas-free model achieves a comparable eccentricity spread with an inclined IMBH alone, an intermediate-mass perturber offers a robust explanation for these kinematics. Additionally, \citet{Zheng2026} show that von Zeipel--Lidov--Kozai oscillations from the inclined companion are specifically responsible for pulling the outer disc stars into highly inclined, high-eccentricity orbits, providing an alternative route to the misaligned, counter-rotating disc components. However, while these individual stars are kicked into extreme orbits, their collective appearance as a counter-clockwise disc is episodic and loosely clustered, in contrast to the coherently precessing fragments that our VRR model predicts.

Some residual differences between the simulations and observations are expected given the idealised nature of the present models and the observational uncertainties. These include the precise location of the middle (F1/CCW) component and the highest eccentricities inferred for some outer structures. Our simulations use a static, spherical background and therefore omit stochastic torques, two-body relaxation, and dynamical friction from a live NSC. These processes would add angular-momentum diffusion, and dynamical friction would gradually drive the IMBH inward potentially accelerating the disc fragmentation (cf. Fig~\ref{fig:t_prec_ma}). Thus, the $\sim10$--$20$~Myr development time in our idealised models may represent an upper limit; as we show in a companion paper (Paper~II, in prep).

The relevant comparison age is also uncertain at the multi-Myr level. The O/WR disc is dated to approximately $\sim6\pm2$~Myr from spectral classification and evolutionary tracks \citep{PaumardEtAl2006, Bartko2010}, while a Bayesian fit of the central cluster prefers a younger age of $2.5$--$5.8$~Myr \citep{Lu2013}. 

The eccentricities are likewise statistical. Unlike the inner S-stars, which have full six-element Keplerian orbits \citep{Gillessen2009, Gillessen2017}, most disc stars have only five of their six phase-space coordinates measured, with the line-of-sight distance undetermined. Their individual eccentricities and semimajor axes are therefore inferred statistically \citep{Bartko2009, Yelda2014, Jia2023, vonFellenberg2022}, so the most meaningful comparisons are between distributions rather than individual orbits.

While our results show that an inclined IMBH can fragment the disc into distinct angular-momentum components, an IMBH is not the only possible mechanism for producing such substructure. Two classes of alternatives can also warp and disrupt a disc without an IMBH: resonant friction from a rotating nuclear star cluster \citep{Levin2024} and torques from residual cluster asphericity \citep{Perets2018}. An important simplification of our present modelling is the stationary spherical (Plummer) background, which suppresses stochastic torques from a live cluster that contribute to VRR and two-body relaxation \citep{Panamarev2022}. We also bracket uncertain birth conditions with the \textsc{stardisc} and \textsc{thermal} initial models while neglecting gas, stellar evolution, and binaries. In Paper~II (in prep.) we will replace the static potential with an isotropic live $N$-body NSC to quantify the role of stochastic torques and two-body relaxation. 

To fully understand stellar dynamics in the inner parsec of the Galactic centre, these models will ultimately need to be extended to aspherical and rotating clusters, and the impact of additional physics (stellar evolution, binaries, and gas) assessed for the long-term survival and observability of the substructures; rotating models will be particularly important for comparisons with the \citet{Levin2024} scenario. Finally, torques from the gaseous torus (circumnuclear disc) may also influence the evolution, and could plausibly alter the orientation of the outer disc structure \citep{Subr+2009,Haas_Subr_Kroupa2011,Kocsis2011}.

\section*{Acknowledgements}

This work was supported by the Science and Technology Facilities Council Grant Number ST/W000903/1. We acknowledge the support of the Project No. BR34836926 "Research and monitoring of near-Earth and deep space through the development and scaling of the Kazakhstan optical telescope network", financed by the Aerospace committee of the Ministry of Artificial Intelligence and Digital Development of the Republic of Kazakhstan.

\section*{Data Availability}
The data underlying this article will be shared on reasonable request to the corresponding author.

\bibliographystyle{mnras}
\bibliography{thesis}

\begin{thebibliography}{}
\makeatletter
\relax
\def\mn@urlcharsother{\let\do\@makeother \do\$\do\&\do\#\do\^\do\_\do\%\do\~}
\def\mn@doi{\begingroup\mn@urlcharsother \@ifnextchar [ {\mn@doi@}
  {\mn@doi@[]}}
\def\mn@doi@[#1]#2{\def\@tempa{#1}\ifx\@tempa\@empty \href
  {http://dx.doi.org/#2} {doi:#2}\else \href {http://dx.doi.org/#2} {#1}\fi
  \endgroup}
\def\mn@eprint#1#2{\mn@eprint@#1:#2::\@nil}
\def\mn@eprint@arXiv#1{\href {http://arxiv.org/abs/#1} {{\tt arXiv:#1}}}
\def\mn@eprint@dblp#1{\href {http://dblp.uni-trier.de/rec/bibtex/#1.xml}
  {dblp:#1}}
\def\mn@eprint@#1:#2:#3:#4\@nil{\def\@tempa {#1}\def\@tempb {#2}\def\@tempc
  {#3}\ifx \@tempc \@empty \let \@tempc \@tempb \let \@tempb \@tempa \fi \ifx
  \@tempb \@empty \def\@tempb {arXiv}\fi \@ifundefined
  {mn@eprint@\@tempb}{\@tempb:\@tempc}{\expandafter \expandafter \csname
  mn@eprint@\@tempb\endcsname \expandafter{\@tempc}}}

\bibitem[\protect\citeauthoryear{{Ali} et~al.,}{{Ali} et~al.}{2020}]{Ali2020}
{Ali} B.,  et~al., 2020, \mn@doi [\apj] {10.3847/1538-4357/ab93ae}, \href
  {https://ui.adsabs.harvard.edu/abs/2020ApJ...896..100A} {896, 100}

\bibitem[\protect\citeauthoryear{{Bartko} et~al.,}{{Bartko}
  et~al.}{2009}]{Bartko2009}
{Bartko} H.,  et~al., 2009, \mn@doi [\apj] {10.1088/0004-637X/697/2/1741},
  \href {http://adsabs.harvard.edu/abs/2009ApJ...697.1741B} {697, 1741}

\bibitem[\protect\citeauthoryear{{Bartko} et~al.,}{{Bartko}
  et~al.}{2010}]{Bartko2010}
{Bartko} H.,  et~al., 2010, \mn@doi [\apj] {10.1088/0004-637X/708/1/834}, \href
  {http://adsabs.harvard.edu/abs/2010ApJ...708..834B} {708, 834}

\bibitem[\protect\citeauthoryear{{Bonnell} \& {Rice}}{{Bonnell} \&
  {Rice}}{2008}]{Bonnell2008}
{Bonnell} I.~A.,  {Rice} W.~K.~M.,  2008, \mn@doi [Science]
  {10.1126/science.1160653}, \href
  {https://ui.adsabs.harvard.edu/abs/2008Sci...321.1060B} {321, 1060}

\bibitem[\protect\citeauthoryear{{Capuzzo-Dolcetta}}{{Capuzzo-Dolcetta}}{1993}]{Capuzzo-Dolcetta1993}
{Capuzzo-Dolcetta} R.,  1993, \mn@doi [\apj] {10.1086/173189}, \href
  {http://adsabs.harvard.edu/abs/1993ApJ...415..616C} {415, 616}

\bibitem[\protect\citeauthoryear{{Capuzzo-Dolcetta} \&
  {Miocchi}}{{Capuzzo-Dolcetta} \& {Miocchi}}{2008}]{Capuzzo-Dolcetta2008}
{Capuzzo-Dolcetta} R.,  {Miocchi} P.,  2008, \mn@doi [\mnras]
  {10.1111/j.1745-3933.2008.00501.x}, \href
  {https://ui.adsabs.harvard.edu/abs/2008MNRAS.388L..69C} {388, L69}

\bibitem[\protect\citeauthoryear{{Chandrasekhar}}{{Chandrasekhar}}{1943}]{Chandra1943}
{Chandrasekhar} S.,  1943, \mn@doi [\apj] {10.1086/144517}, \href
  {https://ui.adsabs.harvard.edu/abs/1943ApJ....97..255C} {97, 255}

\bibitem[\protect\citeauthoryear{{Evans}, {Rasskazov}, {Remmelzwaal},
  {Marchetti}, {Castro-Ginard}, {Rossi}  \& {Bovy}}{{Evans}
  et~al.}{2023}]{Evans2023}
{Evans} F.~A.,  {Rasskazov} A.,  {Remmelzwaal} A.,  {Marchetti} T.,
  {Castro-Ginard} A.,  {Rossi} E.~M.,   {Bovy} J.,  2023, \mn@doi [\mnras]
  {10.1093/mnras/stad2273}, \href
  {https://ui.adsabs.harvard.edu/abs/2023MNRAS.525..561E} {525, 561}

\bibitem[\protect\citeauthoryear{{Generozov}, {Nayakshin}  \&
  {Madigan}}{{Generozov} et~al.}{2022}]{Generozov2022}
{Generozov} A.,  {Nayakshin} S.,   {Madigan} A.~M.,  2022, \mn@doi [\mnras]
  {10.1093/mnras/stac419}, \href
  {https://ui.adsabs.harvard.edu/abs/2022MNRAS.512.4100G} {512, 4100}

\bibitem[\protect\citeauthoryear{{Genzel}, {Eisenhauer}  \&
  {Gillessen}}{{Genzel} et~al.}{2010}]{Genzel2010}
{Genzel} R.,  {Eisenhauer} F.,   {Gillessen} S.,  2010, \mn@doi [Rev. Mod.
  Phys.] {10.1103/RevModPhys.82.3121}, \href
  {http://adsabs.harvard.edu/abs/2010RvMP...82.3121G} {82, 3121}

\bibitem[\protect\citeauthoryear{{Ghez}, {Salim}, {Hornstein}, {Tanner}, {Lu},
  {Morris}, {Becklin}  \& {Duch{\^e}ne}}{{Ghez} et~al.}{2005}]{Ghez2005}
{Ghez} A.~M.,  {Salim} S.,  {Hornstein} S.~D.,  {Tanner} A.,  {Lu} J.~R.,
  {Morris} M.,  {Becklin} E.~E.,   {Duch{\^e}ne} G.,  2005, \mn@doi [\apj]
  {10.1086/427175}, \href {http://adsabs.harvard.edu/abs/2005ApJ...620..744G}
  {620, 744}

\bibitem[\protect\citeauthoryear{{Giersz}, {Leigh}, {Hypki}, {L{\"u}tzgendorf}
  \& {Askar}}{{Giersz} et~al.}{2015}]{Giersz+2015}
{Giersz} M.,  {Leigh} N.,  {Hypki} A.,  {L{\"u}tzgendorf} N.,   {Askar} A.,
  2015, \mn@doi [\mnras] {10.1093/mnras/stv2162}, \href
  {https://ui.adsabs.harvard.edu/abs/2015MNRAS.454.3150G} {454, 3150}

\bibitem[\protect\citeauthoryear{{Gillessen}, {Eisenhauer}, {Trippe},
  {Alexander}, {Genzel}, {Martins}  \& {Ott}}{{Gillessen}
  et~al.}{2009}]{Gillessen2009}
{Gillessen} S.,  {Eisenhauer} F.,  {Trippe} S.,  {Alexander} T.,  {Genzel} R.,
  {Martins} F.,   {Ott} T.,  2009, \mn@doi [\apj]
  {10.1088/0004-637X/692/2/1075}, \href
  {http://adsabs.harvard.edu/abs/2009ApJ...692.1075G} {692, 1075}

\bibitem[\protect\citeauthoryear{{Gillessen} et~al.,}{{Gillessen}
  et~al.}{2017}]{Gillessen2017}
{Gillessen} S.,  et~al., 2017, \mn@doi [\apj] {10.3847/1538-4357/aa5c41}, \href
  {http://adsabs.harvard.edu/abs/2017ApJ...837...30G} {837, 30}

\bibitem[\protect\citeauthoryear{{Ginat} \& {Kocsis}}{{Ginat} \&
  {Kocsis}}{2025}]{Ginat2025}
{Ginat} Y.~B.,  {Kocsis} B.,  2025, \mn@doi [arXiv e-prints]
  {10.48550/arXiv.2502.08709}, \href
  {https://ui.adsabs.harvard.edu/abs/2025arXiv250208709G} {p. arXiv:2502.08709}

\bibitem[\protect\citeauthoryear{{Ginat}, {Panamarev}, {Kocsis}  \&
  {Perets}}{{Ginat} et~al.}{2023}]{Ginat2023}
{Ginat} Y.~B.,  {Panamarev} T.,  {Kocsis} B.,   {Perets} H.~B.,  2023, \mn@doi
  [\mnras] {10.1093/mnras/stad2400}, \href
  {https://ui.adsabs.harvard.edu/abs/2023MNRAS.525.4202G} {525, 4202}

\bibitem[\protect\citeauthoryear{{Goodman} \& {Tan}}{{Goodman} \&
  {Tan}}{2004}]{2004ApJ...608..108G}
{Goodman} J.,  {Tan} J.~C.,  2004, \mn@doi [\apj] {10.1086/386360}, \href
  {http://adsabs.harvard.edu/abs/2004ApJ...608..108G} {608, 108}

\bibitem[\protect\citeauthoryear{{Gravity Collaboration} et~al.,}{{Gravity
  Collaboration} et~al.}{2022}]{Gravity2022}
{Gravity Collaboration} et~al., 2022, \mn@doi [\aap]
  {10.1051/0004-6361/202142465}, \href
  {https://ui.adsabs.harvard.edu/abs/2022A&A...657L..12G} {657, L12}

\bibitem[\protect\citeauthoryear{{Gravity Collaboration} et~al.,}{{Gravity
  Collaboration} et~al.}{2023}]{Gravity2023}
{Gravity Collaboration} et~al., 2023, \mn@doi [\aap]
  {10.1051/0004-6361/202245132}, \href
  {https://ui.adsabs.harvard.edu/abs/2023A&A...672A..63G} {672, A63}

\bibitem[\protect\citeauthoryear{{Greene}, {Strader}  \& {Ho}}{{Greene}
  et~al.}{2020}]{Greene2020}
{Greene} J.~E.,  {Strader} J.,   {Ho} L.~C.,  2020, \mn@doi [\araa]
  {10.1146/annurev-astro-032620-021835}, \href
  {https://ui.adsabs.harvard.edu/abs/2020ARA&A..58..257G} {58, 257}

\bibitem[\protect\citeauthoryear{{Gruzinov}, {Levin}  \& {Zhu}}{{Gruzinov}
  et~al.}{2020}]{Gruzinov2020}
{Gruzinov} A.,  {Levin} Y.,   {Zhu} J.,  2020, \mn@doi [\apj]
  {10.3847/1538-4357/abbfaa}, \href
  {https://ui.adsabs.harvard.edu/abs/2020ApJ...905...11G} {905, 11}

\bibitem[\protect\citeauthoryear{{Gualandris} \& {Merritt}}{{Gualandris} \&
  {Merritt}}{2009}]{GualandrisMerritt2009}
{Gualandris} A.,  {Merritt} D.,  2009, \mn@doi [ApJ]
  {10.1088/0004-637X/705/1/361}, \href
  {http://adsabs.harvard.edu/abs/2009ApJ...705..361G} {705, 361}

\bibitem[\protect\citeauthoryear{{Gualandris}, {Gillessen}  \&
  {Merritt}}{{Gualandris} et~al.}{2010}]{Gualandris2010}
{Gualandris} A.,  {Gillessen} S.,   {Merritt} D.,  2010, \mn@doi [\mnras]
  {10.1111/j.1365-2966.2010.17373.x}, \href
  {https://ui.adsabs.harvard.edu/abs/2010MNRAS.409.1146G} {409, 1146}

\bibitem[\protect\citeauthoryear{{Haas}, {{\v{S}}ubr}  \& {Kroupa}}{{Haas}
  et~al.}{2011}]{Haas_Subr_Kroupa2011}
{Haas} J.,  {{\v{S}}ubr} L.,   {Kroupa} P.,  2011, \mn@doi [\mnras]
  {10.1111/j.1365-2966.2010.18025.x}, \href
  {https://ui.adsabs.harvard.edu/abs/2011MNRAS.412.1905H} {412, 1905}

\bibitem[\protect\citeauthoryear{{Habibi} et~al.,}{{Habibi}
  et~al.}{2017}]{Habibi2017}
{Habibi} M.,  et~al., 2017, \mn@doi [\apj] {10.3847/1538-4357/aa876f}, \href
  {https://ui.adsabs.harvard.edu/abs/2017ApJ...847..120H} {847, 120}

\bibitem[\protect\citeauthoryear{{Harfst}, {Gualandris}, {Merritt}, {Spurzem},
  {Portegies Zwart}  \& {Berczik}}{{Harfst} et~al.}{2007}]{HarfstEtAl2007}
{Harfst} S.,  {Gualandris} A.,  {Merritt} D.,  {Spurzem} R.,  {Portegies Zwart}
  S.,   {Berczik} P.,  2007, \mn@doi [\na] {10.1016/j.newast.2006.11.003},
  \href {http://adsabs.harvard.edu/abs/2007NewA...12..357H} {12, 357}

\bibitem[\protect\citeauthoryear{{Ishchenko}, {Sobolenko}, {Kuvatova},
  {Panamarev}  \& {Berczik}}{{Ishchenko} et~al.}{2023}]{Ishchenko2023}
{Ishchenko} M.,  {Sobolenko} M.,  {Kuvatova} D.,  {Panamarev} T.,   {Berczik}
  P.,  2023, \mn@doi [\aap] {10.1051/0004-6361/202245753}, \href
  {https://ui.adsabs.harvard.edu/abs/2023A&A...674A..70I} {674, A70}

\bibitem[\protect\citeauthoryear{{Ishchenko} et~al.,}{{Ishchenko}
  et~al.}{2024}]{Ishchenko2024}
{Ishchenko} M.,  et~al., 2024, \mn@doi [\aap] {10.1051/0004-6361/202450399},
  \href {https://ui.adsabs.harvard.edu/abs/2024A&A...689A.178I} {689, A178}

\bibitem[\protect\citeauthoryear{{Jia} et~al.,}{{Jia} et~al.}{2023}]{Jia2023}
{Jia} S.,  et~al., 2023, \mn@doi [\apj] {10.3847/1538-4357/acb939}, \href
  {https://ui.adsabs.harvard.edu/abs/2023ApJ...949...18J} {949, 18}

\bibitem[\protect\citeauthoryear{{Kaneko}, {Oka}, {Yokozuka}, {Enokiya},
  {Takekawa}, {Iwata}  \& {Tsujimoto}}{{Kaneko} et~al.}{2023}]{Kaneko2023}
{Kaneko} M.,  {Oka} T.,  {Yokozuka} H.,  {Enokiya} R.,  {Takekawa} S.,  {Iwata}
  Y.,   {Tsujimoto} S.,  2023, \mn@doi [\apj] {10.3847/1538-4357/aca66a}, \href
  {https://ui.adsabs.harvard.edu/abs/2023ApJ...942...46K} {942, 46}

\bibitem[\protect\citeauthoryear{{Kocsis} \& {Tremaine}}{{Kocsis} \&
  {Tremaine}}{2011}]{Kocsis2011}
{Kocsis} B.,  {Tremaine} S.,  2011, \mn@doi [\mnras]
  {10.1111/j.1365-2966.2010.17897.x}, \href
  {http://adsabs.harvard.edu/abs/2011MNRAS.412..187K} {412, 187}

\bibitem[\protect\citeauthoryear{{Kocsis} \& {Tremaine}}{{Kocsis} \&
  {Tremaine}}{2015}]{Kocsis2015}
{Kocsis} B.,  {Tremaine} S.,  2015, \mn@doi [\mnras] {10.1093/mnras/stv057},
  \href {https://ui.adsabs.harvard.edu/abs/2015MNRAS.448.3265K} {448, 3265}

\bibitem[\protect\citeauthoryear{Kroupa}{Kroupa}{2001}]{Kroupa2001}
Kroupa P.,  2001, \mn@doi [Mon Not R Astron Soc]
  {10.1046/j.1365-8711.2001.04022.x}, 322, 231

\bibitem[\protect\citeauthoryear{{Levin}}{{Levin}}{2007}]{Levin2007}
{Levin} Y.,  2007, \mn@doi [\mnras] {10.1111/j.1365-2966.2006.11155.x}, \href
  {http://adsabs.harvard.edu/abs/2007MNRAS.374..515L} {374, 515}

\bibitem[\protect\citeauthoryear{{Levin}}{{Levin}}{2024}]{Levin2024}
{Levin} Y.,  2024, \mn@doi [\apj] {10.3847/1538-4357/ad81f5}, \href
  {https://ui.adsabs.harvard.edu/abs/2024ApJ...975..278L} {975, 278}

\bibitem[\protect\citeauthoryear{{Levin} \& {Beloborodov}}{{Levin} \&
  {Beloborodov}}{2003}]{LevinBeloborodov2003}
{Levin} Y.,  {Beloborodov} A.~M.,  2003, \mn@doi [ApJL] {10.1086/376675}, \href
  {http://adsabs.harvard.edu/abs/2003ApJ...590L..33L} {590, L33}

\bibitem[\protect\citeauthoryear{{Lu}, {Do}, {Ghez}, {Morris}, {Yelda}  \&
  {Matthews}}{{Lu} et~al.}{2013}]{Lu2013}
{Lu} J.~R.,  {Do} T.,  {Ghez} A.~M.,  {Morris} M.~R.,  {Yelda} S.,   {Matthews}
  K.,  2013, \mn@doi [\apj] {10.1088/0004-637X/764/2/155}, \href
  {http://adsabs.harvard.edu/abs/2013ApJ...764..155L} {764, 155}

\bibitem[\protect\citeauthoryear{{Madau} \& {Rees}}{{Madau} \&
  {Rees}}{2001}]{Madau_Rees2001}
{Madau} P.,  {Rees} M.~J.,  2001, \mn@doi [\apjl] {10.1086/319848}, \href
  {https://ui.adsabs.harvard.edu/abs/2001ApJ...551L..27M} {551, L27}

\bibitem[\protect\citeauthoryear{{M{\'a}th{\'e}}, {Sz{\"o}lgy{\'e}n}  \&
  {Kocsis}}{{M{\'a}th{\'e}} et~al.}{2022}]{Mathe2022}
{M{\'a}th{\'e}} G.,  {Sz{\"o}lgy{\'e}n} {\'A}.,   {Kocsis} B.,  2022, arXiv
  e-prints, \href {https://ui.adsabs.harvard.edu/abs/2022arXiv220207665M} {p.
  arXiv:2202.07665}

\bibitem[\protect\citeauthoryear{{McKernan}, {Ford}, {Lyra}  \&
  {Perets}}{{McKernan} et~al.}{2012}]{McKernan2012}
{McKernan} B.,  {Ford} K.~E.~S.,  {Lyra} W.,   {Perets} H.~B.,  2012, \mn@doi
  [\mnras] {10.1111/j.1365-2966.2012.21486.x}, \href
  {https://ui.adsabs.harvard.edu/#abs/2012MNRAS.425..460M} {425, 460}

\bibitem[\protect\citeauthoryear{{McKernan}, {Ford}, {Kocsis}, {Lyra}  \&
  {Winter}}{{McKernan} et~al.}{2014}]{McKernan+2014}
{McKernan} B.,  {Ford} K.~E.~S.,  {Kocsis} B.,  {Lyra} W.,   {Winter} L.~M.,
  2014, \mn@doi [\mnras] {10.1093/mnras/stu553}, \href
  {https://ui.adsabs.harvard.edu/abs/2014MNRAS.441..900M} {441, 900}

\bibitem[\protect\citeauthoryear{{Naoz}}{{Naoz}}{2016}]{2016ARA&A..54..441N}
{Naoz} S.,  2016, \mn@doi [\araa] {10.1146/annurev-astro-081915-023315}, \href
  {http://adsabs.harvard.edu/abs/2016ARA%26A..54..441N} {54, 441}

\bibitem[\protect\citeauthoryear{{Naoz}, {Will}, {Ramirez-Ruiz}, {Hees}, {Ghez}
   \& {Do}}{{Naoz} et~al.}{2020}]{Naoz2020}
{Naoz} S.,  {Will} C.~M.,  {Ramirez-Ruiz} E.,  {Hees} A.,  {Ghez} A.~M.,   {Do}
  T.,  2020, \mn@doi [\apjl] {10.3847/2041-8213/ab5e3b}, \href
  {https://ui.adsabs.harvard.edu/abs/2020ApJ...888L...8N} {888, L8}

\bibitem[\protect\citeauthoryear{{O'Leary}, {Rasio}, {Fregeau}, {Ivanova}  \&
  {O'Shaughnessy}}{{O'Leary} et~al.}{2006}]{O'Leary+2006}
{O'Leary} R.~M.,  {Rasio} F.~A.,  {Fregeau} J.~M.,  {Ivanova} N.,
  {O'Shaughnessy} R.,  2006, \mn@doi [\apj] {10.1086/498446}, \href
  {https://ui.adsabs.harvard.edu/abs/2006ApJ...637..937O} {637, 937}

\bibitem[\protect\citeauthoryear{{Oyama} et~al.,}{{Oyama}
  et~al.}{2024}]{Oyama2024}
{Oyama} T.,  et~al., 2024, \mn@doi [arXiv e-prints]
  {10.48550/arXiv.2401.02312}, \href
  {https://ui.adsabs.harvard.edu/abs/2024arXiv240102312O} {p. arXiv:2401.02312}

\bibitem[\protect\citeauthoryear{{Panamarev} \& {Kocsis}}{{Panamarev} \&
  {Kocsis}}{2022}]{Panamarev2022}
{Panamarev} T.,  {Kocsis} B.,  2022, \mn@doi [\mnras] {10.1093/mnras/stac3050},
  \href {https://ui.adsabs.harvard.edu/abs/2022MNRAS.517.6205P} {517, 6205}

\bibitem[\protect\citeauthoryear{{Panamarev}, {Shukirgaliyev}, {Meiron},
  {Berczik}, {Just}, {Spurzem}, {Omarov}  \& {Vilkoviskij}}{{Panamarev}
  et~al.}{2018}]{Panamarev2018}
{Panamarev} T.,  {Shukirgaliyev} B.,  {Meiron} Y.,  {Berczik} P.,  {Just} A.,
  {Spurzem} R.,  {Omarov} C.,   {Vilkoviskij} E.,  2018, \mn@doi [\mnras]
  {10.1093/mnras/sty459}, \href
  {https://ui.adsabs.harvard.edu/abs/2018MNRAS.476.4224P} {476, 4224}

\bibitem[\protect\citeauthoryear{{Panamarev}, {Ginat}  \& {Kocsis}}{{Panamarev}
  et~al.}{2026}]{Panamarev2025}
{Panamarev} T.,  {Ginat} Y.~B.,   {Kocsis} B.,  2026, \mn@doi [\mnras]
  {10.1093/mnras/stag039}, \href
  {https://ui.adsabs.harvard.edu/abs/2026MNRAS.546ag039P} {546, stag039}

\bibitem[\protect\citeauthoryear{{Paumard} et~al.,}{{Paumard}
  et~al.}{2006}]{PaumardEtAl2006}
{Paumard} T.,  et~al., 2006, \mn@doi [ApJ] {10.1086/503273}, \href
  {http://adsabs.harvard.edu/abs/2006ApJ...643.1011P} {643, 1011}

\bibitem[\protect\citeauthoryear{{Perets}, {Hopman}  \& {Alexander}}{{Perets}
  et~al.}{2007}]{Perets2007}
{Perets} H.~B.,  {Hopman} C.,   {Alexander} T.,  2007, \mn@doi [\apj]
  {10.1086/510377}, \href {http://adsabs.harvard.edu/abs/2007ApJ...656..709P}
  {656, 709}

\bibitem[\protect\citeauthoryear{{Perets}, {Mastrobuono-Battisti}, {Meiron}  \&
  {Gualandris}}{{Perets} et~al.}{2018}]{Perets2018}
{Perets} H.~B.,  {Mastrobuono-Battisti} A.,  {Meiron} Y.,   {Gualandris} A.,
  2018, arXiv e-prints, \href
  {https://ui.adsabs.harvard.edu/abs/2018arXiv180200012P} {p. arXiv:1802.00012}

\bibitem[\protect\citeauthoryear{{Plummer}}{{Plummer}}{1911}]{Plummer1911}
{Plummer} H.~C.,  1911, MNRAS, \href
  {http://adsabs.harvard.edu/abs/1911MNRAS..71..460P} {71, 460}

\bibitem[\protect\citeauthoryear{{Portegies Zwart} \& {McMillan}}{{Portegies
  Zwart} \& {McMillan}}{2002}]{PortegiesZwart_McMillan2002}
{Portegies Zwart} S.~F.,  {McMillan} S. L.~W.,  2002, \mn@doi [\apj]
  {10.1086/341798}, \href
  {https://ui.adsabs.harvard.edu/abs/2002ApJ...576..899P} {576, 899}

\bibitem[\protect\citeauthoryear{{Portegies Zwart}, {Baumgardt}, {McMillan},
  {Makino}, {Hut}  \& {Ebisuzaki}}{{Portegies Zwart}
  et~al.}{2006}]{Portegies_Zwart+2006}
{Portegies Zwart} S.~F.,  {Baumgardt} H.,  {McMillan} S. L.~W.,  {Makino} J.,
  {Hut} P.,   {Ebisuzaki} T.,  2006, \mn@doi [\apj] {10.1086/500361}, \href
  {https://ui.adsabs.harvard.edu/abs/2006ApJ...641..319P} {641, 319}

\bibitem[\protect\citeauthoryear{{Rauch} \& {Tremaine}}{{Rauch} \&
  {Tremaine}}{1996}]{Rauch1996}
{Rauch} K.~P.,  {Tremaine} S.,  1996, \mn@doi [\na]
  {10.1016/S1384-1076(96)00012-7}, \href
  {https://ui.adsabs.harvard.edu/abs/1996NewA....1..149R} {1, 149}

\bibitem[\protect\citeauthoryear{{Reid} \& {Brunthaler}}{{Reid} \&
  {Brunthaler}}{2004}]{Reid2004}
{Reid} M.~J.,  {Brunthaler} A.,  2004, \mn@doi [\apj] {10.1086/424960}, \href
  {https://ui.adsabs.harvard.edu/abs/2004ApJ...616..872R} {616, 872}

\bibitem[\protect\citeauthoryear{{Reid} \& {Brunthaler}}{{Reid} \&
  {Brunthaler}}{2020}]{Reid2020}
{Reid} M.~J.,  {Brunthaler} A.,  2020, \mn@doi [\apj]
  {10.3847/1538-4357/ab76cd}, \href
  {https://ui.adsabs.harvard.edu/abs/2020ApJ...892...39R} {892, 39}

\bibitem[\protect\citeauthoryear{{Rizzuto} et~al.,}{{Rizzuto}
  et~al.}{2021}]{Rizzuto2021}
{Rizzuto} F.~P.,  et~al., 2021, \mn@doi [\mnras] {10.1093/mnras/staa3634},
  \href {https://ui.adsabs.harvard.edu/abs/2021MNRAS.501.5257R} {501, 5257}

\bibitem[\protect\citeauthoryear{{Rose}, {Naoz}, {Sari}  \& {Linial}}{{Rose}
  et~al.}{2022}]{Rose2022}
{Rose} S.~C.,  {Naoz} S.,  {Sari} R.,   {Linial} I.,  2022, \mn@doi [\apjl]
  {10.3847/2041-8213/ac6426}, \href
  {https://ui.adsabs.harvard.edu/abs/2022ApJ...929L..22R} {929, L22}

\bibitem[\protect\citeauthoryear{{Strokov}, {Fragione}  \& {Berti}}{{Strokov}
  et~al.}{2023}]{Strokov2023}
{Strokov} V.,  {Fragione} G.,   {Berti} E.,  2023, \mn@doi [\mnras]
  {10.1093/mnras/stad2002}, \href
  {https://ui.adsabs.harvard.edu/abs/2023MNRAS.524.2033S} {524, 2033}

\bibitem[\protect\citeauthoryear{{Sz{\"o}lgy{\'e}n} \&
  {Kocsis}}{{Sz{\"o}lgy{\'e}n} \& {Kocsis}}{2018}]{Szolgyen2018}
{Sz{\"o}lgy{\'e}n} {\'A}.,  {Kocsis} B.,  2018, \mn@doi [\prl]
  {10.1103/PhysRevLett.121.101101}, \href
  {https://ui.adsabs.harvard.edu/abs/2018PhRvL.121j1101S} {121, 101101}

\bibitem[\protect\citeauthoryear{{Sz{\"o}lgy{\'e}n}, {M{\'a}th{\'e}}  \&
  {Kocsis}}{{Sz{\"o}lgy{\'e}n} et~al.}{2021}]{Szolgyen2021}
{Sz{\"o}lgy{\'e}n} {\'A}.,  {M{\'a}th{\'e}} G.,   {Kocsis} B.,  2021, \mn@doi
  [\apj] {10.3847/1538-4357/ac13ab}, \href
  {https://ui.adsabs.harvard.edu/abs/2021ApJ...919..140S} {919, 140}

\bibitem[\protect\citeauthoryear{{Tagawa}, {Haiman}  \& {Kocsis}}{{Tagawa}
  et~al.}{2020}]{Tagawa+2020}
{Tagawa} H.,  {Haiman} Z.,   {Kocsis} B.,  2020, \mn@doi [\apj]
  {10.3847/1538-4357/ab7922}, \href
  {https://ui.adsabs.harvard.edu/abs/2020ApJ...892...36T} {892, 36}

\bibitem[\protect\citeauthoryear{{Tremaine}}{{Tremaine}}{1976}]{Tremaine1976}
{Tremaine} S.~D.,  1976, \mn@doi [\apj] {10.1086/154085}, \href
  {http://adsabs.harvard.edu/abs/1976ApJ...203..345T} {203, 345}

\bibitem[\protect\citeauthoryear{{Tsuboi}, {Kitamura}, {Tsutsumi}, {Uehara},
  {Miyoshi}, {Miyawaki}  \& {Miyazaki}}{{Tsuboi} et~al.}{2017}]{Tsuboi2017}
{Tsuboi} M.,  {Kitamura} Y.,  {Tsutsumi} T.,  {Uehara} K.,  {Miyoshi} M.,
  {Miyawaki} R.,   {Miyazaki} A.,  2017, \mn@doi [\apjl]
  {10.3847/2041-8213/aa97d3}, \href
  {https://ui.adsabs.harvard.edu/abs/2017ApJ...850L...5T} {850, L5}

\bibitem[\protect\citeauthoryear{{Tsuboi}, {Kitamura}, {Tsutsumi}, {Miyawaki},
  {Miyoshi}  \& {Miyazaki}}{{Tsuboi} et~al.}{2020}]{Tsuboi2020}
{Tsuboi} M.,  {Kitamura} Y.,  {Tsutsumi} T.,  {Miyawaki} R.,  {Miyoshi} M.,
  {Miyazaki} A.,  2020, \mn@doi [\pasj] {10.1093/pasj/psaa016}, \href
  {https://ui.adsabs.harvard.edu/abs/2020PASJ...72L...5T} {72, L5}

\bibitem[\protect\citeauthoryear{{Will}, {Naoz}, {Hees}, {Tucker}, {Zhang},
  {Do}  \& {Ghez}}{{Will} et~al.}{2023}]{Will2023}
{Will} C.~M.,  {Naoz} S.,  {Hees} A.,  {Tucker} A.,  {Zhang} E.,  {Do} T.,
  {Ghez} A.,  2023, \mn@doi [\apj] {10.3847/1538-4357/ad09b3}, \href
  {https://ui.adsabs.harvard.edu/abs/2023ApJ...959...58W} {959, 58}

\bibitem[\protect\citeauthoryear{{Yelda}, {Ghez}, {Lu}, {Do}, {Meyer}, {Morris}
   \& {Matthews}}{{Yelda} et~al.}{2014}]{Yelda2014}
{Yelda} S.,  {Ghez} A.~M.,  {Lu} J.~R.,  {Do} T.,  {Meyer} L.,  {Morris} M.~R.,
    {Matthews} K.,  2014, \mn@doi [\apj] {10.1088/0004-637X/783/2/131}, \href
  {http://adsabs.harvard.edu/abs/2014ApJ...783..131Y} {783, 131}

\bibitem[\protect\citeauthoryear{{Yu} \& {Tremaine}}{{Yu} \&
  {Tremaine}}{2003}]{YuTremaine2003}
{Yu} Q.,  {Tremaine} S.,  2003, \mn@doi [\apj] {10.1086/379546}, \href
  {https://ui.adsabs.harvard.edu/abs/2003ApJ...599.1129Y} {599, 1129}

\bibitem[\protect\citeauthoryear{{Zheng}, {Wang}, {Lin}, {Burkert}  \&
  {Mao}}{{Zheng} et~al.}{2026}]{Zheng2026}
{Zheng} X.,  {Wang} L.,  {Lin} D. N.~C.,  {Burkert} A.,   {Mao} S.,  2026,
  \mn@doi [arXiv e-prints] {10.48550/arXiv.2606.08971}, \href
  {https://ui.adsabs.harvard.edu/abs/2026arXiv260608971Z} {p. arXiv:2606.08971}

\bibitem[\protect\citeauthoryear{{{\v{S}}ubr}, {Schovancov{\'a}}  \&
  {Kroupa}}{{{\v{S}}ubr} et~al.}{2009}]{Subr+2009}
{{\v{S}}ubr} L.,  {Schovancov{\'a}} J.,   {Kroupa} P.,  2009, \mn@doi [\aap]
  {10.1051/0004-6361:200811075}, \href
  {https://ui.adsabs.harvard.edu/abs/2009A&A...496..695S} {496, 695}

\bibitem[\protect\citeauthoryear{{von Fellenberg} et~al.,}{{von Fellenberg}
  et~al.}{2022}]{vonFellenberg2022}
{von Fellenberg} S.,  et~al., 2022, arXiv e-prints, \href
  {https://ui.adsabs.harvard.edu/abs/2022arXiv220507595V} {p. arXiv:2205.07595}

\makeatother
\end{thebibliography}


\onecolumn
\appendix

\section{Skymaps for a prograde model}
\label{App:skymap i45}

\begin{figure}
    \centering
    \includegraphics[width=1\linewidth]{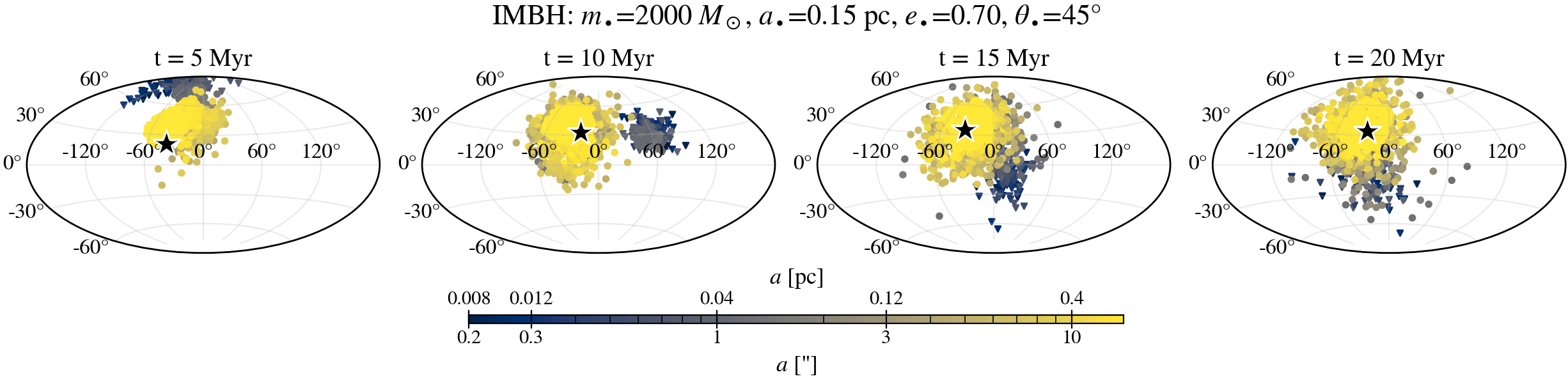}
    \caption{Aitoff projections for a model with an IMBH as stated in the figure title. }
    \label{fig:skymapsi45}
\end{figure}

Figure~\ref{fig:skymapsi45} shows Aitoff projections of the angular momentum vector directions of the disc stars corresponding to 5, 10, 15, and 20 Myr for a model with a prograde IMBH with respect to the net angular momentum vector of the stellar disc. The figure shows formation of an inner stellar disc which is prominently visible at 10 Myr, but tends to align with the outer disc and with the net angular momentum vector of the system by 20 Myr.

\section{Analytical estimates for the stellar disc self--torques}
\label{App:torques analytical}

For a razor--thin disc, the mean nodal precession frequency due to the VRR torque exerted by the disc on a test star of mass $m_\star$, orbiting at radius $r$ with inclination $\theta$ with respect to the disc plane, can be written as \citep{Kocsis2011, Kocsis2015}
\begin{equation}
    \boldsymbol{\Omega}_{\rm disc}(r,\theta)
    = \sum_{j=1}^{N_{\rm disc}} \boldsymbol{\Omega}_{ij}
    \simeq N_{\rm disc}
    \left\langle
      \sum_{\ell}
      \frac{\mathcal{J}_{ij\ell}\,P'_{\ell}(\cos \theta)}{L_\star}
    \right\rangle_j ,
\end{equation}
where $\mathcal{J}_{ij\ell}$ is given by Eq.~\eqref{eq:Jijl} and
$L_\star=\sqrt{G \MSMBH r}$ is the angular momentum of the test star around
the SMBH.

Using Eq.~\eqref{eq:Jijl} and writing the disc surface density as
$\Sigma(r)$, the azimuthally averaged contribution of each even $\ell$
to the precession frequency of a star at $r$ can be written as
\begin{equation}
  I_\ell(r,\theta)
  =
  [P_{\ell}(0)]^2 P'_\ell(\cos\theta)\,
  \frac{G}{\sqrt{G \MSMBH r}}
  \left[
    \frac{1}{r} \int_{0}^{r}
       \left(\frac{r'}{r}\right)^{\ell} 2\pi r'\,\Sigma(r')\,\mathrm{d}r'
    +
      \int_{r}^{r_{\rm d}}
       \left(\frac{r}{r'}\right)^{\ell} 2\pi\,\Sigma(r')\,\mathrm{d}r'
  \right],
\end{equation}
where $r_{\rm d}$ is the outer radius of the disc. The total torque is
\begin{equation}
  \boldsymbol{\Omega}_{\rm disc}(r,\theta)
  =
  \sum_{\ell=2,4,\dots}^{\infty} I_\ell(r,\theta)\,
  \hat{\boldsymbol{n}}(\theta),
\end{equation}
with $\hat{\boldsymbol{n}}(\theta)$ the unit vector along the precession
axis.

Consider a finite disc with outer radius $r_{\rm d}$ and a power--law
surface density profile
\begin{equation}
  \Sigma(r)
  = \frac{(2-\Gamma)M_{\rm d}}{2\pi r_{\rm d}^2}
    \left(\frac{r}{r_{\rm d}}\right)^{-\Gamma},
\end{equation}
where $M_{\rm d}$ is the total disc mass and $\Gamma$ the slope.
Evaluating the radial integrals analytically gives
\begin{align}
    I_\ell(r,\theta)
    &=
    [P_{\ell}(0)]^2 P'_\ell(\cos\theta)\,
    \frac{2\pi(2-\Gamma)}{t_{\rm orb}(r)}\,
    \frac{M_{\rm d}}{\MSMBH}\,
    \left[
      \frac{\left(\frac{r}{r_{\rm d}}\right)^{2-\Gamma}}{\ell+2-\Gamma}
      +
      \frac{
        \left(\frac{r}{r_{\rm d}}\right)^{2-\Gamma}
        -
        \left(\frac{r}{r_{\rm d}}\right)^{\ell+1}
      }{\ell+\Gamma-1}
    \right],
\end{align}
where $t_{\rm orb}(r)=2\pi\sqrt{r^3/(G \MSMBH)}$ is the orbital period.
Summing over even $\ell$ yields
\begin{align}
    \boldsymbol{\Omega}_{\rm disc}(r,\theta)
    &=
    \Omega_0(r)\,
    \Bigg[
      \sum_{\ell=2,4,\dots}^{\infty}
        [P_{\ell}(0)]^2 P'_{\ell}(\cos \theta)
        \left(
          \frac{\left(\frac{r}{r_{\rm d}}\right)^{2-\Gamma}}{\ell+2-\Gamma}
          +
          \frac{
            \left(\frac{r}{r_{\rm d}}\right)^{2-\Gamma}
            -
            \left(\frac{r}{r_{\rm d}}\right)^{\ell+1}
          }{\ell+\Gamma-1}
        \right)
    \Bigg]\hat{\boldsymbol{n}}(\theta),
\label{eq:omega_disc-sum-corrected-final}
\end{align}
where
\begin{equation}
  \Omega_0(r)
  \equiv
  \frac{2\pi(2-\Gamma)}{t_{\rm orb}(r)}\,
  \frac{M_{\rm d}}{\MSMBH}
\end{equation}
sets the characteristic precession frequency.

The sums in Eq.~\eqref{eq:omega_disc-sum-corrected-final} can be evaluated in a closed-form, as we show in Appendix~\ref{App:Legendre}. Identifying
$x=\cos\theta$ and $z=r/r_{\rm d}$, and rewriting the radial factors as a combination of terms independent of $z$
and terms proportional to $z^\ell$, the torque can be written as
\begin{equation}
  \boldsymbol{\Omega}_{\rm disc}(r,\theta)
  \simeq
  \Omega_0(r)\,
  \Big\{
    z^{2-\Gamma}
    \big[
      S(\cos\theta,1;2-\Gamma)
      + S(\cos\theta,1;\Gamma-1)
    \big]
    - z\,S(\cos\theta,z;\Gamma-1)
  \Big\}\hat{\boldsymbol{n}}(\theta),
  \label{eq:omega_disc_hybrid_general_corrected-final}
\end{equation}
where $S$ is defined in Appendix~\ref{app:analytic-legendre}.

In the small-angle limit ($\theta\ll1$), the kernel at $z=1$ behaves
as
\begin{equation}
  S(\cos\theta,1;\alpha)
  \simeq \frac{1}{\pi}\cot\theta + C_\alpha,
  \qquad \theta\ll1,
\end{equation}
where $C_\alpha$ is a constant depending on $\alpha$ and expressible in
terms of $\Delta_1(\alpha)$ and $\Delta_2(\alpha)$
(see Eqs.~\ref{eq:Delta1} and \ref{eq:Delta2}).
For the two sums that appear in Eq.~\eqref{eq:omega_disc_hybrid_general_corrected-final},
\begin{equation}
  S(\cos\theta,1;2-\Gamma)
  +
  S(\cos\theta,1;\Gamma-1)
  \simeq
  \frac{2}{\pi}\cot\theta + \mathcal{C}(\Gamma),
\end{equation}
with
\begin{equation}
\label{eq:C_gamma}
  \mathcal{C}(\Gamma)
  =
  3\big[\Delta_1(2-\Gamma) + \Delta_1(\Gamma-1)\big]
  + 10\big[\Delta_2(2-\Gamma) + \Delta_2(\Gamma-1)\big].
\end{equation}
Here $\Delta_1(\cdot)$ and $\Delta_2(\cdot)$ are defined in Eqs.~\eqref{eq:Delta1}--\eqref{eq:Delta2}.

For $0<z<1$, taking the limit $\theta\to0$ at fixed $z$ yields the finite expression
\begin{equation}
  S(\cos\theta,z;\Gamma-1)
  \simeq
  \frac{1}{\pi}
  \left[
    \frac{z^2}{1-z^2}
    -\frac{1}{2}\ln(1-z^2)
  \right]
  + 3\,\Delta_1(\Gamma-1)\,z^2
  +10\,\Delta_2(\Gamma-1)\,z^4,
  \qquad \theta\ll1.
\end{equation}
However, this approximation is not uniform as $z\to1$: the limits $\theta\to0$ and $z\to1$ do not commute, and for $z$ sufficiently close to unity the dominant contribution comes from a wide range of multipoles up to $\ell\sim \theta^{-1}$, so that the explicit $\theta$-dependence of $S(\cos\theta,z;\Gamma-1)$ must be retained.

A simple practical improvement that remains accurate for $z\lesssim1$ is to apply the cotangent approximation only to the $z=1$ boundary terms while keeping the full hybrid kernel for the $z$-dependent term,
\begin{equation}
  \frac{\Omega_{\rm disc}(r,\theta)}{\Omega_0(r)}
  \approx
    z^{2-\Gamma}
    \left[
      \frac{2}{\pi}\cot\theta + \mathcal{C}(\Gamma)
    \right]
    - z\,S(\cos\theta,z;\Gamma-1),
  \qquad \theta\ll1,
  \label{eq:Omega_small_angle_improved}
\end{equation}
which reduces to the previous expression at fixed $z<1$ and captures the correct crossover behaviour as $z\to1$.

\begin{figure*}
  \centering
  \includegraphics[width=0.49\textwidth]{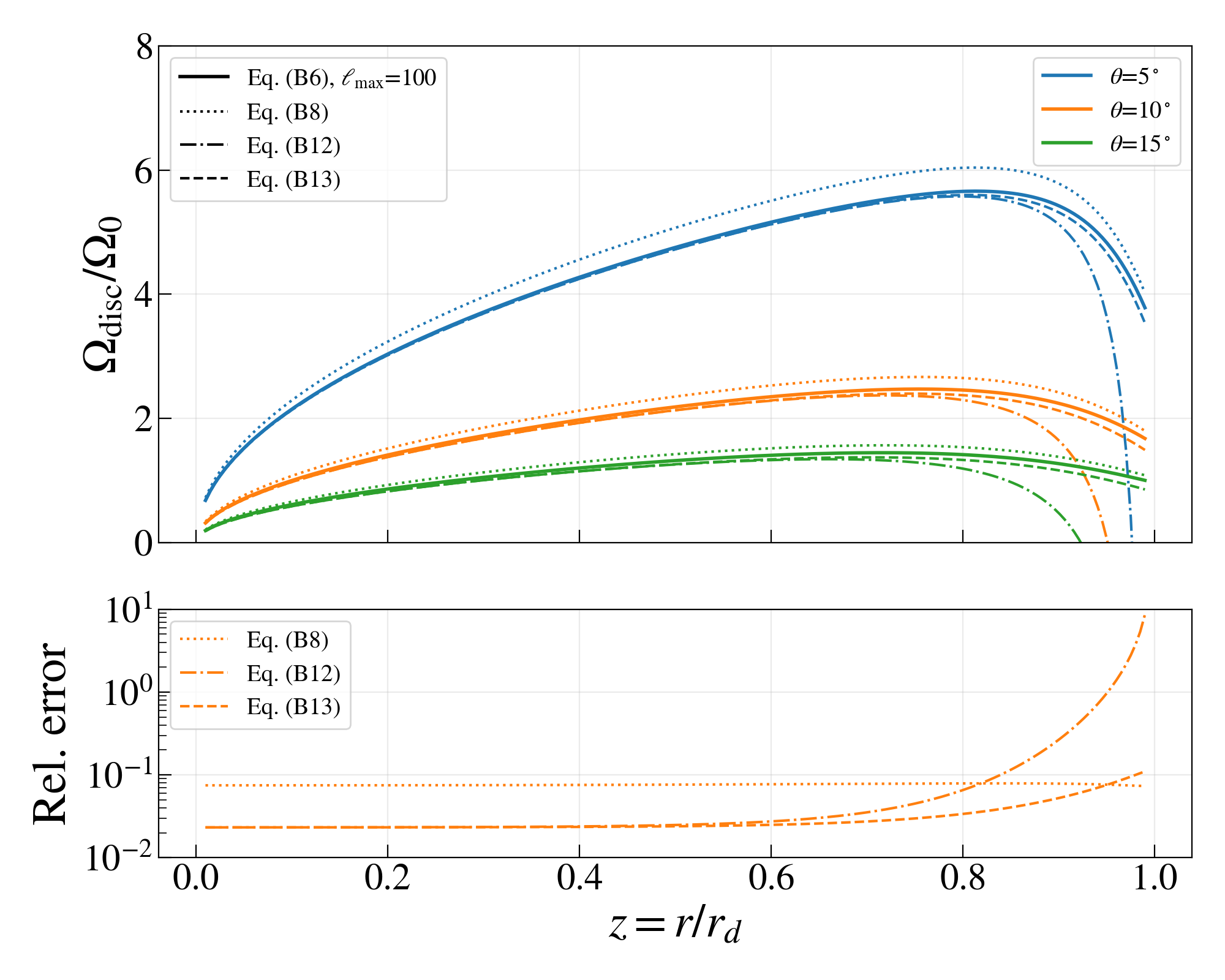}\hfill
  \includegraphics[width=0.49\textwidth]{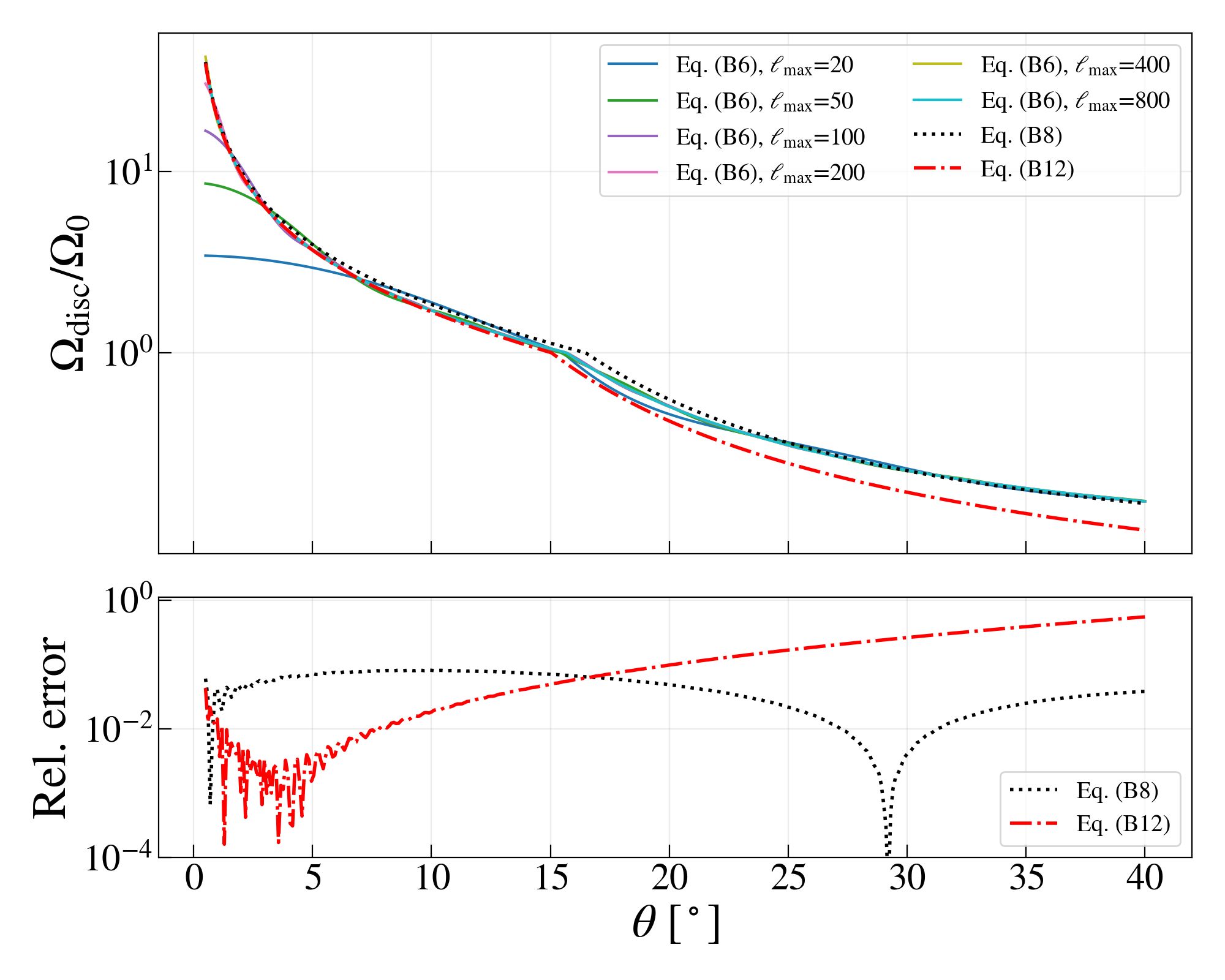}
  \caption{Comparison of the disc self--torque $\Omega_{\rm disc}/\Omega_0$ computed from the truncated even--$\ell$ sum in Eq.~\eqref{eq:omega_disc-sum-corrected-final} with the hybrid closed--form expression in Eq.~\eqref{eq:omega_disc_hybrid_general_corrected-final} and the improved small--angle approximation in Eq.~\eqref{eq:Omega_small_angle_improved}. Left: $\Omega_{\rm disc}/\Omega_0$ versus $z=r/r_{\rm d}$ for several fixed inclinations. Right: $\Omega_{\rm disc}/\Omega_0$ versus $\theta$ for a fixed $z$, showing convergence with $\ell_{\max}$.}
  \label{fig:disc_torque_small_angle_tests}
\end{figure*}

This behaviour is illustrated in Fig.~\ref{fig:disc_torque_small_angle_tests}. The left panel compares $\Omega_{\rm disc}/\Omega_0$ as a function of $z=r/r_{\rm d}$ for several fixed inclinations, showing that the improved small--angle prescription follows the truncated even--$\ell$ sum well up to $z\lesssim1$ while the non-uniform small--angle limit at fixed $z$ overestimates the torque as $z\to1$. The right panel compares $\Omega_{\rm disc}/\Omega_0$ as a function of $\theta$ for a fixed $z = 0.3$, demonstrating the convergence of the truncated sum with increasing $\ell_{\max}$ and the accuracy of the hybrid and improved small--angle approximations over the full range of angles.

\medskip
The hybrid representation of $S$ (Appendix~\ref{App:Legendre}) is accurate to within a few per cent over the full range of angles for $0<z\leq1$, and the relative error decreases rapidly for $z<1$ because the correction $\Delta S$ converges quickly when $z^\ell$ suppresses high $\ell$. The accuracy can be systematically improved by including additional low--$\ell$ terms in $\Delta S_{\rm corr}$ (e.g.\ extending beyond $\ell=4$) or, equivalently, by matching more terms in the large--$\ell$ expansion of $[P_{2n}(0)]^2/(2n+\alpha)$ in the asymptotic tail. In numerical benchmarks based on truncating the sum at $\ell_{\max}$, one must also ensure convergence at small angles: the $z=1$ kernels receive appreciable contributions from multipoles up to $\ell\sim \theta^{-1}$, so the truncation error can dominate for $\ell_{\max}\lesssim \theta^{-1}$ even when the hybrid approximation is accurate.

\medskip
If the disc formally extends to infinity while the local surface density
is well approximated by a power law,
\begin{equation}
  \Sigma(r') = \Sigma(r)\left(\frac{r'}{r}\right)^{-\Gamma},
\end{equation}
it is natural to characterise the disc by the local disc mass per
logarithmic interval in radius,
\begin{equation}
  M_{\rm d,loc}(r)
  \equiv \frac{{\rm d}M_{\rm d}}{{\rm d}\ln r}\Big|_{r}
  = 2\pi r^2\,\Sigma(r),
\end{equation}
and the corresponding local precession frequency scale
\begin{equation}
  \Omega_{\rm loc}(r)
  \equiv \frac{2\pi}{t_{\rm orb}(r)}\,\frac{M_{\rm d,loc}(r)}{\MSMBH}.
\end{equation}
In this case the small--angle torque takes the
explicit form
\begin{equation}
  \frac{\Omega_{\rm disc}^{(\infty)}(r,\theta)}{\Omega_{\rm loc}(r)}
  \approx
  \frac{2}{\pi}\cot\theta + \mathcal{C}(\Gamma),
  \qquad \theta\ll1,
\end{equation}
i.e.\ the same $\cot\theta$ scaling as in the finite--disc case, with the
overall amplitude set by the local disc mass $M_{\rm d,loc}(r)$.

\section{Legendre kernels for circular-disc torques}
\label{App:Legendre}

The VRR Hamiltonian expressed by Eq.~\ref{eq:H_VRR} involves an infinite sum with Legendre polynomials. In \citet{Panamarev2025}, we revisited mathematical techniques on evaluating these infinite sums. In this appendix we show how to use them to evaluate the infinite sums which we encounter when computing the net torque from a disc of stars on circular orbits. 

\subsection{Derivation of the analytic representation}
\label{app:analytic-legendre}

In this subsection we derive an analytic representation for the even-order Legendre sums that appear in the VRR Hamiltonian when the coefficients contain the factors $[P_{2\ell}(0)]^2$. This allows us to replace the full infinite sum by a closed-form expression plus a small number of low-$\ell$ corrections. The method builds on the large-$\ell$ asymptotics of $P_{2\ell}(0)$ and on analytic expressions for even Legendre sums derived using the generating function (see \citealt{Kocsis2015} and \citealt{Panamarev2025}).

We are interested in sums of the form
\begin{equation}
  S(x,z;\alpha)
  = \sum_{n=1}^{\infty}
  \frac{\bigl[P_{2n}(0)\bigr]^2\,P'_{2n}(x)}{2n+\alpha}\,z^{2n},
  \label{eq:S_def}
\end{equation}
where $x=\cos\theta=\mathbf{L}_i\!\cdot\!\mathbf{L}_j$, $0<z\leq1$ encodes the ratio of radii or semi-major axes, and $\alpha$ is related to the power-law slope of the surface density of the stellar disc (cf. Appendix~\ref{App:torques analytical}).

For large even order $\ell=2n$, the values at the origin satisfy 
\begin{equation}
\label{eq:P2n_asympt}
\bigl[P_{2n}(0)\bigr]^2
= \frac{1}{\pi n}
\left( 1 - \frac{1}{4n} + \frac{5}{32 n^{2}} + O\!\left(n^{-3}\right) \right),
\qquad n\to\infty .
\end{equation}
The factor in the denominator can be expanded as
\begin{equation}
  \frac{1}{2n+\alpha}
  = \frac{1}{2n}
    \left(1 - \frac{\alpha}{2n} + O\!\left(n^{-2}\right)\right),
  \qquad n\to\infty,
\end{equation}
so that
\begin{equation}
  \frac{\bigl[P_{2n}(0)\bigr]^2}{2n+\alpha}
  = \frac{1}{2\pi n^2} + O\!\left(\frac{1}{n^3}\right),
  \qquad n\to\infty.
  \label{eq:Coeff_asympt}
\end{equation}
Thus the large-$n$ behaviour of the summand in Eq.~\eqref{eq:S_def} is governed by
\begin{equation}
  \frac{\bigl[P_{2n}(0)\bigr]^2}{2n+\alpha}P'_{2n}(x)z^{2n}
  = \frac{1}{2\pi}\,\frac{P'_{2n}(x)}{n^2}z^{2n}
    + O\!\left(\frac{P'_{2n}(x)\,z^{2n}}{n^3}\right).
\end{equation}

Following \citet{Kocsis2015} and \citet{Panamarev2025}, it is convenient to introduce the even-$\ell$ reference sum
\begin{equation}
  S_{2,\mathrm{even}}(x,z)
  := \sum_{\ell=2,4,\dots}^{\infty}
     \frac{P'_{\ell}(x)}{\ell^2}\,z^{\ell}.
  \label{eq:S2even_def_appendix}
\end{equation}
In terms of $n$ with $\ell=2n$,
\begin{equation}
  \sum_{n=1}^{\infty}\frac{P'_{2n}(x)}{n^2}z^{2n}
  = 4 \sum_{\ell=2,4,\dots}^{\infty}\frac{P'_{\ell}(x)}{\ell^2}z^{\ell}
  = 4\,S_{2,\mathrm{even}}(x,z).
\end{equation}
$S_{2,\mathrm{even}}$ admits the closed form (cf. Appendix~A of \citealt{Panamarev2025})
\begin{equation}
  S_{2,\mathrm{even}}(x,z)
  = \frac{1}{2(1-x)}\ln\!\left[\frac{(1+X+z)(1+Y-z)}{4}\right]
    -\frac{1}{2(1+x)}\ln\!\left(\frac{(1+X-z)(1+Y+z)}{4}\right),
  \label{eq:S2even_closed_appendix}
\end{equation}
where
\begin{equation}
  X=\sqrt{1-2xz+z^2},\qquad
  Y=\sqrt{1+2xz+z^2}.
\end{equation}

Comparing Eq.~\eqref{eq:Coeff_asympt} with Eq.~\eqref{eq:S2even_def_appendix}, we identify the leading large-$n$ contribution to Eq.~\eqref{eq:S_def} as
\begin{equation}
  S_{\rm asymp}(x,z;\alpha)
  := \sum_{n=1}^{\infty}
      \frac{1}{2\pi n^2}\,P'_{2n}(x)z^{2n}
   = \frac{2}{\pi}\,S_{2,\mathrm{even}}(x,z).
  \label{eq:S_asymp_def_appendix}
\end{equation}
We then write
\begin{equation}
  S(x,z;\alpha)
  = S_{\rm asymp}(x,z;\alpha) + \Delta S(x,z;\alpha),
\end{equation}
with the correction
\begin{equation}
  \Delta S(x,z;\alpha)
  := \sum_{n=1}^{\infty}
     \left[
       \frac{\bigl[P_{2n}(0)\bigr]^2}{2n+\alpha}
       - \frac{1}{2\pi n^2}
     \right]P'_{2n}(x)z^{2n}.
\end{equation}
The tail $S_{\rm asymp}$ is fully analytic via Eq.~\eqref{eq:S2even_closed_appendix}, while $\Delta S$ contains only subleading $O(n^{-3})$ terms and converges rapidly for $z<1$.

In practice we approximate $\Delta S$ by keeping only the $\ell=2$ and $\ell=4$ terms, which can be evaluated analytically. Using
\begin{equation}
  P_2(0) = -\frac12,\quad [P_2(0)]^2=\frac14,\quad P'_2(x)=3x,
\end{equation}
\begin{equation}
  P_4(0) = \frac{3}{8},\quad [P_4(0)]^2=\frac{9}{64},\quad
  P'_4(x)=\frac{35x^3-15x}{2},
\end{equation}
we define
\begin{align}\label{eq:Delta1}
  \Delta_1(\alpha)
  &:= \frac{1}{4(2+\alpha)} - \frac{1}{2\pi},\\[2pt]
  \Delta_2(\alpha)
  &:= \frac{9}{64(4+\alpha)} - \frac{1}{8\pi}.\label{eq:Delta2}
\end{align}
The correction term is then
\begin{equation}
  \Delta S_{\rm corr}(x,z;\alpha)
  := \Delta_1(\alpha)\,P'_2(x)\,z^2
    +\Delta_2(\alpha)\,P'_4(x)\,z^4
  = 3x\,\Delta_1(\alpha)\,z^2
    +\frac{35x^3-15x}{2}\,\Delta_2(\alpha)\,z^4.
  \label{eq:DeltaS_corr_appendix}
\end{equation}
We therefore evaluate the kernel using
\begin{equation}
  S(x,z;\alpha)
  \simeq S_{\rm asymp}(x,z;\alpha) + \Delta S_{\rm corr}(x,z;\alpha),
\end{equation}
with $S_{\rm asymp}$ given by Eq.~\eqref{eq:S_asymp_def_appendix}. For our purposes this representation is accurate to better than a few per cent over the full range of angles and $0<z\leq1$.

\subsection{Limiting cases}

For later use we record two simple limiting forms of the derived kernel.

\paragraph{(i) Small-angle approximation.}

At $x=1$ (with $0<z<1$) one has $P_\ell(1)=1$ and $P'_\ell(1)=\ell(\ell+1)/2$. For the even sum \eqref{eq:S2even_def_appendix} this gives
\begin{equation}
  S_{2,\mathrm{even}}(x=1,z)
  = \sum_{n=1}^{\infty}\frac{P'_{2n}(1)}{(2n)^2}z^{2n}
  = \sum_{n=1}^{\infty}\left(\frac{1}{2} + \frac{1}{4n}\right)z^{2n}.
\end{equation}
Summing the geometric and logarithmic series,
\begin{equation}
  \sum_{n=1}^{\infty} z^{2n} = \frac{z^2}{1-z^2},\qquad
  \sum_{n=1}^{\infty} \frac{z^{2n}}{n} = -\ln(1-z^2),
\end{equation}
yields the exact $x\to 1$ limit
\begin{equation}
  S_{2,\mathrm{even}}(x\to 1,z)
  \longrightarrow
  \frac{1}{2}\,\frac{z^2}{1-z^2}
  - \frac{1}{4}\,\ln(1-z^2),
  \qquad 0<z<1.
  \label{eq:S2even_x1_limit}
\end{equation}
Using Eq.~\eqref{eq:S_asymp_def_appendix}, the asymptotic tail becomes
\begin{equation}
  S_{\rm asymp}(x\to 1,z;\alpha)
  \longrightarrow
  \frac{2}{\pi}S_{2,\mathrm{even}}(1,z)
  = \frac{1}{\pi}\left[
      \frac{z^2}{1-z^2}
      - \frac{1}{2}\ln(1-z^2)
    \right].
\end{equation}
Evaluating the low-$\ell$ correction at $x=1$ gives
\begin{equation}
  P'_2(1)=3,\qquad P'_4(1)=10,
\end{equation}
so that
\begin{equation}
  \Delta S_{\rm corr}(x\to 1,z;\alpha)
  \longrightarrow
  3\,\Delta_1(\alpha)\,z^2 + 10\,\Delta_2(\alpha)\,z^4.
\end{equation}
Combining these, the small-angle (i.e.\ $x=\cos\theta\to 1$) kernel at fixed $0<z<1$ is
\begin{equation}
  S(x\to 1,z;\alpha)
  \approx
  \frac{1}{\pi}\left[
      \frac{z^2}{1-z^2}
      - \frac{1}{2}\ln(1-z^2)
    \right]
  + 3\,\Delta_1(\alpha)\,z^2
  + 10\,\Delta_2(\alpha)\,z^4.
  \label{eq:Sapprox_small_angle_zlt1}
\end{equation}
This limit is finite and depends only on $z$ and $\alpha$; the $1/\pi$ prefactor follows directly from the asymptotic behaviour \eqref{eq:P2n_asympt}.

\paragraph{(ii) $z=1$.}

For the case of $z=1$, the even-$\ell$ kernel $S_{2,\mathrm{even}}(x,1)$, as shown in \citet{Panamarev2025}, is very well approximated by a cotangent:
\begin{equation}
  S_{2,\mathrm{even}}(x=\cos\theta,z=1)
  \simeq \frac{1}{2}\cot\theta
  \end{equation}
with relative errors of a few per cent over the range of angles $0<\theta<180^\circ$. Through Eq.~\eqref{eq:S_asymp_def_appendix} this implies
\begin{equation}
  S_{\rm asymp}(x=\cos\theta,z=1;\alpha)
  \simeq \frac{2}{\pi}\,S_{2,\mathrm{even}}(x,1)
  \simeq \frac{1}{\pi}\cot\theta.
\end{equation}
The low-$\ell$ correction at $z=1$ is
\begin{equation}
  \Delta S_{\rm corr}(x,z=1;\alpha)
  = 3x\,\Delta_1(\alpha)
    +\frac{35x^3-15x}{2}\,\Delta_2(\alpha),
\end{equation}
which is regular as $\theta\to 0$ and can be expanded as
\begin{equation}
  \Delta S_{\rm corr}(x=\cos\theta,z=1;\alpha)
  = 3\,\Delta_1(\alpha)
    + 10\,\Delta_2(\alpha)
    + O(\theta^2).
\end{equation}
Thus, for $z=1$ the small-angle behaviour of the kernel is
\begin{equation}
  S(x=\cos\theta,z=1;\alpha)
  \simeq \frac{1}{\pi}\cot\theta
  + 3\,\Delta_1(\alpha)
  + 10\,\Delta_2(\alpha)
  + O(\theta^2),
  \qquad \theta\ll 1,
\end{equation}
i.e.\ the expected $1/\theta$ singularity with an overall amplitude $1/\pi$, plus a finite offset determined entirely by the $\ell=2$ and $\ell=4$ modes.

\section{Intact-fragment regimes across the IMBH parameter space}
\label{App:disruption regimes}

This appendix expands Fig.~\ref{fig:disruption-regimes} of Sec.~\ref{subsec:disc disruption} into a full sweep of the IMBH semimajor axis against eccentricity and inclination. As in the main text, all three precession rates are computed from the exact even-Legendre torque series ($\ell_{\max}=120$) using the disruption criterion of Eq.~\eqref{eqn:fundamental criterion}. Each cell shows the intact-fragment region -- where $\Omega_{i,\mathrm{diff}}<|\mathbf{\Omega}_{i,\mathrm{disc}}\times\hat\L_i|<|\mathbf{\Omega}_{i,\IMBH}\times\hat\L_i|$ (orange) -- as a function of disc radius and IMBH-to-disc mass ratio, between the intact-disc-region (grey) below the boundary $|\mathbf{\Omega}_{i,\mathrm{disc}}\times\hat\L_i|=|\mathbf{\Omega}_{i,\IMBH}\times\hat\L_i|$ (solid) and the dispersed region (purple) above $|\mathbf{\Omega}_{i,\mathrm{disc}}\times\hat\L_i|=\Omega_{i,\mathrm{diff}}$ (dashed). Within each cell the boundaries are also drawn for every disc slope $\Gamma\in[0.75,2.2]$, colour-coded by $\Gamma$, with the cross-hatch marking where a intact fragment forms for all $\Gamma$.

\begin{figure*}
    \centering
    \includegraphics[width=\textwidth]{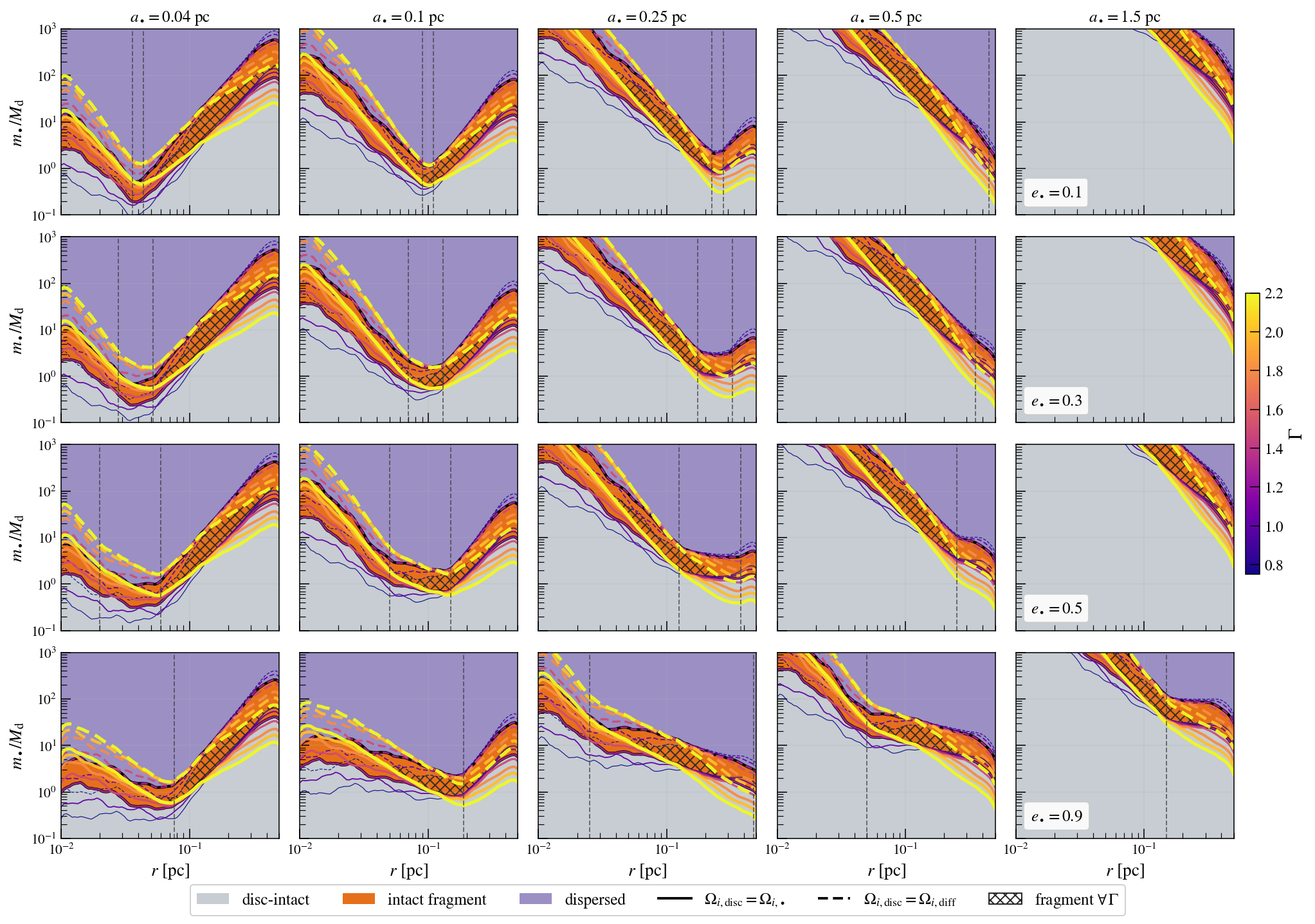}
    \caption{Same as Figure~\ref{fig:disruption-regimes} showing the disruption regimes in the $(r,m_\IMBH/M_{\rm d})$ plane for a grid of IMBH semimajor axes (columns: $a_\IMBH=0.04,0.1,0.25,0.5,1.5$ pc) and eccentricities (rows: $e_\IMBH=0.1,0.3,0.5,0.9$), at fixed inclination $\theta_\IMBH=135^\circ$. In each cell, the shaded regions are computed for the corresponding values of $a_\IMBH$ and $e_\IMBH$ at the fiducial surface-density slope $\Gamma=1.3$. In the grey region the disc remains intact, the orange region is the intact-fragment regime, where $\Omega_{i,\mathrm{diff}}<|\mathbf{\Omega}_{i,\mathrm{disc}}\times\hat\L_i|<|\mathbf{\Omega}_{i,\IMBH}\times\hat\L_i|$, and the purple region is dispersed. The black curves bound the fiducial shaded regions and show $|\mathbf{\Omega}_{i,\mathrm{disc}}\times\hat\L_i|=|\mathbf{\Omega}_{i,\IMBH}\times\hat\L_i|$ (solid) and $|\mathbf{\Omega}_{i,\mathrm{disc}}\times\hat\L_i|=\Omega_{i,\mathrm{diff}}$ (dashed). The colour-coded curves repeat these same two boundaries for $\Gamma\in[0.75,2.2]$, showing how the disruption regimes shift as $\Gamma$ is changed. The cross-hatched regions mark where an intact fragment forms for all values of $\Gamma$ shown, i.e. the intersection of the intact-fragment regions over the $\Gamma$ sweep, similar to Fig.~\ref{fig:disruption-regimes}. The vertical dashed lines show the pericentre and apocentre of the IMBH orbit.}
    \label{fig:regime-grid-e}
\end{figure*}

\begin{figure*}
    \centering
    \includegraphics[width=\textwidth]{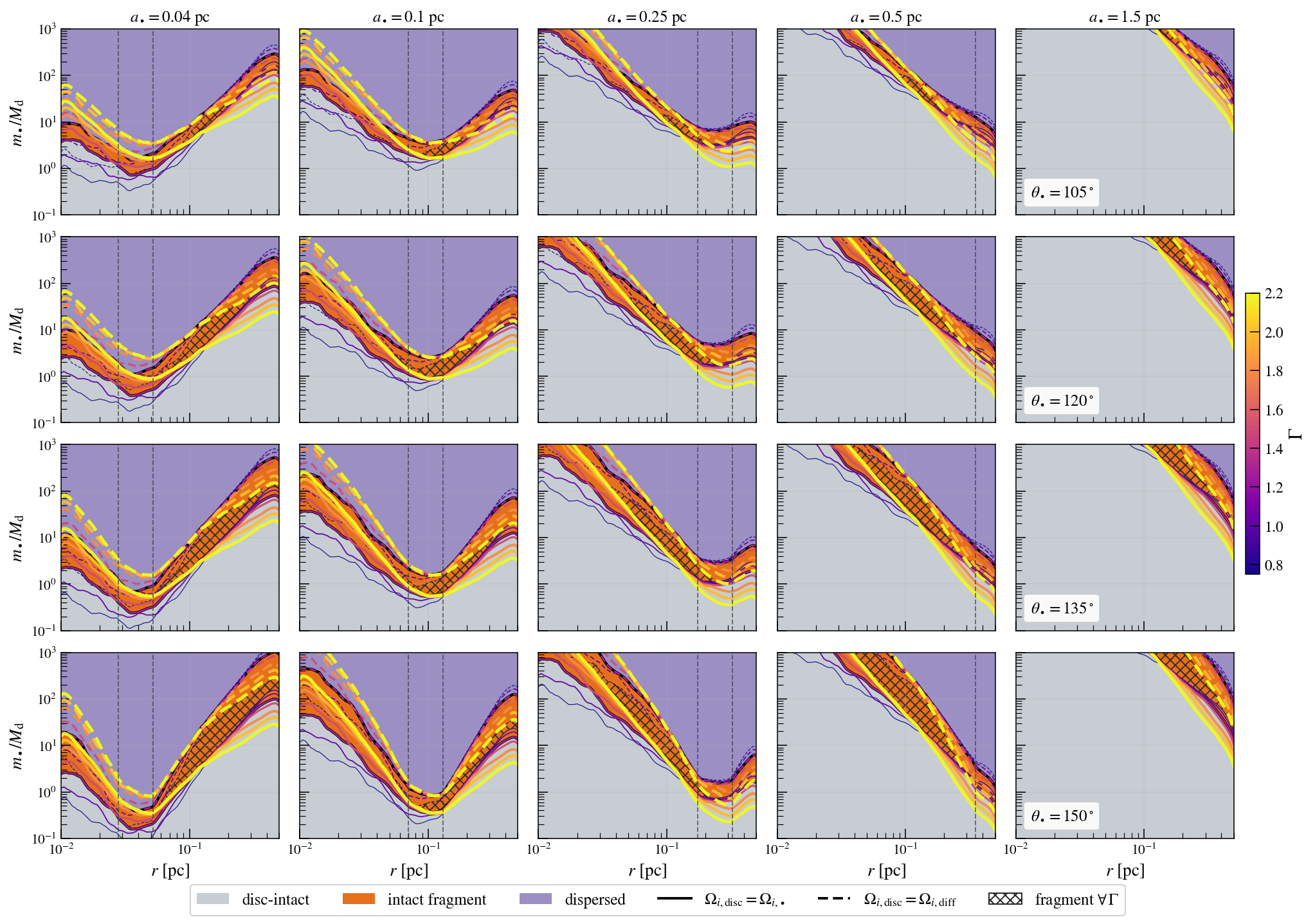}
    \caption{As Fig.~\ref{fig:regime-grid-e}, but for a grid of IMBH semimajor axes (columns: $a_\IMBH=0.04,0.1,0.25,0.5,1.5$ pc) and inclinations (rows: $\theta_\IMBH=105^\circ,120^\circ,135^\circ,150^\circ$), at fixed eccentricity $e_\IMBH=0.3$. In each cell, the shaded regions are computed for the corresponding values of $a_\IMBH$ and $\theta_\IMBH$ at the fiducial surface-density slope $\Gamma=1.3$. The grey, orange, and purple regions show the globally intact disc, the intact-fragment, and the dispersed regimes, respectively. The colour-coded curves show how the same two regime boundaries shift for $\Gamma\in[0.75,2.2]$, and the cross-hatched regions mark where an intact fragment forms for all values of $\Gamma$ shown.}
    \label{fig:regime-grid-i}
\end{figure*}

In both figures the columns step through the IMBH semimajor axis $a_\IMBH\in\{0.04,0.1,0.25,0.5,1.5\}$ pc. Figure~\ref{fig:regime-grid-e} fixes the inclination at $\theta_\IMBH=135^\circ$ and steps the rows through eccentricity $e_\IMBH\in\{0.1,0.3,0.5,0.9\}$; Figure~\ref{fig:regime-grid-i} fixes $e_\IMBH=0.3$ and steps the rows through inclination $\theta_\IMBH\in\{105^\circ,120^\circ,135^\circ,150^\circ\}$. Across a row the intact-fragment region tracks the IMBH orbit, moving outward and narrowing as $a_\IMBH$ grows; down a column it shrinks and moves to higher mass ratios as the eccentricity increases or the orbit becomes more nearly polar, and is widest near anti-coplanar. Full dispersal at every radius requires an IMBH far more massive than the disc across almost the whole grid.

\clearpage

\clearpage

\bsp	
\label{lastpage}
\end{document}